\documentclass[%
 aip,
 jmp,%
 amsmath,amssymb,
 reprint,%
]{revtex4-1}

\usepackage{graphicx}
\usepackage{dcolumn}
\usepackage{bm}
\usepackage[cp1251]{inputenc}
\usepackage{empheq}
\newcommand{\nc}{\newcommand}
\newcommand{\rnc}{\renewcommand}

\nc{\be}{\begin{equation}}
\nc{\ee}{\end{equation}}
\nc{\bse}{\begin{equation*}}
\nc{\ese}{\end{equation*}}
\nc{\ba}{\begin{array}}
\nc{\ea}{\end{array}}
\nc{\bea}{\begin{eqnarray}}
\nc{\eea}{\end{eqnarray}}
\nc{\bi}{\begin{itemize}}
\nc{\ei}{\end{itemize}}
\nc{\bn}{\begin{enumerate}}
\nc{\en}{\end{enumerate}}
\nc{\bt}{\begin{tabular}}
\nc{\et}{\end{tabular}}
\nc{\bb}{\begin{equation}\begin{array}{|c|}\hline } \nc{\eb}{\\
\hline\end{array}\end{equation}}
\nc{\bw}{\begin{widetext}}
\nc{\ew}{\end{widetext}}
\nc{\disp}{\displaystyle}

\newcommand{\lb}{\left[}
\newcommand{\rb}{\right]}
\newcommand{\lp}{\left(}
\newcommand{\rp}{\right)}
\newcommand{\lf}{\left\{}
\newcommand{\rf}{\right\}}

\newcommand{\rv}{\right|}
\newcommand{\ld}{\left.}
\newcommand{\rd}{\right.}

\nc{\Def}{\stackrel{def}{=}} \nc{\sign}{\mathrm{sign}\;}
\nc{\diag}{\mathrm{diag}\;}
\nc{\eq}{\equiv} \nc{\we}{\wedge} \nc{\ra}{\rightarrow}
\nc{\bfrac}{\disp\frac}
\nc{\bdel}{\bar{\del}}

\nc{\bfpa}{{\bm{\partial}}}          
\nc{\pa}[1]{{\partial_{#1}}{}}       
\nc{\pau}[1]{{\partial^{#1}}{}}      
\nc{\alp}{\alpha}
\nc{\bet}{\beta}
\nc{\gam}{\gamma}
\nc{\del}{\delta}
\nc{\eps}{\epsilon}
\nc{\veps}{\varepsilon}
\nc{\zet}{\zeta}
\nc{\tet}{\theta}
\nc{\vtet}{\vartheta}
\nc{\iot}{\iota}
\nc{\kap}{\kappa}
\nc{\lam}{\lambda}
\nc{\vpi}{\varpi}
\nc{\vrho}{\varrho}
\nc{\sig}{\sigma}
\nc{\vsig}{\varsigma}
\nc{\ups}{\upsilon}
\nc{\vphi}{\varphi}
\nc{\ome}{\omega}

\nc{\Gam}{\Gamma}
\nc{\Del}{\Delta}
\nc{\Tet}{\Theta}
\nc{\Lam}{\Lambda}
\nc{\Sig}{\Sigma}
\nc{\Ups}{\Upsilon}
\nc{\Ome}{\Omega}

\nc{\BL}[1]{\mathbf{#1}}

\nc{\bfa}{\BL{a}} \nc{\bfb}{\BL{b}} \nc{\bfc}{\BL{c}}
\nc{\bfd}{\BL{d}} \nc{\bfe}{\BL{e}} \nc{\bff}{\BL{f}}
\nc{\bfg}{\BL{g}} \nc{\bfh}{\BL{h}} \nc{\bfi}{\BL{i}}
\nc{\bfj}{\BL{j}} \nc{\bfk}{\BL{k}} \nc{\bfl}{\BL{l}}
\nc{\bfm}{\BL{m}} \nc{\bfn}{\BL{n}} \nc{\bfo}{\BL{o}}
\nc{\bfp}{\BL{p}} \nc{\bfq}{\BL{q}} \nc{\bfr}{\BL{r}}
\nc{\bfs}{\BL{s}} \nc{\bft}{\BL{t}} \nc{\bfu}{\BL{u}}
\nc{\bfv}{\BL{v}} \nc{\bfw}{\BL{w}} \nc{\bfx}{\BL{x}}
\nc{\bfy}{\BL{y}} \nc{\bfz}{\BL{z}}

\nc{\bfA}{\BL{A}}
\nc{\bfB}{\BL{B}}
\nc{\bfC}{\BL{C}}
\nc{\bfD}{\BL{D}}
\nc{\bfE}{\BL{E}}
\nc{\bfF}{\BL{F}}
\nc{\bfG}{\BL{G}}
\nc{\bfH}{\BL{H}}
\nc{\bfI}{\BL{I}}
\nc{\bfJ}{\BL{J}}
\nc{\bfK}{\BL{K}}
\nc{\bfL}{\BL{L}}
\nc{\bfM}{\BL{M}}
\nc{\bfN}{\BL{N}}
\nc{\bfO}{\BL{O}}
\nc{\bfP}{\BL{P}}
\nc{\bfQ}{\BL{Q}}
\nc{\bfR}{\BL{R}}
\nc{\bfS}{\BL{S}}
\nc{\bfT}{\BL{T}}
\nc{\bfU}{\BL{U}}
\nc{\bfV}{\BL{V}}
\nc{\bfW}{\BL{W}}
\nc{\bfX}{\BL{X}}
\nc{\bfY}{\BL{Y}}
\nc{\bfZ}{\BL{Z}}

\nc{\bfalp}{\bm{\alp}}
\nc{\bfbet}{\bm{\bet}}
\nc{\bfgam}{\bm{\gam}}
\nc{\bfdel}{\bm{\del}}
\nc{\bfeps}{\bm{\eps}}
\nc{\bfveps}{\bm{\veps}}
\nc{\bfzet}{{\bm{\zet}}}
\nc{\bfeta}{\bm{\eta}}
\nc{\bftet}{\bm{\tet}}
\nc{\bfvtet}{\bm{\vtet}}
\nc{\bfiot}{\bm{\iot}}
\nc{\bfkap}{\bm{\kap}}
\nc{\bflam}{\bm{\lam}}
\nc{\bfmu}{\bm{\mu}}
\nc{\bfnu}{\bm{\nu}}
\nc{\bfxi}{\bm{\xi}}
\nc{\bfpi}{\bm{\pi}}
\nc{\bfvpi}{\bm{\vpi}}
\nc{\bfrho}{\bm{\rho}}
\nc{\bfvrho}{\bm{\vrho}}
\nc{\bfsig}{\bm{\sig}}
\nc{\bfvsig}{\bm{\sig}}
\nc{\bftau}{\bm{\tau}}
\nc{\bfups}{\bm{\ups}}
\nc{\bfphi}{\bm{\phi}}
\nc{\bfvphi}{\bm{\vphi}}
\nc{\bfchi}{\bm{\chi}}
\nc{\bfpsi}{\bm{\psi}}
\nc{\bfome}{\bm{\ome}}

\nc{\bfGam}{\bm{\Gam}}
\nc{\bfDel}{\bm{\Del}}
\nc{\bfTet}{\bm{\Tet}}
\nc{\bfLam}{\bm{\Lam}}
\nc{\bfXi}{\bm{\Xi}}
\nc{\bfPi}{\bm{\Pi}}
\nc{\bfSig}{\bm{\Sig}}
\nc{\bfUps}{\bm{\Ups}}
\nc{\bfPhi}{\bm{\Phi}}
\nc{\bfPsi}{\bm{\Psi}}
\nc{\bfOme}{\bm{\Ome}}

\DeclareFontFamily{OT1}{rsfs}{}
\DeclareFontShape{OT1}{rsfs}{m}{n}{<5> rsfs5 <7> rsfs7 <10>
rsfs10}{} \DeclareSymbolFont{mathrsfs}{OT1}{rsfs}{m}{n}
\DeclareSymbolFontAlphabet{\mathrsfs}{mathrsfs}

\nc{\rsA}{\mathrsfs{A}}
\nc{\rsB}{\mathrsfs{B}}
\nc{\rsC}{\mathrsfs{C}}
\nc{\rsD}{\mathrsfs{D}}
\nc{\rsE}{\mathrsfs{E}}
\nc{\rsF}{\mathrsfs{F}}
\nc{\rsG}{\mathrsfs{G}}
\nc{\rsH}{\mathrsfs{H}}
\nc{\rsI}{\mathrsfs{I}}
\nc{\rsJ}{\mathrsfs{J}}
\nc{\rsK}{\mathrsfs{K}}
\nc{\rsL}{\mathrsfs{L}}
\nc{\rsM}{\mathrsfs{M}}
\nc{\rsN}{\mathrsfs{N}}
\nc{\rsO}{\mathrsfs{O}}
\nc{\rsP}{\mathrsfs{P}}
\nc{\rsQ}{\mathrsfs{Q}}
\nc{\rsR}{\mathrsfs{R}}
\nc{\rsS}{\mathrsfs{S}}
\nc{\rsT}{\mathrsfs{T}}
\nc{\rsU}{\mathrsfs{U}}
\nc{\rsV}{\mathrsfs{V}}
\nc{\rsW}{\mathrsfs{W}}
\nc{\rsX}{\mathrsfs{X}}
\nc{\rsY}{\mathrsfs{Y}}
\nc{\rsZ}{\mathrsfs{Z}}

\nc{\CA}[1]{\mathcal{#1}}
\nc{\caA}{\CA{A}}
\nc{\caB}{\CA{B}}
\nc{\caC}{\CA{C}}
\nc{\caD}{\CA{D}}
\nc{\caE}{\CA{E}}
\nc{\caF}{\CA{F}}
\nc{\caG}{\CA{G}}
\nc{\caH}{\CA{H}}
\nc{\caI}{\CA{I}}
\nc{\caJ}{\CA{J}}
\nc{\caK}{\CA{K}}
\nc{\caL}{\CA{L}}
\nc{\caM}{\CA{M}}
\nc{\caN}{\CA{N}}
\nc{\caO}{\CA{O}}
\nc{\caP}{\CA{P}}
\nc{\caQ}{\CA{Q}}
\nc{\caR}{\CA{R}}
\nc{\caS}{\CA{S}}
\nc{\caT}{\CA{T}}
\nc{\caU}{\CA{U}}
\nc{\caV}{\CA{V}}
\nc{\caW}{\CA{W}}
\nc{\caX}{\CA{X}}
\nc{\caY}{\CA{Y}}
\nc{\caZ}{\CA{Z}}

\nc{\lag}{\rsL}      
\nc{\bflag}{{\bf\rsL}} 
\nc{\lagG}{\lag^G}   
\nc{\lagM}{\lag^M}   

\nc{\kro}[2]{\del^{#1}_{#2}}

\nc{\bfmet}{\bfg}                     
\nc{\met}[2]{g_{#1 #2}}               
\nc{\metu}[2]{g^{#1 #2}}              
\nc{\rmet}{\sqrt{-g}}                 
\nc{\bfcon}{\bfGam}                   
\nc{\con}[3]{\Gam^{#1}{}_{#2 #3}}     
\nc{\cond}[3]{\Gam_{#1,\, #2 #3}}        
\nc{\bftor}{\bfT}                     
\nc{\tor}[3]{T^{#1}{}_{#2 #3}}        
\nc{\tord}[3]{T_{#1,\, #2 #3}}        
\nc{\toru}[3]{T^{#1,\, #2 #3}}        
\nc{\bfsT}{{\stackrel{*}{\mathbf{T}}}{}}
\nc{\bfstor}{\bfsT}                   
\nc{\sT}{{\stackrel{*}{T}}{}}         
\nc{\stor}[3]{\sT^{#1}{}_{#2 #3}}     
\nc{\stord}[3]{\sT_{#1,\, #2 #3}}     
\nc{\storu}[3]{\sT^{#1,\, #2 #3}}     
\nc{\bfctor}{\bfK}                    
\nc{\ctor}[3]{K^{#1}{}_{#2 #3}}       
\nc{\ctord}[3]{K_{#1,\, #2 #3}}       
\nc{\ctoru}[3]{K^{#1,\, #2 #3}}       
\nc{\bfcur}{\bfR}                     
\nc{\cur}[4]{R^{#1}{}_{#2 #3 #4}}     
\nc{\curd}[4]{R_{#1 #2 #3 #4}}        
\nc{\curud}[4]{R^{#1 #2}{}_{#3 #4}}   
\nc{\ric}[2]{R_{#1 #2}}               
\nc{\ricu}[2]{R^{#1 #2}}              
\nc{\bfein}{\bfE}                     
\nc{\ein}[2]{E_{#1 #2}}               
\nc{\einu}[2]{E^{#1 #2}}              
\nc{\einud}[2]{E^{#1}{}_{#2}}         
\nc{\bfna}{{\bm{\nabla}}}             
\nc{\na}[1]{{\nabla_{#1}}{}}          
\nc{\nau}[1]{{\nabla^{#1}}{}}         
\nc{\bfsna}{{\stackrel{*}{{\bm{\nabla}}}}{}}           
\nc{\sna}[1]{{\stackrel{*}{\nabla}_{#1}}{}}            
\nc{\snau}[1]{{\stackrel{*}{\nabla}{}^{#1}}{}}         
\nc{\bfdome}{\bfd\bfome\,}                 
\nc{\dome}{dx\,}                           
\nc{\intOme}{\int\limits_{\Ome}}           
\nc{\intb}{\int\limits_{\Sig_1}^{\Sig_2}}  
\nc{\bfdsig}[1]{\bfd\bfsig_{#1}\,}         
\nc{\dsig}[1]{d\sig_{#1}\,}                
\nc{\intdOme}{\oint\limits_{\partial\Ome}} 
\nc{\intSig}{\int\limits_{\Sig}}           
\nc{\intSiga}{\int\limits_{\Sig_1}}        
\nc{\intSigb}{\int\limits_{\Sig_2}}        
\nc{\bfds}[2]{\bfd\bfs_{#1 #2}\,}          
\nc{\ds}[2]{ds_{#1 #2}\,}                  
\nc{\intdSig}{\oint\limits_{\partial\Sig}} 

\nc{\bfdx}[1]{\bfd\bfx^{#1}}          

\nc{\bfgfi}{\bfPhi}                   
\nc{\gfi}[1]{\Phi^{#1}}               
\nc{\gfiA}{\Phi^A}                    
\nc{\gfiB}{\Phi^B}                    
\nc{\bfjfi}{\bfphi}                   
\nc{\jfi}[1]{\phi^{#1}}               
\nc{\jfiA}{\jfi{a}}                    
\nc{\jfiB}{\jfi{b}}                    
\nc{\bfffi}{\bfvphi}                   
\nc{\ffi}[1]{\vphi^{#1}}               
\nc{\ffiA}{\ffi{a}}                    
\nc{\ffiB}{\ffi{b}}                    
\nc{\bfpara}{\bfxi}              
\nc{\para}[1]{\xi^{#1}}          
\nc{\dbfpara}{\del\bfpara}              
\nc{\dpara}[1]{\del\para{#1}}          
\nc{\bfparK}{\bfchi}              
\nc{\parK}[1]{\chi^{#1}}          
\nc{\bfparKa}{\bfchi_{(a)}}              
\nc{\parKa}[1]{\chi_{(a)}{}^{#1}}          
\nc{\dbfparK}{\del\bfparK}         
\nc{\dx}[1]{\del x^{#1}}
\nc{\dbfLam}{\del\bfLam}
\nc{\dbfLamdpara}{\dbfLam[\dbfpara]}
\nc{\dLam}[1]{\del\Lam^{#1}}
\nc{\dSig}{\del\Sig}
\nc{\dIdgfi}[1]{\bfrac{\del I}{\del\gfi{#1}}}
\nc{\dIdgfiA}{\dIdgfi{A}}
\nc{\DIDgfi}[1]{\bfrac{\Del I}{\Del\gfi{#1}}}
\nc{\DIDgfiA}{\DIDgfi{A}}
\nc{\DIDjfi}[1]{\bfrac{\Del I}{\Del\jfi{#1}}}
\nc{\DIDjfiA}{\DIDjfi{a}}
\nc{\DsIDjfi}[1]{\bfrac{\Del^* I}{\Del\jfi{#1}}}
\nc{\DsIDjfiA}{\DsIDjfi{a}}
\nc{\DIDnajfi}[2]{\bfrac{\Del I}{\Del(\na{#1}\jfi{#2})}}
\nc{\DIDnajfiA}[1]{\DIDnajfi{#1}{a}}
\nc{\DsIDnajfi}[2]{\bfrac{\Del^* I}{\Del(\na{#1}\jfi{#2})}}
\nc{\DsIDnajfiA}[1]{\DsIDnajfi{#1}{a}}
\nc{\DIDffi}[1]{\bfrac{\Del I}{\Del\ffi{#1}}}
\nc{\DIDffiA}{\DIDffi{a}}
\nc{\DIMDffi}[1]{\bfrac{\Del I^M}{\Del\ffi{#1}}}
\nc{\DIMDffiA}{\DIMDffi{a}}
\nc{\DIDnaffi}[2]{\bfrac{\Del I}{\Del(\na{#1}\ffi{#2})}}
\nc{\DIDnaffiA}[1]{\DIDnaffi{#1}{a}}
\nc{\DIDmet}[2]{\bfrac{\Del I}{\Del \met{#1}{#2}}}
\nc{\DIMDmet}[2]{\bfrac{\Del I^M}{\Del \met{#1}{#2}}}
\nc{\DIGDmet}[2]{\bfrac{\Del I^G}{\Del \met{#1}{#2}}}
\nc{\DIDtor}[3]{\bfrac{\Del I}{\Del \tor{#1}{#2}{#3}}}
\nc{\DsIDtor}[3]{\bfrac{\Del^* I}{\Del \tor{#1}{#2}{#3}}}
\nc{\DIDnator}[4]{\bfrac{\Del I}{\Del(\na{#1}\tor{#2}{#3}{#4})}}
\nc{\DsIDnator}[4]{\bfrac{\Del^* I}{\Del(\na{#1}\tor{#2}{#3}{#4})}}
\nc{\DIMDtor}[3]{\bfrac{\Del I^M}{\Del \tor{#1}{#2}{#3}}}
\nc{\DIGDtor}[3]{\bfrac{\Del I^G}{\Del \tor{#1}{#2}{#3}}}
\nc{\DIDcon}[3]{\bfrac{\Del I}{\Del\con{#1}{#2}{#3}}}
\nc{\K}{K}
\nc{\Kg}[2]{\K^{#1}|_{#2}} \nc{\KgA}[1]{\Kg{#1}{A}}
\nc{\KgB}[1]{\Kg{#1}{B}}
\nc{\Kj}[2]{\K^{#1}|_{#2}} \nc{\KjA}[1]{\Kj{#1}{a}}
\nc{\KjB}[1]{\Kj{#1}{b}}
\nc{\sKj}[2]{{}^*\K^{#1}|_{#2}} \nc{\sKjA}[1]{\sKj{#1}{a}}
\nc{\sKjB}[1]{\sKj{#1}{b}}
\nc{\Kf}[2]{\K^{#1}|_{#2}} \nc{\KfA}[1]{\Kf{#1}{a}}
\nc{\KfB}[1]{\Kf{#1}{b}}
\nc{\Km}[3]{\K^{#1}|^{#2 #3}} \nc{\Kt}[4]{\K^{#1}|_{#2}{}^{#3 #4}}
\nc{\sKt}[4]{{}^*\K^{#1}|_{#2}{}^{#3 #4}}
\rnc{\L}{L}
\nc{\Lg}[3]{\L^{#1 #2}|_{#3}} \nc{\LgA}[2]{\Lg{#1}{#2}{A}}
\nc{\LgB}[2]{\Lg{#1}{#2}{B}}
\nc{\Lj}[3]{\L^{#1 #2}|_{#3}} \nc{\LjA}[2]{\Lj{#1}{#2}{a}}
\nc{\LjB}[2]{\Lj{#1}{#2}{b}}
\nc{\Lf}[3]{\L^{#1 #2}|_{#3}} \nc{\LfA}[2]{\Lf{#1}{#2}{a}}
\nc{\LfB}[2]{\Lf{#1}{#2}{b}}
\nc{\Lm}[4]{\L^{#1 #2}|^{#3 #4}} \nc{\Lt}[5]{\L^{#1 #2}|_{#3}{}^{#4
#5}}
\nc{\gfia}[2]{\gfi{}_{#1}|^{#2}} \nc{\gfiaA}[1]{\gfia{#1}{A}}
\nc{\gfib}[3]{\gfi{}_{#1}{}^{#2}|^{#3}}
\nc{\gfibA}[2]{\gfib{#1}{#2}{A}} \nc{\gfic}[4]{\gfi{}_{#1}{}^{#2
#3}|^{#4}} \nc{\gficA}[3]{\gfic{#1}{#2}{#3}{A}}
\nc{\jfia}[2]{\jfi{}_{#1}|^{#2}} \nc{\jfiaA}[1]{\jfia{#1}{a}}
\nc{\jfib}[3]{\jfi{}_{#1}{}^{#2}|^{#3}}
\nc{\jfibA}[2]{\jfib{#1}{#2}{a}} \nc{\jfic}[4]{\jfi{}_{#1}{}^{#2
#3}|^{#4}} \nc{\jficA}[3]{\jfic{#1}{#2}{#3}{a}}
\nc{\ffia}[2]{\ffi{}_{#1}|^{#2}} \nc{\ffiaA}[1]{\ffia{#1}{a}}
\nc{\ffib}[3]{\ffi{}_{#1}{}^{#2}|^{#3}}
\nc{\ffibA}[2]{\ffib{#1}{#2}{a}} \nc{\ffic}[4]{\ffi{}_{#1}{}^{#2
#3}|^{#4}} \nc{\fficA}[3]{\ffic{#1}{#2}{#3}{a}}

\nc{\bfJpara}{\bfJ[\bfpara]}
\nc{\Jpara}[1]{J^{#1}[\bfpara]}
\nc{\bfJdpara}{\bfJ[\dbfpara]}
\nc{\Jdpara}[1]{J^{#1}[\dbfpara]}
\nc{\bfJdparK}{\bfJ[\dbfparK]}
\nc{\JdparK}[1]{J^{#1}[\dbfparK]}

\nc{\QdxiSig}{Q[\dbfpara;\Sig]}
\nc{\U}[2]{U_{#1}{}^{#2}} \nc{\Uu}[2]{U^{#1 #2}} \nc{\Ud}[2]{U_{#1
#2}} 
\nc{\M}[3]{M_{#1}{}^{#2 #3}} \nc{\Mu}[3]{M^{#1 #2 #3}}
\nc{\Md}[3]{M_{#1 #2 #3}}
\nc{\N}[4]{N_{#1}{}^{#2 #3 #4}} \nc{\Nu}[4]{N^{#1 #2 #3 #4}}
\nc{\Nud}[4]{N^{#1 #2 #3}{}_{#4}}
\nc{\Ia}[1]{I_{#1}} \nc{\Ib}[2]{I_{#1}{}^{#2}}
\nc{\bfpot}{\bftet}
\nc{\bfpotpara}{\bftet[\bfpara]} \nc{\potpara}[2]{\tet^{#1
#2}[\bfpara]}
\nc{\bfpotdpara}{\bftet[\dbfpara]} \nc{\potdpara}[2]{\tet^{#1
#2}[\dbfpara]}
\nc{\bfpotpdpara}{\bftet'[\dbfpara]} \nc{\potpdpara}[2]{\tet^{'#1
#2}[\dbfpara]}
\nc{\pota}[2]{\tet^{#1 #2}} \nc{\potb}[3]{\tet_{#1}{}^{#2 #3}}
\nc{\potbu}[3]{\tet^{#1 #2 #3}} \nc{\potc}[4]{\tet_{#1}{}^{#2 #3
#4}}
\nc{\ppota}[2]{\tet'^{#1 #2}} \nc{\dpota}[2]{\Del\tet^{#1 #2}}
\nc{\rsJdpara}[1]{\rsJ^{#1}[\dbfpara]}          
\nc{\bfrsJ}{{\bm{\rsJ}}}
\nc{\bfrsJdpara}{\bfrsJ[\dbfpara]}
\nc{\bfrsJpara}{\bfrsJ[\bfpara]}
\nc{\bfJsdpara}{\stackrel{sym}{\bfJ}[\dbfpara]}          
\nc{\Jsdpara}[1]{\stackrel{sym}{J}{}^{#1}[\dbfpara]}          
\nc{\bfrsB}{{\bm{\rsB}}}
\nc{\bfpotBdpara}{\bfrsB[\dbfpara]} \nc{\potBdpara}[2]{\rsB^{#1
#2}[\dbfpara]}
\nc{\bfpotsdpara}{\stackrel{sym}{\bftet}[\dbfpara]}
\nc{\potsdpara}[2]{\stackrel{sym}{\tet^{#1 #2}}[\dbfpara]}
\nc{\bfUs}{\stackrel{sym}{\bfU}}
\nc{\Us}[2]{\stackrel{sym}{U}_{#1}{}^{#2}}
\nc{\bfJsdparK}{\stackrel{sym}{\bfJ}[\dbfparK]}          
\nc{\JsdparK}[1]{\stackrel{sym}{J^{#1}}[\dbfparK]}          

\nc{\A}[3]{A_{#1}{}^{#2 #3}} \nc{\B}[4]{B_{#1}{}^{#2 #3 #4}}
\nc{\C}[4]{C_{#1}{}^{#2 #3 #4}} 
\nc{\youtaa}[2]{\ba{|c|c|}\hline #1 & #2\\ \hline \ea}
\nc{\youtab}[2]{\ba{|c|}\hline 1 \\ \hline 2\\ \hline \ea}
\nc{\youtba}[3]{\ba{|c|c|c|}\hline #1 & #2 & #3\\ \hline \ea}
\nc{\youtbb}[3]{\ba{|c|c|c}\hline #1 & #2\\ \hline #3\\ \cline{1-1}
\ea} \nc{\youtbc}[3]{\ba{|c|}\hline #1 \\ \hline #2\\ \hline #3\\
\hline \ea}
\nc{\yous}[1]{\hat{s}\lp\; #1 \;\rp} \nc{\youa}[1]{\hat{a}\lp\; #1
\;\rp}
\nc{\Na}[4]{a_{#1}{}^{#2 #3 #4}} \nc{\bfNb}{\bfb}
\nc{\Nb}[4]{b_{#1}{}^{#2 #3 #4}} \nc{\bfNc}{\bfc}
\nc{\Nc}[4]{c_{#1}{}^{#2 #3 #4}} \nc{\Nd}[4]{d_{#1}{}^{#2 #3 #4}}
\nc{\krob}[4]{\del^{{#1} {#2}}_{{#3} {#4}}}
\nc{\Dd}[6]{\Del^{#1 #2 #3}_{\underline{#4 #5 #6}}}
\nc{\Du}[6]{\Del^{\overline{#1 #2 #3}}_{#4 #5 #6}}
\nc{\dbfgfi}{\del\bfgfi}                 
\nc{\dgfi}[1]{\del\gfi{#1}}              
\nc{\dgfiA}{\dgfi{A}}                    
\nc{\dbfjfi}{\del\bfjfi}                 
\nc{\djfi}[1]{\del\jfi{#1}}              
\nc{\djfiA}{\djfi{a}}                    
\nc{\djfiB}{\djfi{b}}                    
\nc{\dbfffi}{\del\bfffi}                 
\nc{\dffi}[1]{\del\ffi{#1}}              
\nc{\dffiA}{\dffi{a}}                    
\nc{\dffib}{\dffi{b}}                    
\nc{\dmet}[2]{\del\met{#1}{#2}}          
\nc{\dmetu}[2]{\del\metu{#1}{#2}}        
\nc{\drmet}{\del\rmet}                   
\nc{\dlag}{\del\lag}                     
\nc{\dcon}[3]{\del\con{#1}{#2}{#3}}      
\nc{\dtor}[3]{\del\tor{#1}{#2}{#3}}      
\nc{\dcur}[4]{\del\cur{#1}{#2}{#3}{#4}}  
\nc{\bdbfgfi}{\bar{\del}\bfPhi}              
\nc{\bdgfiA}{\bar{\del}\Phi^A}               
\nc{\Dbrf}[4]{(\Del^{#1}{}_{#2})\ld^{#3}\rv_{#4}}  
\nc{\DbrfAB}[2]{\Dbrf{#1}{#2}{a}{b}}               
\nc{\Dbrfd}[4]{(\Del_{#1 #2})\ld^{#3}\rv_{#4}}  
\nc{\DbrfdAB}[2]{\Dbrfd{#1}{#2}{a}{b}}               
\nc{\Dbrfu}[4]{(\Del^{#1 #2})\ld^{#3}\rv_{#4}}  
\nc{\DbrfuAB}[2]{\Dbrfu{#1}{#2}{a}{b}}               
\nc{\Dbrj}[4]{(\Del^{#1}{}_{#2})\ld^{#3}\rv_{#4}} 
\nc{\DbrjAB}[2]{\Dbrj{#1}{#2}{a}{b}}              
\nc{\Dbrju}[4]{(\Del^{#1#2})\ld^{#3}\rv_{#4}} 
\nc{\DbrjuAB}[2]{\Dbrju{#1}{#2}{a}{b}}              
\nc{\Dbrjd}[4]{(\Del_{#1#2})\ld^{#3}\rv_{#4}} 
\nc{\DbrjdAB}[2]{\Dbrjd{#1}{#2}{a}{b}}              
\nc{\Dbrg}[4]{(\Del^{#1}{}_{#2})\ld^{#3}\rv_{#4}} 
\nc{\DbrgAB}[2]{\Dbrg{#1}{#2}{A}{B}}              
\nc{\Dbrm}[6]{(\Del^{#1}{}_{#2})\ld_{#3 #4}\rv^{#5 #6}} 
\nc{\Dbrt}[8]{(\Del^{#1}{}_{#2})\ld^{#3}{}_{#4 #5}\rv_{#6}{}^{#7 #8}\,} 
\nc{\Dbrtd}[8]{(\Del_{#1 #2})\ld^{#3}{}_{#4 #5}\rv_{#6}{}^{#7 #8}\,} 
\nc{\Dbrtu}[8]{(\Del^{#1 #2})\ld^{#3}{}_{#4 #5}\rv_{#6}{}^{#7 #8}\,} 
\nc{\Dbrc}[9]{(\Del^{#1}{}_{#2})\ld^{#3}{}_{#4 #5 #6}\rv_{#7}{}^{#8 #9}{}} 

\nc{\dparabfgfi}{\del_{\xi}\bfPhi}    
\nc{\dparagfiA}{\del_{\xi}\Phi^A}     
\nc{\pax}{\partial x}
\nc{\ten}{P}                          
\nc{\meta}[3]{g_{#1}|_{#2 #3}}                
\nc{\metb}[4]{g_{#1}{}^{#2}|_{#3 #4}}         
\nc{\tora}[4]{T_{#1}|^{#2}{}_{#3 #4}}         
\nc{\torb}[5]{T_{#1}{}^{#2}|^{#3}{}_{#4 #5}}  
\nc{\bfnator}{\bfna\bfT} \nc{\bfnanator}{\bfna\bfna\bfT}
\nc{\bfnaffi}{\bfna\bfvphi} \nc{\bfnanaffi}{\bfna\bfna\bfvphi}
\nc{\bfnajfi}{\bfna\bfphi} \nc{\bfnanajfi}{\bfna\bfna\bfphi}
\nc{\dslagdmet}[2]{\bfrac{\partial^*\lag}{\partial\met{#1}{#2}}}
\nc{\dlagdcur}[4]{\bfrac{\partial\lag}{\partial\cur{#1}{#2}{#3}{#4}}}
\nc{\dslagdtor}[3]{\bfrac{\partial^*\lag}{\partial\tor{#1}{#2}{#3}}}
\nc{\dlagdnator}[4]{\bfrac{\partial\lag}{\partial(\na{#1}
\tor{#2}{#3}{#4})}}
\nc{\dlagdnanator}[5]{\bfrac{\partial\lag}{\partial(\na{#1}\na{#2}
\tor{#3}{#4}{#5})}}
\nc{\dslagdjfi}[1]{\bfrac{\partial^*\lag}{\partial\jfi{#1}}}
\nc{\dslagdjfiA}{\dslagdjfi{a}}
\nc{\dlagdnajfi}[2]{\bfrac{\partial\lag}{\partial(\na{#1}\jfi{#2})}}
\nc{\dlagdnajfiA}[1]{\dlagdnajfi{#1}{a}}
\nc{\dlagdnanajfi}[3]{\bfrac{\partial\lag}{\partial(\na{#1}\na{#2}\jfi{#3})}}
\nc{\dlagdnanajfiA}[2]{\dlagdnanajfi{#1}{#2}{a}}
\nc{\dslagdffi}[1]{\bfrac{\partial^*\lag}{\partial\ffi{#1}}}
\nc{\dslagdffiA}{\dslagdffi{a}}
\nc{\dlagdnaffi}[2]{\bfrac{\partial\lag}{\partial(\na{#1}\ffi{#2})}}
\nc{\dlagdnaffiA}[1]{\dlagdnaffi{#1}{a}}
\nc{\dlagdnanaffi}[3]{\bfrac{\partial\lag}{\partial(\na{#1}\na{#2}\ffi{#3})}}
\nc{\dlagdnanaffiA}[2]{\dlagdnanaffi{#1}{#2}{a}}
\nc{\G}[4]{G_{#1}{}^{#2 #3 #4}}          
\nc{\Gu}[4]{G^{#1 #2 #3 #4}}             
\nc{\Gd}[4]{G_{#1 #2}{}^{#3 #4}}          
\nc{\bfsem}{\bft} \nc{\sem}[2]{t^{#1}{}_{#2}} \nc{\semu}[2]{t^{#1
#2}} \nc{\semd}[2]{t_{#1 #2}}
\nc{\bfsems}{\stackrel{sym}{\bft}}
\nc{\sems}[2]{\stackrel{sym}{t}{}^{#1}{}_{#2}}
\nc{\semsu}[2]{\stackrel{sym}{t}{}^{#1 #2}}
\nc{\semsd}[2]{\stackrel{sym}{t}{}_{#1 #2}}
\nc{\bfsemm}{\stackrel{met}{\bft}}
\nc{\semm}[2]{\stackrel{met}{t}{}^{#1}{}_{#2}}
\nc{\semmu}[2]{\stackrel{met}{t}{}^{#1 #2}}
\nc{\semmd}[2]{\stackrel{met}{t}{}_{#1 #2}}
\nc{\bfsema}{\stackrel{add}{\bft}}
\nc{\sema}[2]{\stackrel{add}{t}{}^{#1}{}_{#2}}
\nc{\semau}[2]{\stackrel{add}{t}{}^{#1 #2}}
\nc{\semad}[2]{\stackrel{add}{t}{}_{#1 #2}}
\nc{\bfsemi}{\stackrel{mod}{\bft}}
\nc{\semi}[2]{\stackrel{mod}{t}{}^{#1}{}_{#2}}
\nc{\semiu}[2]{\stackrel{mod}{t}{}^{#1 #2}}
\nc{\semid}[2]{\stackrel{mod}{t}{}_{#1 #2}}
\nc{\bfsemM}{\bfT} \nc{\semM}[2]{T^{#1}{}_{#2}} \nc{\semMu}[2]{T^{#1
#2}} \nc{\semMd}[2]{T_{#1 #2}}
\nc{\bfsemMs}{\stackrel{sym}{\bfT}}
\nc{\semMs}[2]{\stackrel{sym}{T}{}^{#1}{}_{#2}}
\nc{\semMsu}[2]{\stackrel{sym}{T}{}^{#1 #2}}
\nc{\semMsd}[2]{\stackrel{sym}{T}{}_{#1 #2}}
\nc{\bfsemMm}{\stackrel{met}{\bfT}}
\nc{\semMm}[2]{\stackrel{met}{T}{}^{#1}{}_{#2}}
\nc{\semMmu}[2]{\stackrel{met}{T}{}^{#1 #2}}
\nc{\semMmd}[2]{\stackrel{met}{T}{}_{#1 #2}}
\nc{\bfsemMa}{\stackrel{add}{\bfT}}
\nc{\semMa}[2]{\stackrel{add}{T}{}^{#1}{}_{#2}}
\nc{\semMau}[2]{\stackrel{add}{T}{}^{#1 #2}}
\nc{\semMad}[2]{\stackrel{add}{T}{}_{#1 #2}}
\nc{\bfsemMi}{\stackrel{mod}{\bfT}}
\nc{\semMi}[2]{\stackrel{mod}{T}{}^{#1}{}_{#2}}
\nc{\semMiu}[2]{\stackrel{mod}{T}{}^{#1 #2}}
\nc{\semMid}[2]{\stackrel{mod}{T}{}_{#1 #2}}
\nc{\bfspi}{\bfs} \nc{\spi}[3]{s^{#1}{}_{#2 #3}}
\nc{\spiu}[3]{s^{#1,\, #2 #3}} \nc{\spid}[3]{s_{#1,\, #2 #3}}
\nc{\spiud}[3]{s^{#1, #2}{}_{#3}}
\nc{\spir}{{}^{(R)}\spi} \nc{\spiru}{{}^{(R)}\spiu}
\nc{\spird}{{}^{(R)}\spid} \nc{\spirud}{{}^{(R)}\spiud}
\nc{\spij}{{}^{(\jfi{})}\spi} \nc{\spijud}{{}^{(\jfi{})}\spiud}
\nc{\fj}[3]{{}^{(\jfi{})}f^{#1}{}_{#2 #3}}
\nc{\fju}[3]{{}^{(\jfi{})}f^{#1,\, #2 #3}}
\nc{\fjd}[3]{{}^{(\jfi{})}f_{#1,\, #2 #3}}
\nc{\fjud}[3]{{}^{(\jfi{})}f^{#1,\, #2}{}_{#3}}
\nc{\bfspia}{\stackrel{add}{\bfs}}
\nc{\spia}[3]{\stackrel{add}{s}{}^{#1}{}_{#2 #3}}
\nc{\spiau}[3]{\stackrel{add}{s}{}^{#1,\, #2 #3}}
\nc{\spiad}[3]{\stackrel{add}{s}{}_{#1,\, #2 #3}}
\nc{\spiaud}[3]{\stackrel{add}{s}{}^{#1, #2}{}_{#3}}
\nc{\bfspii}{\stackrel{mod}{\bfs}}
\nc{\spii}[3]{\stackrel{mod}{s}{}^{#1}{}_{#2 #3}}
\nc{\spiiu}[3]{\stackrel{mod}{s}{}^{#1,\, #2 #3}}
\nc{\spiid}[3]{\stackrel{mod}{s}{}_{#1,\, #2 #3}}
\nc{\spiiud}[3]{\stackrel{mod}{s}{}^{#1, #2}{}_{#3}}
\nc{\bfspiM}{\bfS} \nc{\spiM}[3]{S^{#1}{}_{#2 #3}}
\nc{\spiMu}[3]{S^{#1,\, #2 #3}} \nc{\spiMd}[3]{S_{#1,\, #2 #3}}
\nc{\spiMud}[3]{S^{#1, #2}{}_{#3}}
\nc{\bfspiMa}{\stackrel{add}{\bfS}}
\nc{\spiMa}[3]{\stackrel{add}{S}{}^{#1}{}_{#2 #3}}
\nc{\spiMau}[3]{\stackrel{add}{S}{}^{#1,\, #2 #3}}
\nc{\spiMad}[3]{\stackrel{add}{S}{}_{#1,\, #2 #3}}
\nc{\spiMaud}[3]{\stackrel{add}{S}{}^{#1, #2}{}_{#3}}
\nc{\bfspiMi}{\stackrel{mod}{\bfS}}
\nc{\spiMi}[3]{\stackrel{mod}{S}{}^{#1}{}_{#2 #3}}
\nc{\spiMiu}[3]{\stackrel{mod}{S}{}^{#1,\, #2 #3}}
\nc{\spiMid}[3]{\stackrel{mod}{S}{}_{#1,\, #2 #3}}
\nc{\spiMiud}[3]{\stackrel{mod}{S}{}^{#1, #2}{}_{#3}}
\nc{\bfbel}{\bfb} \nc{\bel}[3]{b^{#1}{}_{#2 #3}} \nc{\belu}[3]{b^{#1
#2 #3}} \nc{\beld}[3]{b_{#1 #2 #3}} \nc{\belud}[3]{b^{#1 #2}{}_{#3}}
\nc{\bfbela}{\stackrel{add}{\bfbel}}
\nc{\bela}[3]{\stackrel{add}{b}{}^{#1}{}_{#2 #3}}
\nc{\belau}[3]{\stackrel{add}{b}{}^{#1 #2 #3}}
\nc{\belad}[3]{\stackrel{add}{b}{}_{#1 #2 #3}}
\nc{\belaud}[3]{\stackrel{add}{b}{}^{#1 #2}{}_{#3}}
\nc{\bfbeli}{\stackrel{mod}{\bfbel}}
\nc{\beli}[3]{\stackrel{mod}{b}{}^{#1}{}_{#2 #3}}
\nc{\beliu}[3]{\stackrel{mod}{b}{}^{#1 #2 #3}}
\nc{\belid}[3]{\stackrel{mod}{b}{}_{#1 #2 #3}}
\nc{\beliud}[3]{\stackrel{mod}{b}{}^{#1 #2}{}_{#3}}
\nc{\bfbelM}{\bfB} \nc{\belM}[3]{B^{#1}{}_{#2 #3}}
\nc{\belMu}[3]{B^{#1 #2 #3}} \nc{\belMd}[3]{B_{#1 #2 #3}}
\nc{\belMud}[3]{B^{#1 #2}{}_{#3}}
\nc{\bfbelMa}{\stackrel{add}{\bfbelM}}
\nc{\belMa}[3]{\stackrel{add}{B}{}^{#1}{}_{#2 #3}}
\nc{\belMau}[3]{\stackrel{add}{B}{}^{#1 #2 #3}}
\nc{\belMad}[3]{\stackrel{add}{B}{}_{#1 #2 #3}}
\nc{\belMaud}[3]{\stackrel{add}{B}{}^{#1 #2}{}_{#3}}
\nc{\bfbelMi}{\stackrel{mod}{\bfbel}}
\nc{\belMi}[3]{\stackrel{mod}{B}{}^{#1}{}_{#2 #3}}
\nc{\belMiu}[3]{\stackrel{mod}{B}{}^{#1 #2 #3}}
\nc{\belMid}[3]{\stackrel{mod}{B}{}_{#1 #2 #3}}
\nc{\belMiud}[3]{\stackrel{mod}{B}{}^{#1 #2}{}_{#3}}
\nc{\ogG}{\rv_{\lag = \lagG}} \nc{\ogM}{\rv_{\lag = \lagM}}

\nc{\bfCar}{\rsC} \nc{\Car}[3]{\rsC^{#1}{}_{#2 #3}}
\nc{\Caru}[3]{\rsC^{#1 #2 #3}} \nc{\Carud}[3]{\rsC^{#1 #2}{}_{#3}}
\nc{\bfEin}{\rsE} \nc{\Ein}[2]{\rsE^{#1}{}_{#2}}
\nc{\Einu}[2]{\rsE^{#1 #2}} \nc{\Eind}[2]{\rsE_{#1 #2}}

\nc{\bfpaffi}{\bfpa\bfffi} \nc{\bfpapaffi}{\bfpa\bfpa\bfffi}

\nc{\dIdpaffi}[2]{\bfrac{\del I}{\del(\pa{#1}\ffi{#2})}}
\nc{\dIdpaffiA}[1]{\dIdpaffi{#1}{a}}
\nc{\dlagdpaffi}[2]{\bfrac{\partial\lag}{\partial(\pa{#1}\ffi{#2})}}
\nc{\dlagdpaffiA}[1]{\dlagdpaffi{#1}{a}}
\nc{\dlagdpapaffi}[3]{\bfrac{\partial\lag}{\partial(\pa{#1}\pa{#2}\ffi{#3})}}
\nc{\dlagdpapaffiA}[2]{\dlagdpapaffi{#1}{#2}{a}}

\nc{\bfMin}{\bfeta}                      
\nc{\Min}[2]{\eta_{#1 #2}}               
\nc{\Minu}[2]{\eta^{#1 #2}}              

\nc{\bftmet}{\tilde{\bfg}}                      
\nc{\tmet}[2]{\tilde{g}{}_{#1 #2}}               
\nc{\tmetu}[2]{\tilde{g}{}^{#1 #2}}              

\nc{\bftna}{\tilde{{\bm{\nabla}}}}               
\nc{\tna}[1]{{\tilde{\nabla}{}_{#1}}{}}          
\nc{\tnau}[1]{{\tilde{\nabla}{}^{#1}}{}}         

\nc{\DIDtnaffi}[2]{\bfrac{\Del I}{\del(\tna{#1}\ffi{#2})}}
\nc{\DIDtnaffiA}[1]{\DIDtnaffi{#1}{a}}
\nc{\dlagdtnaffi}[2]{\bfrac{\partial\lag}{\partial(\tna{#1}\ffi{#2})}}
\nc{\dlagdtnaffiA}[1]{\dlagdtnaffi{#1}{a}}
\nc{\dlagdtnatnaffi}[3]{\bfrac{\partial\lag}{\partial(\tna{#1}\tna{#2}\ffi{#3})}}
\nc{\dlagdtnatnaffiA}[2]{\dlagdtnatnaffi{#1}{#2}{a}}

\nc{\bftcur}{\tilde{\bfR}}                     
\nc{\tcur}[4]{\tilde{R}{}^{#1}{}_{#2 #3 #4}}     
\nc{\tcurd}[4]{\tilde{R}{}_{#1 #2 #3 #4}}        
\nc{\tcurud}[4]{\tilde{R}{}^{#1 #2}{}_{#3 #4}}   
\nc{\bfttor}{\tilde{\bfT}}                     
\nc{\ttor}[3]{\tilde{T}{}^{#1}{}_{#2 #3}}        
\nc{\ttord}[3]{\tilde{T}{}_{#1,\, #2 #3}}        
\nc{\ttoru}[3]{\tilde{T}{}^{#1,\, #2 #3}}        

\nc{\bfzsem}{{}^{(0)}\bft} \nc{\zsem}[2]{{}^{(0)}t^{#1}{}_{#2}}
\nc{\zsemu}[2]{{}^{(0)}t^{#1 #2}} \nc{\zsemd}[2]{{}^{(0)}t_{#1 #2}}
\nc{\bfosem}{{}^{(1)}\bft} \nc{\osem}[2]{{}^{(1)}t^{#1}{}_{#2}}
\nc{\osemu}[2]{{}^{1)}t^{#1 #2}} \nc{\osemd}[2]{{}^{(1)}t_{#1 #2}}
\nc{\bfzspi}{{}^{(0)}\bfs} \nc{\zspi}[3]{{}^{(0)}s^{#1}{}_{#2 #3}}
\nc{\zspiu}[3]{{}^{(0)}s^{#1,\, #2 #3}}
\nc{\zspid}[3]{{}^{(0)}s_{#1,\, #2 #3}}
\nc{\zspiud}[3]{{}^{(0)}s^{#1, #2}{}_{#3}}
\nc{\bfospi}{{}^{(1)}\bfs} \nc{\ospi}[3]{{}^{(1)}s^{#1}{}_{#2 #3}}
\nc{\ospiu}[3]{{}^{(1)}s^{#1,\, #2 #3}}
\nc{\ospid}[3]{{}^{(1)}s_{#1,\, #2 #3}}
\nc{\ospiud}[3]{{}^{(1)}s^{#1, #2}{}_{#3}}
\nc{\bftspi}{{}^{(2)}\bfs} \nc{\tspi}[3]{{}^{(2)}s^{#1}{}_{#2 #3}}
\nc{\tspiu}[3]{{}^{(2)}s^{#1,\, #2 #3}}
\nc{\tspid}[3]{{}^{(2)}s_{#1,\, #2 #3}}
\nc{\tspiud}[3]{{}^{(2)}s^{#1, #2}{}_{#3}}
\nc{\bfzbel}{{}^{(0)}\bfb} \nc{\zbel}[3]{{}^{(0)}b^{#1}{}_{#2 #3}}
\nc{\zbelu}[3]{{}^{(0)}b^{#1 #2 #3}} \nc{\zbeld}[3]{{}^{(0)}b_{#1 #2
#3}} \nc{\zbelud}[3]{{}^{(0)}b^{#1 #2}{}_{#3}}
\nc{\bfobel}{{}^{(1)}\bfb} \nc{\obel}[3]{{}^{(1)}b^{#1}{}_{#2 #3}}
\nc{\obelu}[3]{{}^{(1)}b^{#1 #2 #3}} \nc{\obeld}[3]{{}^{(1)}b_{#1 #2
#3}} \nc{\obelud}[3]{{}^{(1)}b^{#1 #2}{}_{#3}}

\begin{document}

\preprint{AIP/123-QED}

\title{Covariant Differential Identities and Conservation Laws in Metric-Torsion Theories of Gravitation. I. General Consideration}


\author{Robert R. Lompay}
\affiliation{Department of Physics, Uzhgorod National University,
Voloshyna str., 54, Uzhgorod, 88000, UKRAINE}
\email{rlompay@gmail.com}

\author{Alexander N. Petrov}
\affiliation{Moscow M. V. Lomonosov State University, Sternberg
Astronomical institute, Universitetskii pr., 13, Moscow, 119992,
RUSSIA} \email{alex.petrov55@gmail.com}
\date{\today}

\begin{abstract}
Arbitrary diffeomorphically invariant metric-torsion theories
of gravity are considered. It is assumed that Lagrangians of such
theories contain derivatives of field variables
(tensor densities of arbitrary ranks and weights) up to a second
order only. The generalized Klein-Noether methods for constructing
manifestly covariant identities and conserved quantities are
developed. Manifestly covariant expressions are constructed
 without including auxiliary structures like a background
metric. In the Riemann-Cartan space, the following \emph{manifestly
generally covariant results} are presented: (a) The complete generalized
system of differential identities (the Klein-Noether identities) is
obtained. (b) The generalized currents of three types depending on
an arbitrary vector field displacements are constructed: they are
the canonical Noether current, symmetrized Belinfante current and
identically conserved Hilbert-Bergmann current. In particular, it is
stated that the symmetrized Belinfante current  does not depend on divergences in the Lagrangian.
(c) The generalized boundary Klein theorem (third Noether theorem) is proved. (d) The
construction of the generalized superpotential is presented in
details, and questions related to its ambiguities are analyzed.
\end{abstract}

\maketitle

\section{Introduction}\label{sec_01-00}

Last decades, one can see an unprecedented active development of
alternative theories of gravity, which modify general relativity (GR) in
various ways\cite{Hinterbichler_2012,
Clifton_Ferreira_Padilla_Skordis_2012, Capozziello_DeLaurentis_2011,
Hammond_2002}. Among them there are scalar-tensor theories
\cite{Fujii_Maeda_2004}, the Einstein-Cartan theory
\cite{Hehl_Heyde_Kerlick_Nester_1976}, the Lovelock theory in the
general form \cite{Lovelock_1971} as well as its special cases, such as very popular Einstein-Gauss-Bonnet gravity
\cite{Zwiebach_1985, Boulware_Deser_1985}, metric-affine theories
\cite{Ponomarev_Barvinsky_Obukhov_1985_en, Hehl_McCrea_Mielke_Neeman_1995}, supergravity
\cite{Binetruy_Girardi_Grimm_2001}, $f(R)$-theories
\cite{DeFelice_Tsujikawa_2010}, Chern-Simons modifications of GR
\cite{Alexander_Yunes_2009}, Lovelock-Cartan theories
\cite{Chandia_Zanelli_1998}, topologically massive gravity
\cite{Deser_Jackiw_Templeton_1982_b_rep}, topologically massive
supergravity \cite{Deser_Kay_1983}, new massive gravity
\cite{Bergshoeff_Hohm_Townsend_2009_b}, critical gravity
\cite{Deser_Liu_Lu_Pope_Sisman_Tekin_2011}, chiral gravity
\cite{Li_Song_Strominger_2008}, various topological gauge theories
of gravity and supergravity \cite{Chamseddine_1990,
Stelle_West_1980, Salgado_Cataldo_delCampo_2002,
Troncoso_Zanelli_2000, Zanelli_2008, Zanelli_2012}, etc.

Constructing the conservation laws (CLs) and conserved quantities (CQs) in an arbitrary field theory, including gravitational theories, is a main problem. Many above listed theories, presented in the second order formalism, are the metric-torsion theories. Therefore, there is a demand in \emph{universal} expressions for CLs and CQs. Thus, in the present paper, we consider the metric-torsion theories only. It is the main goal of the current work to construct in a \emph{manifestly generally covariant} form and analyze differential identities and conserved quantities, existing due to a diffeomorphic invariance of metric-torsion theories of gravity in the most general formulation. Besides being self-sufficient, the present paper (Paper~I \cite{Lompay_Petrov_2013_a}) is the first one in a series of the works. In the second work (Paper~II \cite{Lompay_Petrov_2013_b}), we plan to apply the developed here formalism to construct and study the conserved quantities, and to examine the structure of the field equations in metric-torsion theories, which have \emph{manifestly generally covariant Lagrangians} (see the definition below). In the third work (Paper~III \cite{Lompay_Petrov_2013_c}), we plan (1) to analyze a physical and geometrical meaning of conserved quantities (with taking into account surface terms), constructed in the first and second works; (2) to apply the obtained results to study some important solutions in the Lovelock-Cartan gravity and other theories with torsion.

To avoid ambiguities, let us state the definitions utilized hereinafter. We call a theory as \emph{generally covariant} one if
it is invariant with respect to general diffeomorphisms, unlike a
\emph{gauge covariant} theory that is invariant with respect to
internal gauge transformations. Hence, it is clear that for both
of these types of theories conserved currents have a definite
\emph{universal} structure\cite{Deser_1972, Abbott_Deser_1982_b, Julia_Silva_1998, Silva_1999, Barnich_Brandt_2002, Barnich_2003}. Therefore, for the sake of universality
and uniformity of the presentation, we call a theory as a \emph{gauge-invariant
theory} in wide sense if it is invariant under
continuous transformations, parameters of which are \emph{functions}
of spacetime points. Such transformations we call as \emph{gauge
transformations}. On the other hand, the usual gauge theories with an internal gauge group we call the gauge theories of Utiyama-Yang-Mills type. We call an expression as \emph{manifestly generally
covariant} one if it is constructed (as a rule, by contractions)
from explicitly covariant quantities (tensors, spinors, covariant
derivatives), which are transformed in correspondence with linear
homogeneous representations of the diffeomorphism group. Thus, it is
evidently that a \emph{manifestly generally covariant} expression is
a \emph{generally covariant} one. But, the converse is generally not true. For example, pseudotensors can be interpreted as
\emph{generally covariant} quantities because their expressions hold in arbitrary coordinate systems, but they cannot be presented in a
\emph{manifestly generally covariant} form because they are
non-tensorial quantities.

In metric theories of gravity a construction of energy-momentum
tensors and spin tensors of pure gravitational field meets well known obstacles -- ambiguities
appear unavoidably. The reason is the existence of the
equivalence principle. During nearly hundred year history of GR
--- basic metric theory of gravity --- numerous variants of expressions for energy,
momentum and angular momentum of gravitational field were put forth. As a rule,
these expressions are \emph{generally covariant}. However, among
them there are both tensorial expressions and non-tensorial ones
(for example, pseudotensors). The latter are not so desirable,
therefore they or methods of their construction
are usually covariantized (i.e., reconstructed into \emph{manifestly generally
covariant} form). Frequently, such a covariantization is based on
including an auxiliary structure, like a background metric, see, for
example, Refs.~\cite{Ray_1968, Mitskievich_Efremov_Nesterov_1985,
Petrov_2008} and also recent works (Refs.~\cite{Petrov_2011,
Petrov_Lompay_2013}). It is impossible to present more or less
complete bibliography even in GR\footnote{In future we plan to
close this gap in a separate work of the bibliographic
character both for GR and for another theories of gravitation.}, particulary
one can find reviews \cite{Petrov_2008, Szabados_2009, Pitts_Schive_2001_b}.
Significantly less attention was paid to constructing \emph{manifestly generally covariant} CQs, where auxiliary structures are not used. Concerning earlier works, \emph {only} Komar \cite{Komar_1959} has suggested a manifestly generally covariant superpotential in GR that has been modified in Refs.~\cite{Winicour_Tamburino_1965_a, Winicour_Tamburino_1965_b, Winicour_1980} and generalized in Refs.~\cite{Lee_Wald_1990, Wald_1993, Iyer_Wald_1994, Iyer_Wald_1995, Wald_Zoupas_2000, Giachetta_Sardanashvily_1995_c, Giachetta_Sardanashvily_1996}. In the last years, up to our knowledge, \emph{only} manifestly generally covariant charges are constructed in asymptotically anti-de Sitter gravity (see, e.g. Refs.~\cite{Aros_Contreras_Olea_Troncoso_Zanelli_2000_a, Aros_Contreras_Olea_Troncoso_Zanelli_2000_b, Olea_2007, Kofinas_Olea_2007}, and references there in).

One of the main methods for constructing conserved quantities is the
procedure suggested by Noether in 1918 \cite{Noether_1918,
Noether_1918_en_c} (for alternative methods see, e.g., review in Ref.~\cite{Szabados_2009} and references therein). It is well known that
Noether has proved two
general theorems in her seminal work \cite{Noether_1918}. The first theorem states the existence of $r$ currents
$\bfJ_{(a)}$, $a=\overline{1,r}$ conserved on field equations. This follows from the invariance of the action functional under transformations presenting a finite $r$-parameters Lie group, and vice versa. The method to prove the first
theorem gives a recipe for constructing such currents. The second
theorem -- the existence of a set of differential identities between
the left hand sides of the field equations of motion (the Noether
identities) follows from the invariance of the action functional
under the gauge transformations, and vice versa.

It is not widely known that in the same work Noether has proved the statement
that has not been formulated as a separate theorem. However,
sometimes it is called as the third Noether theorem, or the boundary
theorem (see, e.g., Refs.~\cite{Konopleva_Popov_1980_en,
Byers_1998, Brading_Brown_2000, Brading_Brown_2003, Brading_2005}).
In 1915, almost three years prior to Noether, Hilbert in his known work \cite{Hilbert_1915,
Hilbert_1915_en} has constructed the energy-momentum vector
$\bfJpara$ for the system of interacting gravitational and
electromagnetic fields depending on an arbitrary vector field
$\bfpara$. Klein, examinating this current \cite{Klein_1917}, and
little earlier Noether (see comments in Ref.~\cite{Klein_1917}),
have found that the Hilbert current transfers into a divergence of
an antisymmetric tensor $\bfpotpara$ if the field equations of
motion hold. Thus, the current is conserved identically. Therefore, according to Klein's and Noether's opinion, the Hilbert conservation law cannot
be thought as a usual conservation law for the energy-momentum. As
an answer, Hilbert has supposed (see comments in Ref.~\cite{Klein_1917}) that an analogous situation could take place
in all the generally covariant theories. The Hilbert assumption has been proved right.
Noether has generalized the properties of the Hilbert current
$\bfJpara$ on arbitrary gauge-invariant theories. Combining the
results of the first and second theorems, she has shown that in an
\emph{arbitrary} gauge-invariant theory the Noether current $\bfJ$,
constructed according to the first theorem and with the
use of the Noether identities, always can be completed up to the
\emph{identically} conserved current $\bfrsJ$. Thus, unlike $\bfJ$, $\bfrsJ$ is
conserved independently of satisfying equations of motion. From here the boundary theorem follows directly: in
an  \emph{arbitrary} gauge-invariant theory the Noether current $\bfJ$,
constructed by the first theorem for a finite (global) subgroup of a
gauge group is presented as a sum of two terms: the first
vanishes on the equations of motion, whereas the second is
expressed through a divergence of an antisymmetric tensor $\bftet$
--- superpotential. At the same time, Noether did not give a rule
for the superpotential construction.

It is also not widely known that the Noether identities are not a complete
system of differential identities following from a gauge invariance
of a theory. Noether studied the problem in an active
collaboration with Klein, who independently obtained the results
analogous to Noether's \cite{Klein_1917, Klein_1918_a,
Klein_1918_b}. Noether remarked that her work
\cite{Noether_1918} and Klein's work \cite{Klein_1918_a}
"were mutually influential" \cite{Noether_1918,Noether_1918_en_c} (see also
comments by Klein and Hilbert in Ref.~\cite{Klein_1917}). In
work \cite{Klein_1918_a}, Klein, considering an example of generally
covariant metric theories, obtained a \emph{complete} system of
differential identities, from which the Noether identities follow. One of his identities is in fact
the boundary theorem, whereas the others give recipe for
constructing a superpotential.

Later, unfortunately, the above famous results by Hilbert,
Klein and Noether have been almost forgotten. The studies of the
identically conserved current ${\bf\rsI}$ in generally covariant theories has
been re-stated by Bergman \cite{Bergmann_1949} 30 years later.
The existence of the superpotential $\bfpot$
corresponding to the identically conserved Bergmann current has been
stated by Zatzkis \cite{Zatzkis_1951}. The existence of the
identically conserved current $\bfrsJpara$ in generally covariant
theories, which depends on an arbitrary vector $\bfpara$ has been
rediscovered by Bergmann and Shiller \cite{Bergmann_Schiller_1953}
in a special case. In the general case it was rediscovered by
Mitskievich  \cite{Mizkjewitsch_1957, Mitskievich_1969_en,
Mitskievich_Efremov_Nesterov_1985}, who systematically studied the
generalized current and have constructed the correspondent
(generalized) superpotential $\bfpotpara$.

The complete Klein system of identities has been rediscovered and
studied in detail for
constructing CQs by Trautman  \cite{Trautman_1962_b, Trautman_1966_en} (see also
Refs.~\cite{Logunov_Folomeshkin_1977_b_en, Logunov_Folomeshkin_1977_c_en},
\cite{Mizkjewitsch_1957, Mitskievich_1969_en,
Mitskievich_Efremov_Nesterov_1985}, \cite{Trautman_1996}). Little
earlier the Klein-like identities has been stated by Utiyama
\cite{Utiyama_1956, Utiyama_1959} in $SU(N)$-invariant gauge
theories. Just the above Trautman's and Utiyama's results became the
basis for studying differential identities in gauge theories of
gravity and the Einstein-Cartan theory  \cite{Kibble_1961,
Utiyama_Fukuyama_1971, Trautman_1972_a, Trautman_1972_b,
Trautman_1972_c, Trautman_1973, Hehl_Heyde_Kerlick_Nester_1976}, in
supergravity \cite{Henneaux_Julia_Silva_1999, Brandt_2002}, in metric-affine
theories of gravity \cite{Hehl_McCrea_Mielke_Neeman_1995}. The
Klein-Noether theorem in the general form, probably independently,
has been rediscovered by Francaviglia and coauthors \cite{Ferraris_Francaviglia_1992, Fatibene_Ferraris_Francaviglia_1994, Fatibene_Ferraris_Francaviglia_1995, Fatibene_Ferraris_Francaviglia_1997}, Julia and Silva \cite{Julia_Silva_1998, Silva_1999}, Barnich and Brandt
\cite{Barnich_Brandt_2002, Barnich_2003}.
Recently, using the jet stratification technique and the variational
bi-complex technique, the theorem has been stated in a very
generic case (non-closed algebras, Grassmannian fields, graduate
groups) in the works by Francaviglia {\it et. al.} \cite{Fatibene_Ferraris_Francaviglia_McLeneghan_2002, Fatibene_Ferraris_Francaviglia_2004_b, Fatibene_Francaviglia_2004} and by Sardanashvily {\it et. al.}
\cite{Bashkirov_Giachetta_Mangiarotti_Sardanashvily_2005_b, Sardanashvily_2009_a}. To finalize a short historical
discourse, we remark that the conclusion that a superpotential has
to exist in GR directly follows from Einstein's work of 1916
\cite{Einstein_1916_b, Einstein_1916_b_en}.

The novelty of our results is in the following:
\bi
\item \emph{Universality}. We consider an \emph{arbitrary}
diffeomorphically invariant classical field theories, Lagrangians of
which contain derivatives of field variables (tensor densities of
\emph{arbitrary}, but fixed ranks and weights) up to the second order;
\item \emph{Manifest general covariance}. We develop manifestly generally
covariant formalism, first, using \emph{initially} generally
covariant expressions (without using auxiliary structures, such as a
background metric); second, all of our calculations, unlike many of
aforementioned works, are manifestly generally covariant at
\emph{each and every steps};
\item \emph{The torsion field is taken into account}. A spacetime under consideration is
presented by an \emph{arbitrary Riemann-Cartan space}. Both the
torsion tensor and the metric tensor are the dynamical
fields, the torsion coupling in the Lagrangian can be both
minimal (through connection) and non-minimal (explicit).
\ei

A technique developed in the present work to analyze
diffeomorphic invariance can be directly applied to both
\emph{manifestly covariant} and \emph{gauge invariant} studies of
 gauge invariance properties of \emph{arbitrary} nature field
theories given in \emph{Riemann-Cartan spacetime}, which could be
classical gauge theories or theories with a local supersymmetries.

In the most of the present-day works related to analyzing general
gauge theories (see, e.g., works by Julia and Silva
\cite{Julia_Silva_1998, Silva_1999}, by Barnich and Brandt {\it et. al.} \cite{Barnich_Brandt_2002, Barnich_2003}, by Obukhov {\it et. al.} \cite{Obukhov_Rubilar_2007, Obukhov_Rubilar_2008,
Gratus_Obukhov_Tucker_2012}, by Baykal and Delice \cite{Baykal_Delice_2011}, by Sardanashvily
and Giachetta {\it et. al.} \cite{Sardanashvily_2004,
Bashkirov_Giachetta_Mangiarotti_Sardanashvily_2005_a, Bashkirov_Giachetta_Mangiarotti_Sardanashvily_2005_b,
Giachetta_Mangiarotti_Sardanashvily_2005, Giachetta_Mangiarotti_Sardanashvily_2009_a, Sardanashvily_2009_a}, by Francaviglia {\it et al.} \cite{Fatibene_Francaviglia_Palese_2001, Fatibene_Ferraris_Francaviglia_McLeneghan_2002, Fatibene_Ferraris_Francaviglia_2004_b, Palese_Winterroth_2004, Fatibene_Francaviglia_Mercadante_2010_a}), unfortunately, authors frequently use rarely known formalisms for physicists, such as
the aforementioned variational bi-complex, jet stratification,
and also differential form technique, etc. Unlike them, we
perform all the calculations and present the final results in the usual
tensorial language.

The rest of the paper is organized as follows: In section \ref{sec_01_a-00}, we
suggest the general Noether relation in a \emph{manifestly generally
covariant form}, find general expressions for the generalized
(depending on an arbitrary infinitesimal vector field $\dbfpara$ ---
displacement field) conserved current $\bfJdpara$ and Noether charge
$\QdxiSig$.

In section \ref{sec_01_b-00}, we develop the \emph{complete manifestly
covariant universal} system of differential identities, which take place in an \emph{arbitrary} diffeomorphically invariant theory of
the class under consideration. Thus, the results in the second part
of section \ref{sec_01_a-00} and in section \ref{sec_01_b-00}
present the \emph{covariant} generalization of the Klein approach
\cite{Klein_1917, Klein_1918_a, Klein_1918_b} (see also the work
\cite{Bergmann_1949, Trautman_1962_b, Trautman_1966_en} and, especially, books
\cite{Mitskievich_1969_en, Mitskievich_Efremov_Nesterov_1985}).

In section \ref{sec_01_c-00} based on the generalized Noether
current $\bfJdpara$, we obtain the identically conserved current
$\bfrsJdpara$. Using the latter, we prove the generalized boundary
Klein-Noether theorem in the manifestly generally covariant form.
After that the generalized superpotential $\bfpotdpara$ is
constructed and a problem of its ambiguity is analyzed. A physical meaning of the generalized Noether
current $\bfJdpara$ and its connection to the usual Noether current
$\bfJ$, a numerical value of the conserved generalized charge $\QdxiSig$, a
physical meaning of the superpotential $\bfpotdpara$ are also
discussed.

In section \ref{sec_01_d-00}, utilizing the generalized
Belinfante procedure, the generalized symmetrized Noether current
$\bfJsdpara$ is constructed. As the result, we show that this current
is a linear combination of the Lagrangian derivatives of the
action functional. Thus, it vanishes on the equations of motion.
This means that \emph{divergences in the Lagrangian do not contribute to the current}. This conclusion is a wide generalization of the claims made in Refs.~\cite{Szabados_1991, Szabados_1992}.

Intermediate and cumbersome calculations are left in the appendixes. In Appendix
\ref{app_01_a-00}, we review (without a proof) basic facts of the Riemann-Cartan geometry. This could be used as a introduction into the world of the
Riemann-Cartan geometry, assuming the reader is fluent in more simple Riemannian geometry.

In Appendix \ref{app_01_b-00}, basic notions of irreducible
representations of a symmetric group (group of permutations) for two- and three-indexes
quantities are given. Using the Young projectors, we develop a \emph{new} technique, which is employed in the main text.

In Appendix \ref{app_01_c-00}, some general geometrical identities
are proved. They are used for a simplification of the Klein system
of identities and in analyzing ambiguities in the superpotential.

In Appendix \ref{app_01_d-00}, the technique of Appendix
\ref{app_01_b-00} is used to solve a system of equations defining
a superpotential and determining its general representations.

In the paper we use the following notations: Greek indexes $\alp$,
$\bet$, \dots, $\mu$, $\nu$, \dots take values of $0$, $1$, \dots,
$D$ and numerate spacetime coordinates $x\Def\{x^\alp\}$, partial
$\bfpa \Def\{\pa\alp\}\Def\{\pa{}/\pa{} x^\alp\}$ and covariant
$\bfna \Def\{\na\alp\}$, $\bfsna\Def\{\sna\alp\}$ derivatives, and
spacetime tensor components of fields also. Small Latin indexes from
the middle of alphabet $i$, $j$, \dots, $z$ take values of $1$, $2$,
\dots, $D$ and numerate space components. Coordinate $x^0$ is a time
one, whereas coordinates $\vec{x}\Def\{x^i\}$ are space ones.
Capital Latin indexes $A$, $B$, \dots, are collective and numerate
components of the full set of the physical fields
$\bfgfi\Def\{\gfiA(x)\}$ (containing both gravitational and matter fields) and
are related to $1$, $2$, \dots, $N$.  At last, small Latin indexes
from the beginning of the alphabet $a$, $b$, \dots, $h$ numerate
components of matter (non-gravitational) fields
$\bfvphi\Def\{\vphi^a(x)\}$ and take values of $1$, $2$, \dots, $n$.

As usual, for a twice repeated index, the Einstein summation rule is assumed.  Indexes in parentheses need to be symmetrized;
whereas, indexes in brackets needs to be antisymmetrized, for
example,
\bse
A_{(\alp\bet)}=\frac{1}{2}\lp A_{\alp\bet} + A_{\bet\alp} \rp, \qquad A_{[\alp\bet]}=\frac{1}{2}\lp A_{\alp\bet} - A_{\bet\alp} \rp.
\ese
Two vertical lines inside the brackets $()$ and  $[]$ mean that
indexes between them do not participate in
symmetrization/antisymmetrization, for example,
\bse
\ba{l}
A_{(\alp|\bet\gam|\del)} = \bfrac{1}{2} \lp A_{\alp\bet\gam\del} + A_{\del\bet\gam\alp} \rp,\\
A_{[\alp|\bet|\gam]} = \bfrac{1}{2} \lp A_{\alp\bet\gam} -
A_{\gam\bet\alp} \rp. \ea
\ese

Covariant derivatives $\bfna$ and $\bfsna$, and a sign
convention for the curvature tensor
$\bfcur\Def\{\cur\alp\bet\gam\del\}$ and the torsion tensor
$\bftor\Def\{\tor\alp\bet\gam\}$ are derived in Appendix
\ref{app_01_a-00}.

The speed of light in vacuum is set to one.

\section{The general Noether identity. Generalized Noether's current and charge}\label{sec_01_a-00}

We consider a classical field theory determined by the action
functional
\be \ba{c} I[\bfgfi;\Sig_{1,2}] = \intb\dome\rmet\lag, \ea \ee
in space-time $\caC(1,D)$ (see Appendix \ref{app_01_a-00}). Here,
$\dome\Def dx^0 dx^1\dots dx^D$; integration is provided over an
arbitrary $(D+1)$-dimensional volume in  $\caC(1,D)$ restricted by
two spacelike $D$-dimensional hypersurfaces $\Sig_1$ and $\Sig_2$;
Lagrangian $\lag$ is a local function of a set of field variables
$\bfgfi(x)=\{\gfiA(x); A=\overline{1,N}\}$ and their first and
second derivatives.

Consider a \emph{total variation of the action} $\bdel
I[\bfgfi;\Sig_{1,2}]$ initiated by both the general variations of
the field variables $\dbfgfi$ and the boundary hypersurfaces
$\dSig_{1,2}$:
\be\label{sec_01_a-03} \lf \ba{rcl}
\bfgfi(x) & \ra & \bfgfi'(x) = \bfgfi(x) + \dbfgfi(x);\\
\Sig_{1,2}(x) & \ra & \Sig_{1,2}'(x) = \Sig_{1,2}(x) +
\dSig_{1,2}(x). \ea \rd \ee
By the definition, one has
\be\label{sec_01_a-02} \ba{l}
\bdel I[\bfgfi;\Sig_{1,2}] \Def I[\bfgfi+\dbfgfi;\Sig_{1,2}+\dSig_{1,2}] - I[\bfgfi;\Sig_{1,2}]\\
=\lp I[\bfgfi+\dbfgfi;\Sig_{1,2}+\dSig_{1,2}] - I[\bfgfi+\dbfgfi;\Sig_{1,2}] \rp\\
+\lp I[\bfgfi+\dbfgfi;\Sig_{1,2}] - I[\bfgfi;\Sig_{1,2}] \rp. \ea
\ee
We assume that the field variables and their
derivatives vanish sufficiently fast at spatial infinity. Then, up to the
first order terms in variations, for the first parenthesis in
\eqref{sec_01_a-02} we obtain
%
\be \ba{l}
\del_\Sig I[\bfgfi;\Sig_{1,2}] \Def I[\bfgfi;\Sig_{1,2}+\dSig_{1,2}] - I[\bfgfi;\Sig_{1,2}]\\
= \lp \int\limits_{\Sig_1+\dSig_1}^{\Sig_2+\dSig_2} - \intb \rp \dome\rmet\lag\\
= \lp \intSigb-\intSiga \rp \dsig\mu \lag \dx\mu =
\intb\dome\rmet\sna\mu \lp\lag\dx\mu\rp \ea \ee
%
where the generalized Gauss theorem \eqref{sec_01_a-01} was employed.
The second parenthesis in \eqref{sec_01_a-02} is the \emph{functional
variation of the action}:
\be \ba{l}
\del_{\gfi{}} I[\bfgfi;\Sig_{1,2}] \Def I[\bfgfi+\dbfgfi;\Sig_{1,2}] - I[\bfgfi;\Sig_{1,2}]\\
= \intb\dome \del\lp\rmet\lag\rp. \ea \ee
We consider generally covariant theories of the most popular type,
when $\del_{\gfi{}} I$ is present always as
%
\be\label{sec_01_a-16} \ba{l}
\del_{\gfi{}} I = \intb\dome\rmet\DIDgfiA\dgfiA\\
+ \intb\dome\rmet \sna\mu \lf \KgA\mu \dgfiA + \LgA\bet\mu \na\bet
\dgfiA \rf. \ea \ee
%
Hereinafter, $\Del I/\Del\gfiA$ is defined by the variational
derivative $\del I/\del\gfiA$, which is the operator of equations of motion,
\be \label{}\DIDgfiA \Def \frac{1}{\rmet}\dIdgfiA, \ee
$\bfK \Def \{\KgA\mu\}$ and $\bfL \Def \{\LgA\bet\mu\}$ are local
functions of the field variables $\bfgfi$ and their first and second
derivatives, and are defined in an unique way (without ambiguities)
by the Lagrangian $\lag$. Combining the expressions \eqref{sec_01_a-02} -- \eqref{sec_01_a-16}, one finds
%
\be\label{sec_01_a-04} \ba{l} \bdel I = \intb\dome
\rmet\DIDgfiA\dgfiA\\ + \intb\dome\rmet \sna\mu \lp \KgA\mu\dgfiA +
\LgA\bet\mu\na\bet\dgfiA + \lag\dx\mu \rp. \ea \ee
%

We name \eqref{sec_01_a-03} as \emph{the symmetry transformation}
(see Refs.~\cite{Pons_2011, Brading_Brown_2003, Brandt_2002, Trautman_1962_b}), if it
induces the total variation of the action functional $\bdel I$ in
the form
\be\label{sec_01_a-05} \ba{c} \bdel I = \intb\dome \rmet
\sna\mu\lp\dLam\mu\rp \ea \ee
where $\dbfLam \Def \{\dLam\mu\}$ are infinitesimal local functions
of $\bfgfi$, $\dbfgfi$ and their derivatives ($\dbfLam$ is not a
variation). Equating \eqref{sec_01_a-04} with
\eqref{sec_01_a-05} and taking into account that the volume of
integration is arbitrary, one finds the relation
\be\label{sec_01_a-07}
\sna\mu J^\mu \lb\dbfgfi,\del x,\dbfLam\rb+ \DIDgfiA\dgfiA \eq 0,
\ee
which is called \emph{the general Noether identity} (the main
identity). Here,
\be\label{sec_01_a-09} \ba{l} J^\mu \lb\dbfgfi,\del x,\dbfLam\rb\\
\Def \KgA\mu\dgfiA + \LgA\bet\mu\na\bet\dgfiA + \lag\dx\mu -
\dLam\mu. \ea \ee
If equations of motion $\Del I/\Del\gfiA =0$ hold then the identity
\eqref{sec_01_a-07} transforms into the continuity equation
\be \sna\mu J^\mu \lb\dbfgfi,\del x,\dbfLam\rb = 0 \ee
where $\dbfgfi$, $\del x$ and $\dbfLam$ denote the symmetry
transformation.

Let us consider the infinitesimal diffeomorphisms
\be\label{sec_01_a-08} \lf\ba{rcl}
\dx\mu & = & \dpara\mu(x);\\
\dgfiA(x) & = & \dparagfiA(x) \ea\rd \ee
as the symmetry transformation,
$\dbfpara=\{\dpara\mu(x)\}$ is an arbitrary infinitesimal vector
 (\emph{displacement vector}).
Hereafter, we assume the Lagrangian $\lag$ is a generally covariant scalar. Then, one has
\be\label{sec_01_a-15}
\dbfLamdpara \Def \ld \dbfLam \rv_{\disp\{\dbfgfi=\dparabfgfi;\,\del
x=\dbfpara\}}=0.
\ee
The variations of the field variables have the general
form
\be\label{sec_01_a-06} \ba{l} \dparagfiA(x)\\ = \gfiaA\alp
\dpara\alp + \gfibA\alp\bet \na\bet\dpara\alp + \gficA\alp\bet\gam
\na{(\gam}\na{\bet)}\dpara\alp + \dots \ea \ee
where $\{\gfiaA\alp\}$, $\{\gfibA\alp\bet\}$,
$\{\gficA\alp{(\bet}{\gam\dots)}=\gficA\alp{\bet}{\gam\dots}\}$ are
local functions of $\bfgfi$ and their derivatives, which are defined uniquely by the transformation properties of $\bfgfi$. We
consider only the case when $\delta_\xi\bfgfi$ contains the first
two terms on the right hand side of \eqref{sec_01_a-06}. However,
the discussion can be easily extended to a more generic case, say,
of metric-affine theories of gravity, for which the third term in
\eqref{sec_01_a-06} is also nonzero.

A vector
\be\label{sec_01_b-18} \bfJdpara \Def \ld\bfJ\lb\dbfgfi,\del
x,\dbfLam\rb\rv_{\disp\{\dbfgfi=\dparabfgfi;\,\del
x=\dbfpara;\,\dbfLam=0\}}, \ee
whose components are obtained after the substitution of
\eqref{sec_01_a-08}, \eqref{sec_01_a-15} and \eqref{sec_01_a-06} into
\eqref{sec_01_a-09}:
\be\label{sec_01_b-17} \ba{l}
\Jdpara\mu = \lf \KgA\mu\gfiaA\alp + \lag\kro\mu\alp + \LgA\nu\mu\na\nu\gfiaA\alp \rf \dpara\alp\\
+ \lf \KgA\mu\gfibA\alp\bet + \LgA\bet\mu\gfiaA\alp + \LgA\nu\mu\na\nu\gfibA\alp\bet \rf \na\bet\dpara\alp\\
+ \lf \LgA\gam\mu\gfibA\alp\bet \rf \na\gam\na\bet\dpara\alp \ea \ee
we will call as the \emph{generalized Noether current}.

It is insightful to compare the results obtained here in
the tensorial formalism with the corresponding results in a popular
formalism of differential forms. It is impossible to perform a full
comparison; fortunately, there is no needs for this. To show how the comparison
could be done it is enough to show this for any particular example. The formulae
\eqref{sec_01_b-18} and \eqref{sec_01_b-17}, which are among the main formulae of the
formalism, are well suited for this goal. Thus, the Noether current
$D$-form $\bfj[\dbfpara] = \bfTet [\bfgfi, \dparabfgfi] - \dbfpara
\cdot \bflag$ constructed in works \cite{Wald_1993,
Iyer_Wald_1994, Iyer_Wald_1995, Wald_Zoupas_2000} coincides with the
current \eqref{sec_01_b-18}, \eqref{sec_01_b-17} up to a sign: $\bfj[\dbfpara] \Def
-\Jdpara\mu\, \bfdsig\mu$. The symplectic potential $D$-form
$\bfTet$ in the tensorial notations is presented as $\bfTet [\bfgfi,
\dparabfgfi] = - \ld\lf \KgA\mu\dgfiA + \LgA\bet\mu\na\bet\dgfiA
\rf\rv_{\disp\{\dbfgfi=\dparabfgfi\}} \, \bfdsig\mu$. When a detailed structure of the last expression is analyzed, one finds the
exact correspondence with \eqref{sec_01_b-17}.

 Next,
transform the last term in the expression \eqref{sec_01_b-17}
following the formula \eqref{sec_01_a-10} in Appendix
\ref{app_01_b-02}:
\be\label{sec_01_b-04} \ba{l}
\lf \LgA\gam\mu\gfibA\alp\bet \rf \na\gam\na\bet\dpara\alp\\ =
\lf \bfrac{1}{2} \cur\veps\alp\kap\lam \LgA\kap\mu \gfibA\veps\lam \rf \dpara\alp\\
+ \lf -\bfrac{1}{2} \tor\bet\kap\lam \LgA\kap\mu \gfibA\alp\lam \rf
\na\bet\dpara\alp\\ + \lf \LgA\gam\mu \gfibA\alp\bet \rf
\na{(\gam}\na{\bet)}\dpara\alp. \ea \ee
Taking this into account, find
\be\label{sec_01_b-01} \Jdpara\mu = \U\alp\mu \dpara\alp +
\M\alp\bet\mu \na\bet\dpara\alp + \N\alp\bet\gam\mu
\na{(\gam}\na{\bet)}\dpara\alp \ee
where
\footnote{Remark that the position of indexes in definitions
of $\bfM$ \eqref{sec_01_a-14} and $\bfN$ \eqref{sec_01_a-12} differs
from that of Refs.~\cite{Petrov_2004_b_en, Petrov_2009_a, Petrov_2010_a,
Petrov_2011, Petrov_Lompay_2013}. Thus $\{\M\alp\bet\mu\}$ and
$\{\N\alp\bet\gam\mu\}$ defined here correspond to
$\{\hat{m}_\alp{}^{\mu\bet}\}$ and
$\{\hat{n}_\alp{}^{\mu\bet\gam}\}$ in the cited works. The definition of the tensor $\bfU$ \eqref{sec_01_a-13} is also changed. Thus, the tensor $\{\U\alp\mu+I_\alp{}^\mu\}$ corresponds to the tensor density $\{\hat{u}_\alp{}^\mu\}$. At last, the current $\{\rsJ^\mu  = J^\mu + I^\mu\}$ \eqref{sec_01_c-03} corresponds to the current $\{\hat{i}^\mu\}$.}
\bw
\begin{empheq}[left=\empheqlbrace]{flalign}
\U\alp\mu & \Def \lag\kro\mu\alp + \KgA\mu \gfiaA\alp + \LgA\kap\mu
\lp \na\kap\gfiaA\alp
 + \bfrac{1}{2} \cur\veps\alp\kap\lam \gfibA\veps\lam \rp;\label{sec_01_a-13}\\
\quad \M\alp\bet\mu & \Def \KgA\mu \gfibA\alp\bet + \LgA\bet\mu \gfiaA\alp +\LgA\kap\mu \lp \na\kap\gfibA\alp\bet - \bfrac{1}{2} \tor\bet\kap\lam \gfibA\alp\lam \rp;\label{sec_01_a-14}\\
\quad \N\alp\bet\gam\mu & \Def \LgA{(\gam|}{\mu}
\gfibA{\alp}{|\bet)}\label{sec_01_a-12}.
\end{empheq}
\ew
Note that after the symmetrization in \eqref{sec_01_a-12}, we get
\be\label{sec_01_b-05} \N\alp{(\bet}{\gam)}\mu = \N\alp\bet\gam\mu.
\ee
Also, for the diffeomorphisms \eqref{sec_01_a-08}, \eqref{sec_01_a-15} and
\eqref{sec_01_a-06} the general Noether identity \eqref{sec_01_a-07}
can be rewritten in the form:
\be\label{sec_01_b-02} \sna\mu \Jdpara\mu \eq - \Ia\alp \dpara\alp -
\Ib\alp\bet \na\bet \dpara\alp \ee
where
\begin{empheq}[left=\empheqlbrace]{align}
\Ia\alp \Def \DIDgfiA\gfiaA\alp;\\
\Ib\alp\bet \Def \DIDgfiA\gfibA\alp\bet.\label{sec_01_d-12}
\end{empheq}

Now, we define the \emph{generalized Noether charge} as
\be \label{sec_01_b-02+} \ba{c} \QdxiSig \Def \intSig\dsig\mu \Jdpara\mu \ea \ee
where $\Sig$ is a spacelike $D$-dimensional hypersurface in
$\caC(1,D)$.

Let the equations of motion $\Del I/\Del\gfiA=0$ be satisfied, then the
relation \eqref{sec_01_b-02} acquires the form of the continuity
equation $\sna\mu \Jdpara\mu = 0$ for the current $\bfJdpara =
\{\Jdpara\mu\}$. Next, if additionally the field variables
and their derivatives vanish fast enough at a spatial infinity, then
\eqref{sec_01_b-02} leads to conservation of the generalized charge
\be\label{sec_01_b-19}
\DIDgfiA = 0 \qquad \Rightarrow \qquad Q[\dbfpara;\Sig_1] =
Q[\dbfpara;\Sig_2]
\ee
meaning that its value is the same on each of hypersurfaces
$\Sig$.

Note that the above conclusions are valid for \emph{arbitrary}
vectors $\dbfpara$, \emph{not} just for Killing vectors. Therefore, the
aforementioned conservation laws are not connected with existence or
absence of a spacetime group of motions. After series of works by
Bergmann's group \cite{Bergmann_Schiller_1953,
Bergmann_Goldberg_Janis_Newman_1956, Bergmann_1958}, who have
studied this situation in general relativity, the conclusion was reached that the
charge $\QdxiSig$ is the \emph{generator of infinitesimal
diffeomorphisms} \eqref{sec_01_a-08}, \eqref{sec_01_a-06}. Bergmann {\it et al.} have
utilized the canonical formalism where on the equations of motion
one has
\be \ba{l} \bdel I|_{eq.mot.} = \intb\dome \sna\mu \Jdpara\mu =
\lp \intSigb - \intSiga \rp \dsig\mu \Jdpara\mu\\ \qquad =
\ld\QdxiSig\rv_{\Sig_2} - \ld\QdxiSig\rv_{\Sig_1}. \ea \ee
In the framework of the Lagrangian formalism, the same conclusion
follows from the \emph{Schwinger dynamical principle}
\cite{Schwinger_1951} (see also Refs.~\cite{Schwinger_2000, Toms_2007}).

In the case when a spacetime has a continuous group of motion
$\rsK_r$ with $r$ independent parameters, there are $r$ linearly
independent Killing vector fields $\bfparKa \Def \{
\parKa\mu; a=\overline{1,r}\}$. Then, defining infinitesimal
displacement vectors $\dbfpara$ as
\be \dbfpara(x) = \dbfparK(x) \Def \del\veps^{(a)} \bfparKa(x), \ee
one obtains for the generalized current $\bfJdpara$
\eqref{sec_01_b-01}:
\be \bfJdparK = \del\veps^{(a)} \bfJ_{(a)} \ee
where $\{ \del\veps^{(a)} \}$ is the set of infinitesimal \emph{constant}
transformation parameters of $\rsK_r$. The quantities
\be \bfJ_{(a)} \Def \{ J_{(a)}{}^\mu \}; \ee
\be J_{(a)}{}^\mu \Def \U\alp\mu \parKa\alp + \M\alp\bet\mu
\na\bet\parKa\alp + \N\alp\bet\gam\mu
\na{(\gam}\na{\bet)}\parKa\alp \ee
are just the conserved currents constructed according to the
\emph{first Noether theorem} for the group of motions $\rsK_r$.

\section{The Klein and Noether identities}\label{sec_01_b-00}

Using the generalized Leibnitz rule \eqref{sec_01_b-03}, let us open
explicitly the left hand side of the general Noether identity
\eqref{sec_01_b-02} where the current is introduced in
\eqref{sec_01_b-01}:
\be\label{sec_01_b-08} \ba{l}
\sna\mu \Jdpara\mu = \lf \sna\mu \U\alp\mu \rf \dpara\alp\\ + \lf \U\alp\bet +
\sna\mu \M\alp\bet\mu \rf \na\bet\dpara\alp\\
+ \lf \M\alp\bet\gam + \sna\mu \N\alp\bet\gam\mu \rf
\na\gam\na\bet\dpara\alp\\ + \lf \N\alp\bet\gam\del \rf
\na\del\na\gam\na\bet\dpara\alp. \ea \ee
Using the formula \eqref{sec_01_a-10}, we transform the third term on the R.H.S. in the same way as in \eqref{sec_01_b-04}. Keeping in mind the
property \eqref{sec_01_b-05}, one obtains
\be\label{sec_01_b-06} \ba{l}
\lf \M\alp\bet\gam + \sna\mu \N\alp\bet\gam\mu \rf \na\gam\na\bet\dpara\alp\\
= \lf \bfrac{1}{2} \cur\veps\alp\kap\lam \M\veps\lam\kap \rf
\dpara\alp + \lf -\bfrac{1}{2} \tor\bet\kap\lam \M\alp\lam\kap \rf
\na\bet\dpara\alp\\ + \lf \M{\alp}{(\bet}{\gam)} + \sna\mu
\N{\alp}{(\bet}{\gam)}{\mu} \rf \na{(\gam}\na{\bet)}\dpara\alp. \ea
\ee
Transformation of the fourth term on the R.H.S. in \eqref{sec_01_b-08} is more
complicated, which is worked out it in Appendix \ref{app_01_b-04}. The
finalized result is presented in \eqref{sec_01_a-11} as well as
\bw
\be\label{sec_01_b-07} \ba{l}
\lf \N\alp\bet\gam\del \rf \na\del\na\gam\na\bet\dpara\alp =
\lf -\bfrac{1}{3} \N\kap\lam\mu\nu \lp \na\lam \cur\kap\alp\mu\nu +
\bfrac{1}{2} \tor\sig\mu\nu \cur\kap\alp\lam\sig \rp \rf \dpara\alp\\
+ \lf \bfrac{1}{3} \N\alp\lam\mu\nu \lp 2\cur\bet\lam\mu\nu +
\na\lam\tor\bet\mu\nu + \bfrac{1}{2} \tor\sig\mu\nu \tor\bet\lam\sig \rp -
\N\kap\bet\mu\nu \cur\kap\alp\mu\nu \rf \na\bet\dpara\alp\\
+ \lf \N\alp\bet\mu\nu \tor\gam\mu\nu \rf
\na{(\gam}\na{\bet)}\dpara\alp + \lf \N\alp\bet\gam\del \rf
\na{(\del}\na\gam\na{\bet)}\dpara\alp. \ea \ee
\ew
Substituting \eqref{sec_01_b-06} and \eqref{sec_01_b-07} into
formula \eqref{sec_01_b-08}, then into \eqref{sec_01_b-02}, one
obtains an equivalent representation of the general Noether identity:
\bw
\be\label{sec_01_b-09}
\ba{l}
\lf \sna\mu \U\alp\mu - \bfrac{1}{2} \M\lam\mu\nu \cur\lam\alp\mu\nu - \bfrac{1}{3} \N\kap\lam\mu\nu \lp \na\lam \cur\kap\alp\mu\nu + \bfrac{1}{2} \tor\sig\mu\nu \cur\kap\alp\lam\sig \rp \rf \dpara\alp\\
+ \lf \U\alp\bet + \lp \sna\mu \M\alp\bet\mu + \bfrac{1}{2} \M\alp\mu\nu \tor\bet\mu\nu \rp + \bfrac{1}{3} \N\alp\lam\mu\nu \lp 2\cur\bet\lam\mu\nu + \na\lam \tor\bet\mu\nu + \bfrac{1}{2} \tor\sig\mu\nu \tor\bet\lam\sig \rp - \N\kap\bet\mu\nu \cur\kap\alp\mu\nu \rf \na\bet\dpara\alp\\
+ \lf \M{\alp}{(\bet}{\gam)} + \sna\mu \N\alp\bet\gam\mu + \N{\alp}{(\bet|}{\mu}{\nu} \tor{|\gam)}{\mu}{\nu} \rf \na{(\gam}\na{\bet)}\dpara\alp + \lf \N{\alp}{(\bet}{\gam}{\del)} \rf \na{(\del}\na{\gam}\na{\bet)}\dpara\alp \eq - \Ia\alp \dpara\alp - \Ib\alp\bet \na\bet \dpara\alp.
\ea
\ee
\ew
Notice that this identity is valid, when each function from the set
$\{ \dpara\alp, \pa\bet\dpara\alp, \pa\gam\pa\bet\dpara\alp,
\pa\del\pa\gam\pa\bet\dpara\alp \}$ has an arbitrary
values at every world point. Then, opening \eqref{sec_01_b-09}
explicitly, one can equate to zero the coefficients in front of each function
independently and obtain the system of identities. Such a system
is not manifestly covariant. However, one can transfer to another set of
arbitrary functions $\{ \dpara\alp, \na\bet\dpara\alp,
\na{(\gam}\na{\bet)}\dpara\alp,
\na{(\del}\na\gam\na{\bet)}\dpara\alp \}$. Because the Jacobian of
the transformation is not degenerated, one can use the second set as
equivalent instead to the first one. Thus, equating to zero the
coefficients at the functions of the second set in identity
\eqref{sec_01_b-09}, one obtain the covariant system of identities
equivalent to \eqref{sec_01_b-09}:
\bw
\begin{empheq}[left=\empheqlbrace]{flalign}
\sna\mu \U\alp\mu - \bfrac{1}{2} \M\lam\mu\nu \cur\lam\alp\mu\nu -
\bfrac{1}{3} \N\kap\lam\mu\nu \lp \na\lam \cur\kap\alp\mu\nu +
\bfrac{1}{2} \tor\sig\mu\nu \cur\kap\alp\lam\sig \rp \eq - \Ia\alp; \label{sec_01_b-10}\\
\ba{lr}
\U\alp\bet + \lp \sna\mu \M\alp\bet\mu + \bfrac{1}{2} \M\alp\mu\nu \tor\bet\mu\nu \rp & \\
\qquad  + \bfrac{1}{3} \N\alp\lam\mu\nu \lp 2\cur\bet\lam\mu\nu +
\na\lam \tor\bet\mu\nu + \bfrac{1}{2} \tor\sig\mu\nu
\tor\bet\lam\sig \rp - \N\kap\bet\mu\nu \cur\kap\alp\mu\nu & \eq
-\Ib\alp\bet;
\ea \label{sec_01_b-14}\\
\M{\alp}{(\bet}{\gam)} + \sna\mu \N\alp\bet\gam\mu + \N{\alp}{(\bet|}{\mu}{\nu} \tor{|\gam)}{\mu}{\nu} \eq 0; \label{sec_01_b-13}\\
\N{\alp}{(\bet}{\gam}{\del)} \eq 0.\label{sec_01_b-11}
\end{empheq}
\ew
The equations \eqref{sec_01_b-10}--\eqref{sec_01_b-11} present
\emph{complete manifestly covariant universal} system of differential
identities, which is valid in an \emph{arbitrary} diffeomorfically
invariant field theory. Originally the system, analogous to the
above, has been obtained in a non-covariant form by Klein
\cite{Klein_1918_a} for purely metric theories of gravity. Therefore
we will name system \eqref{sec_01_b-10} -- \eqref{sec_01_b-11} as
the \emph{Klein identities}.

In Appendixes \ref{app_01_c-02} and \ref{app_01_c-03}, we show that
the Klein identities \eqref{sec_01_b-10} and \eqref{sec_01_b-14} can
be rewritten in the form
\bw \be\label{sec_01_b-15} \sna\mu \lp \U\alp\mu - \bfrac{1}{3}
\N\lam\mu\rho\sig \cur\lam\alp\rho\sig \rp - \bfrac{1}{2} \lp
\M{\lam}{[\rho}{\sig]} - \bfrac{2}{3} \sna\mu
\N{\lam}{\mu}{[\rho}{\sig]} + \bfrac{1}{3}
\N{\lam}{[\rho|}{\mu}{\nu} \tor{|\sig]}{\mu}{\nu} \rp
\cur\lam\alp\rho\sig \eq -\Ia\alp \ee
and
\be\label{sec_01_b-12} \ba{l}
\lp \U\alp\bet - \bfrac{1}{3} \N\lam\bet\rho\sig \cur\lam\alp\rho\sig \rp\\
+ \sna\mu \lp \M{\alp}{[\bet}{\mu]} - \bfrac{2}{3} \sna\lam
\N{\alp}{\lam}{[\bet}{\mu]} + \bfrac{1}{3}
\N{\alp}{[\bet|}{\rho}{\sig} \tor{|\mu]}{\rho}{\sig} \rp +
\bfrac{1}{2} \lp \M{\alp}{[\rho}{\sig]} - \bfrac{2}{3} \sna\lam
\N{\alp}{\lam}{[\rho}{\sig]} +
\bfrac{1}{3} \N{\alp}{[\rho|}{\kap}{\lam} \tor{|\sig]}{\kap}{\lam} \rp \tor\bet\rho\sig\\
+ \sna\mu \lp \M{\alp}{(\bet}{\mu)} + \sna\lam \N\alp\bet\mu\lam +
\N{\alp}{(\bet|}{\rho}{\sig} \tor{|\mu)}{\rho}{\sig} \rp -
\sna\mu\sna\lam \N{\alp}{(\bet}{\mu}{\lam)} \eq -\Ib\alp\bet, \ea
\ee \ew
respectively. At the beginning, note that due to identities
\eqref{sec_01_b-13} and \eqref{sec_01_b-11} the last two terms on the
left hand side of \eqref{sec_01_b-12} are equal to zero. Next,
subtract the divergence $\sna\bet$ of \eqref{sec_01_b-12} from the
identity \eqref{sec_01_b-15}, taking into account the identity
\eqref{sec_01_b-16} where one sets
\be \tet_\alp{}^{\bet\mu} = \M{\alp}{[\bet}{\mu]} - \bfrac{2}{3}
\sna\lam\N{\alp}{\lam}{[\bet}{\mu]}
 + \bfrac{1}{3} \N{\alp}{[\bet|}{\rho}{\sig} \tor{|\mu]}{\rho\sig}.
\ee
After, we obtain the new identity
\be \sna\mu \Ib\alp\mu - \Ia\alp \eq 0 \ee
that is the \emph{Noether identity} rewritten in a manifestly
covariant form. All of these mean that instead of the Klein system \eqref{sec_01_b-10}--\eqref{sec_01_b-11},
one can use the equivalent \emph{Klein-Noether system of
identities}:
\bw
\begin{empheq}[left=\empheqlbrace]{flalign}
\sna\mu \Ib\alp\mu \eq \Ia\alp;\label{sec_01_c-01}\\
\ba{rr}
\lp \U\alp\bet - \bfrac{1}{3} \N\lam\bet\rho\sig \cur\lam\alp\rho\sig \rp + \sna\mu \lp \M{\alp}{[\bet}{\mu]} - \bfrac{2}{3} \sna\lam \N{\alp}{\lam}{[\bet}{\mu]} + \bfrac{1}{3} \N{\alp}{[\bet|}{\rho}{\sig} \tor{|\mu]}{\rho}{\sig} \rp & \\
+ \bfrac{1}{2} \lp \M{\alp}{[\rho}{\sig]} - \bfrac{2}{3} \sna\lam
\N{\alp}{\lam}{[\rho}{\sig]} + \bfrac{1}{3}
\N{\alp}{[\rho|}{\kap}{\lam} \tor{|\sig]}{\kap}{\lam} \rp
\tor\bet\rho\sig & \eq -\Ib\alp\bet;
\ea\label{app_01_d-26}\\
\M{\alp}{(\bet}{\gam)} + \sna\mu \N\alp\bet\gam\mu + \N{\alp}{(\bet|}{\mu}{\nu} \tor{|\gam)}{\mu}{\nu} \eq 0;\label{app_01_d-22}\\
\N{\alp}{(\bet}{\gam}{\del)} \eq 0.\label{app_01_d-14}
\end{empheq}
\ew

\section{The generalized Noether superpotential. The boundary Klein-Noether theorem}\label{sec_01_c-00}

Substituting $\Ia\alp $ from \eqref{sec_01_c-01} into the general
Noether identity \eqref{sec_01_b-02}, one obtains another identity
\be\label{sec_01_c-03+} \sna\mu \rsJdpara\mu \eq 0, \ee
which has a meaning of the continuity equation for the current defined
as $\bfrsJdpara \Def \lf \rsJdpara\mu \rf$, where
\be\label{sec_01_c-03} \ba{l}
\rsJdpara\mu \Def \lp \U\alp\mu + \Ib\alp\mu \rp \dpara\alp\\
+ \M\alp\bet\mu \na\bet\ \dpara\alp + \N\alp\bet\gam\mu
\na{(\gam}\na{\bet)} \dpara\alp. \ea \ee
It is evidently that the current $\bfrsJdpara$ is connected with the
generalized Noether current $\bfJdpara$ \eqref{sec_01_b-01} by the
relation:
\be\label{sec_01_c-11} \Jdpara\mu = -\Ib\alp\mu \dpara\alp +
\rsJdpara\mu . \ee

Note that identity \eqref{sec_01_c-03+} takes place
\emph{independently of equations of
motion}. Then, keeping in mind identity \eqref{app_01_c-07}, one
should conclude that the current in \eqref{sec_01_c-03+} can be
represented in the form
\be\label{sec_01_c-02} \rsJdpara\mu = \sna\nu \potdpara\mu\nu +
\frac{1}{2} \potdpara\rho\sig \tor\mu\rho\sig \ee
where
\be \bfpotdpara \Def \lf \potdpara\mu\nu \rf; \qquad
\potdpara{[\mu}{\nu]} = \potdpara\mu\nu \ee
is an antisymmetric tensor --- the \emph{generalized Noether
superpotential}. \footnote{Formula \eqref{sec_01_c-02} is manifestly
covariant generalization of the known Poincar\'e lemma to the case
of the spacetime $\caC(1,D)$. The lemma has a local
character. Attempts to extend the lemma to a global derivation
meets difficulties connected with a topology of a spacetime
$\caC(1,D)$, which is defined by homotopic and cohomologic
properties of $\caC(1,D)$. In fact, this problem requires a further investigation.
Here, for the sake of simplicity, we assume that formula
\eqref{sec_01_c-02} is valid in the global sense also; thus, the current
$\bfrsJdpara$ \eqref{sec_01_c-03} is cohomological to zero.} The
formulae \eqref{sec_01_c-11} and \eqref{sec_01_c-02} represented in
the formalism of the differential forms are equivalent (on the
equations of motion) to the relation $\bfj[\dbfpara] = \bfd \bfQ
[\dbfpara]$ \cite{Wald_1993, Iyer_Wald_1994, Iyer_Wald_1995,
Wald_Zoupas_2000}, where $\bfj[\dbfpara]$ is the Noether current
$D$-form (see discussion above after formula \eqref{sec_01_b-17});
the Noether charge $(D-1)$-form $\bfQ$ in the tensorial notations is
presented as $\bfQ [\dbfpara] = - \frac{1}{2!} \potdpara\mu\nu
\bfds\mu\nu$.

A superpotential in \eqref{sec_01_c-02} is not defined uniquely. Indeed, if
\be\label{sec_01_c-04} \ppota\mu\nu \Def \pota\mu\nu + \lp \sna\lam
\potbu\mu\nu\lam + \potbu{[\mu|}{\rho}{\sig} \tor{|\nu]}{\rho}{\sig}
\rp \Def \pota\mu\nu + \dpota\mu\nu \ee
is another superpotential where
\be\label{app_01_c-16} \potbu{[\mu}{\nu}{\lam]} = \potbu\mu\nu\lam,
\ee
then
\bw \be\label{app_01_c-15} \ba{l}
\rsJ'^\mu = \sna\nu \ppota\mu\nu + \bfrac{1}{2} \ppota\rho\sig \tor\mu\rho\sig\\
= \lb \sna\nu \pota\mu\nu + \bfrac{1}{2} \pota\rho\sig
\tor\mu\rho\sig \rb + \lb \sna\nu \lp \sna\lam \potbu\mu\nu\lam +
\potbu{[\mu|}{\rho}{\sig} \tor{|\nu]}{\rho}{\sig} \rp + \bfrac{1}{2}
\lp \sna\lam \potbu\rho\sig\lam +
\potbu{[\rho|}{\kap}{\lam} \tor{|\sig]}{\kap}{\lam} \rp \tor\mu\rho\sig \rb\\
\Def \rsJ^\mu + \Del \rsJ^\mu. \ea \ee \ew
However, it is easily to show (see Appendix \ref{app_01_c-04}) that
\be\label{app_01_c-14} \Del \rsJ^\mu \eq 0, \ee
therefore
\be \rsJ'^\mu[\dbfpara] = \rsJ^\mu[\dbfpara]. \ee

Now, let us construct the superpotential $\bfpotdpara$ corresponding
to the current \eqref{sec_01_c-03}. We assume that it has the form
\be\label{sec_01_c-06} \potdpara\mu\nu = \A\alp\mu\nu \dpara\alp +
\B\alp\bet\mu\nu \na\bet \dpara\alp \ee
where coefficients $\A{\alp}{[\mu}{\nu]} = \A\alp\mu\nu$ and
$\B{\alp}{\bet}{[\mu}{\nu]} = \B\alp\bet\mu\nu$ do not depend on
$\dbfpara$ and its derivatives. Thus, one has to find the tensors
$\bfA \Def \{\A\alp\mu\nu\}$ and $\bfB \Def \{\B\alp\bet\mu\nu\}$.
Substituting \eqref{sec_01_c-06} into \eqref{sec_01_c-02}, one obtains an expression for the current $\bfrsJdpara$:
\bw \be \ba{l}
\rsJdpara\mu = \lf \lp \sna\nu \A\alp\mu\nu + \bfrac{1}{2} \A\alp\rho\sig
\tor\mu\rho\sig \rp + \bfrac{1}{2} \B\lam\rho\sig\mu \cur\lam\alp\rho\sig \rf \dpara\alp\\
+ \lf -\A\alp\bet\mu + \lp \sna\lam \B\alp\bet\mu\lam + \bfrac{1}{2}
\B\alp\bet\rho\sig \tor\mu\rho\sig - \bfrac{1}{2} \B\alp\rho\sig\mu
\tor\bet\rho\sig \rp \rf \na\bet \dpara\alp + \lf -\B\alp\bet\gam\mu
\rf \na{(\gam}\na{\bet)} \dpara\alp. \ea \ee \ew
Equating this expression to the current \eqref{sec_01_c-03},
we get the system of equations defining $\bfA$ and
$\bfB$:
\bw
\begin{empheq}[left=\empheqlbrace]{flalign}
\lp \sna\nu \A\alp\mu\nu + \bfrac{1}{2} \A\alp\rho\sig \tor\mu\rho\sig \rp + \bfrac{1}{2} \B\lam\rho\sig\mu \cur\lam\alp\rho\sig & = \U\alp\mu + \Ib\alp\mu;\label{sec_01_c-07}\\
-\A\alp\bet\mu + \lp \sna\lam \B\alp\bet\mu\lam + \bfrac{1}{2} \B\alp\bet\rho\sig \tor\mu\rho\sig - \bfrac{1}{2} \B\alp\rho\sig\mu \tor\bet\rho\sig \rp & = \M\alp\bet\mu;\label{sec_01_c-08}\\
-\B{\alp}{(\bet}{\gam)}{\mu} & =
\N\alp\bet\gam\mu.\label{sec_01_c-09}
\end{empheq}
\ew
A general solution of this system (see Appendix \ref{app_01_d-00}) reads
\begin{align}
\A\alp\mu\nu & = -\M{\alp}{[\mu}{\nu]}
 + \bfrac{2}{3} \lp \sna\lam \N{\alp}{\lam}{[\mu}{\nu]} + \bfrac{1}{2}
\tor{[\mu}{\rho}{\sig} \N{\alp}{\nu]}{\rho}{\sig} \rp \nonumber\\
& + \lp \sna\lam \Nc\alp\mu\nu\lam + \Nc{\alp}{[\mu|}{\rho}{\sig}
\tor{|\nu]}{\rho}{\sig}\rp ; \label{sec_01_d-07}\\
\B\alp\lam\mu\nu & = -\bfrac{4}{3} \N{\alp}{\lam}{[\mu}{\nu]} +
\Nc\alp\lam\mu\nu. \label{sec_01_d-08}
\end{align}
where  $\{ \Nc{\alp}{[\lam}{\mu}{\nu]} = \Nc\alp\lam\mu\nu \}$ is an
undefined antisymmetrical tensor.

Now, recall the ambiguity in the superpotential definition \eqref{sec_01_c-04}. We set there
\be \potbu\mu\nu\lam[\dbfpara] = \C\alp\mu\nu\lam \dpara\alp. \ee
where $\{\C{\alp}{[\mu}{\nu}{\lam]} = \C\alp\mu\nu\lam\}$ is an
arbitrary antisymmetrical tensor.  Then it is easily to find that an
ambiguity presented by \eqref{sec_01_c-04} appears in $\bfA$ and
$\bfB$ in the form:
\begin{align}
 A'_\alp{}^{\mu\nu} & = \A\alp\mu\nu + \lp
\sna\lam \C\alp\mu\nu\lam + \C{\alp}{[\mu|}{\rho}{\sig}
\tor{|\nu]}{\rho}{\sig} \rp; \label{app_01_d-27}\\
B'_\alp{}^{\bet\mu\nu} & = \B\alp\bet\mu\nu +
\C\alp\bet\mu\nu.\label{sec_01_c-05}
\end{align}
It is not surprisingly that the ambiguity in \eqref{sec_01_d-07} and
\eqref{sec_01_d-08} is the same as in \eqref{app_01_d-27} and
\eqref{sec_01_c-05}, respectively. But the latter does not contribute
to the current, see \eqref{app_01_c-14}, and consequently, it does not
contribute to the charge. Then, without loss of a generality, one
can set $\Nc\alp\lam\mu\nu = 0$, after that \eqref{sec_01_d-07} and
\eqref{sec_01_d-08} transfer to
\begin{align}
\A\alp\mu\nu & = -\M{\alp}{[\mu}{\nu]} + \frac{2}{3} \lp \sna\lam \N{\alp}{\lam}{[\mu}{\nu]} +
\frac{1}{2} \tor{[\mu}{\rho}{\sig} \N{\alp}{\nu]}{\rho}{\sig} \rp;\label{sec_01_d-07+}\\
\B\alp\lam\mu\nu & = -\frac{4}{3}
\N{\alp}{\lam}{[\mu}{\nu]}.\label{sec_01_d-08+}
\end{align}

Thus,
\be\label{sec_01_c-10} \ba{l}
\potdpara\mu\nu\\
= \lf -\M{\alp}{[\mu}{\nu]} + \bfrac{2}{3} \lp \sna\lam \N{\alp}{\lam}{[\mu}{\nu]} + \bfrac{1}{2} \tor{[\mu}{\rho}{\sig} \N{\alp}{\nu]}{\rho}{\sig} \rp \rf \dpara\alp\\
 + \lf -\bfrac{4}{3} \N{\alp}{\bet}{[\mu}{\nu]} \rf \na\bet \dpara\alp.
\ea \ee
In a more simple case of a field theory in a Riemannian spacetime, a
covariant formula of the \emph{same form} \eqref{sec_01_c-10}
originally was presented at the workshop\cite{Petrov_2000} (see
also Ref.~\cite{Petrov_2008} and references there in). The difference is
that the superpotential \eqref{sec_01_c-10} is constructed without a
background metric for initial variables of the theory, whereas the
superpotential suggested in Ref.~\cite{Petrov_2000} is
intended for perturbations in a curved \emph{background} spacetime.

Substituting the expression \eqref{sec_01_c-02} into formula
\eqref{sec_01_c-11}, integrating it over a spacelike $D$-dimensional
hypersurface $\Sig$ and using the Stockes rule \eqref{sec_01_c-12},
we rewrite the generalized Noether charge \eqref{sec_01_b-02+} in
the form
\be\label{sec_01_c-13}
\ba{c} \QdxiSig = -\intSig\dsig\mu \Ib\alp\mu
\dpara\alp + {\bfrac{1}{2!}} \intdSig \ds\mu\nu \potdpara\mu\nu. \ea
\ee
The above relation is a special case (in an integral form) of
a more general statement: \emph{the boundary Klein
theorem}\cite{Klein_1918_a} or \emph{the third Noether
theorem}\cite{Noether_1918} (see also Refs.~\cite{Brading_2005,
Brading_Brown_2003, Brading_Brown_2000, Julia_Silva_1998,
Silva_1999, Fatibene_Ferraris_Francaviglia_1994, Fatibene_Ferraris_Francaviglia_1995, Fatibene_Ferraris_Francaviglia_1997, Bashkirov_Giachetta_Mangiarotti_Sardanashvily_2005_b, Sardanashvily_2009_a, Trautman_1962_b}) that reads as
\emph{in an arbitrary gauge-invariant theory the Noether current is
presented by a sum of two terms, the first vanishes on
equations of motion, the second is a divergence of a
superpotential.} Note that the ambiguity in
definition of the superpotential presented in \eqref{sec_01_c-04}
disappears in definition of the charge \eqref{sec_01_c-13} by the
Stockes theorem. It is not surprisingly because as we already know that
the ambiguity does not contribute to the generalized Noether
current.

The structure of the charge \eqref{sec_01_c-13} in the Lagrangian
formulation is analogous to the structure of the diffeomorphism
generators in the Hamiltonian formulation of general relativity (GR).
Indeed, the first term on the right hand side of \eqref{sec_01_c-13}
vanishes on equations of motion, see \eqref{sec_01_d-12}. Thus,
the value of the charge \eqref{sec_01_c-13} is defined by the second term on the
right hand side (surface integral of a superpotential) only. The
Hamiltonians in GR have the same property: the first its part
presents integrals of constraints over hypersurface $\Sig$ and
disappears. Then the value of the Hamiltonians in GR is defined by
the second part: a surface integral over the boundary $\partial\Sig$
of $\Sig$. An assumption that surface terms and their contributions
in the Lagrangian formalism are equivalent to the correspondent ones in the
Hamiltonian formalism in an \emph{arbitrary gauge-invariant theory} has been
formulated in an explicit form in Ref.~\cite{Silva_1999}.

In earlier works in the Hamiltonian GR, all the boundary terms were
ignored which led to the problem of ``zero Hamiltonian'' (or
``frozen formalism'') \cite{Bergmann_Brunings_1949,
Bergmann_Penfield_Schiller_Zatzkis_1950}. The role of the boundary
terms has been studied and clarified in Refs.~\cite{Regge_Teitelboim_1974_b, Beig_Murchadha_1987} based on the
requirement of well defined variation of the Hamiltonian action and
well defined Poisson brackets (see also Ref.~\cite{Jamsin_2008}). A
consideration of the boundary terms in the generators of canonical
transformations initiates an extension of the canonical formalism
with inclusion of fields at the boundary $\partial\Sig$. This
important problem has been formulated and studied in Refs.~\cite{Soloviev_1985_en, Soloviev_1993, Soloviev_1996_b, Soloviev_1997_a,
Soloviev_1997_b_en, Soloviev_1999, Soloviev_2002_a, Soloviev_2002_b}.

\section{The generalized symmetrized Noether current}\label{sec_01_d-00}

From now we call the generalized Noether current $\bfJdpara$
\eqref{sec_01_b-01} and the generalized superpotential $\bfpotdpara$
\eqref{sec_01_c-10} as the generalized \emph{canonical} Noether
current and the generalized \emph{canonical} superpotential,
respectively.

Recall that the canonical current $\bfJdpara$ contains derivatives
of a displacement vector $\bfna\dbfpara$, $\bfna\bfna\dbfpara$. In
this section, we construct a new current $\Jsdpara\mu$, instead of
the canonical one $\bfJdpara$, with the property that it does not
contain derivatives of $\dbfpara$. In other words, we search for
\be\label{sec_01_d-01} \Jsdpara\mu = \Us\alp\mu \dpara\alp. \ee
Why is such a property important?

\emph{First}, as we have remarked in Sec. \ref{sec_01_a-00} and in Sec. \ref{sec_01_c-00} (see discussions after formulae \eqref{sec_01_b-19} and \eqref{sec_01_c-13}, respectivelly), under the
transition to canonical formalism the current $\bfJdpara$
becomes a generator of infinitesimal diffeomorphisms with the
parameters $\dbfpara$. Therefore this generator, like generators of
infinitesimal canonical transformations in a field theory,
\emph{should be} proportional to parameters of transformations
\emph{only} (like in \eqref{sec_01_d-01}), and not should be proportional to the derivatives.
In the cases when derivatives appear, one has to
suppress them.

\emph{Second}, the form \eqref{sec_01_d-01} is more compact.
In the case when a spacetime belongs to a group of motion (exact or
asymptotic), dynamic quantities presented by $\bfJdparK$ and based
on Killing vectors $\dbfparK$ of this group are constructed with
using all the tensors $\bfU$, $\bfM$ and $\bfN$. At the
same time, dynamic quantities based on \eqref{sec_01_d-01},
presented by $\bfJsdparK$, are constructed with using $\bfUs \Def \{
\Us\alp\mu \}$ only. Of course, according to the Ockham's razor
argument, the latter is preferred.

In Paper~II \cite{Lompay_Petrov_2013_b} of the current series of works, we will show that in
manifestly generally covariant theories a tensor $\bfU$ contains the canonical
energy-momentum tensor (EMT): $\bfsem \Def \lf \sem\mu\nu \rf$, and
a tensor $\bfM$ contains the spin tensor (ST): $\bfspi \Def \lf
\spi\pi{[\rho}{\sig]} = \spi\pi\rho\sig \rf$. It is well known that
in a general case canonical EMT is not symmetrical: $\semu\nu\mu
\neq \semu\mu\nu$. Owing to this property, even for a field theory in
Minkowski space, it is not possible to construct a \emph{total}
conserved angular momentum with using EMT only (it is necessary to
use ST as well). Thus, a total conserved angular momentum is constructed as
a sum of two terms representing orbital and spin momenta. For a symmetrical EMT the converse is true:  a \emph{total} conserved angular momentum is constructed by using EMT \emph{only} (without an additional ST). Therefore, it
is desirable to construct a symmetrical EMT. Originally a procedure
reconstructing a canonical EMT into a symmetrical EMT in a field
theory in Minkowski space (\emph{symmetrization procedure}) has been
suggested in Belinfante's works \cite{Belinfante_1939,
Belinfante_1940}\footnote{remark that
Belinfante himself \cite{Belinfante_1939} credits Dr.~Podo1anski for
construction of the symmetrization procedure}.

In the terms of the current, under the Belinfante symmetrization
procedure a spin term disappears from the explicit consideration.
Therefore, the reconstruction of the canonical current $\bfJdpara$
into the current  \eqref{sec_01_d-01} without derivatives of
$\dbfpara$ (i.e., without the spin term) is just a generalization
of the Belinfante procedure. By the requirement \eqref{sec_01_d-01}
the tensor $\bfUs$ is equal to the \emph{generalized symmetrized}
EMT $\bfsems$. However, one has to keep in mind that in a general
case a symmetrized EMT need not be symmetrical (see, e.g.,
a symmetrized EMT for perturbation in GR on curved backgrounds in
Refs.~\cite{Petrov_Katz_1999, Petrov_Katz_2002}).

Now, let us search for a \emph{generalized symmetrized Noether
current} $\bfJsdpara$. Because a new current has to be also
differentially conserved, we construct it by adding
an antisymmetrical tensor $\bfpotBdpara \Def \lf
\potBdpara{[\mu}{\nu]} = \potBdpara\mu\nu \rf$, similarly to
\eqref{sec_01_c-02}:
\be\label{sec_01_d-02} \Jsdpara\mu \Def \Jdpara\mu - \lp \sna\nu
\potBdpara\mu\nu + \frac{1}{2} \potBdpara\rho\sig \tor\mu\rho\sig
\rp \ee
We call this formula as a \emph{generalized Belinfante relation},
and a tensor $\bfpotBdpara$ --- as a \emph{generalized Belinfante
tensor}.

Assume that
\be\label{sec_01_d-10} \potBdpara\mu\nu = \A\alp\mu\nu \dpara\alp +
\B\alp\bet\mu\nu \na\bet \dpara\alp \ee
where
\be \A\alp{[\mu}{\nu]} = \A\alp\mu\nu; \qquad \B\alp\bet{[\mu}{\nu]}
= \B\alp\bet\mu\nu \ee
are tensors, which are to be determined. Then,
\bw \be\label{sec_01_d-03} \ba{l} \disp \sna\mu \potBdpara\mu\nu +
\frac12 \potBdpara\rho\sig \tor\mu\rho\sig =
\lf \sna\nu \A\alp\mu\nu + \frac12 \A\alp\rho\sig \tor\mu\rho\sig + \frac12 \B\lam\rho\sig\mu \cur\lam\alp\rho\sig \rf \dpara\alp\\
\disp + \lf -\A\alp\bet\mu + \sna\nu \B\alp\bet\mu\nu + \frac12
\B\alp\bet\rho\sig \tor\mu\rho\sig - \frac12 \B\alp\rho\sig\mu
\tor\bet\rho\sig \rf \na\bet \dpara\alp + \lf - \B\alp\bet\gam\mu
\rf \na{(\gam} \na{\bet)} \dpara\alp. \ea \ee \ew
Substituting \eqref{sec_01_d-01}, \eqref{sec_01_b-01},
\eqref{sec_01_d-03} into \eqref{sec_01_d-02} and equating
coefficients at $\lf \dpara\alp \rf$, $\lf \na\bet \dpara\alp \rf$,
$\lf \na{(\gam}\na{\bet)} \dpara\alp \rf$, one obtains the system of
equations for determining the tensors $\bfA$, $\bfB$, $\bfUs$:
\bw
\begin{empheq}[left=\empheqlbrace]{flalign}
\Us\alp\mu & = \U\alp\mu - \lp \sna\nu \A\alp\mu\nu + \frac{1}{2} \A\alp\rho\sig \tor\mu\rho\sig \rp - \frac{1}{2} \B\lam\rho\sig\mu \cur\lam\alp\rho\sig;\label{sec_01_d-04}\\
0 & = \M\alp\bet\gam + \A\alp\bet\gam - \lp \sna\nu \B\alp\bet\mu\nu + \frac{1}{2} \B\alp\bet\rho\sig \tor\mu\rho\sig - \frac{1}{2} \B\alp\rho\sig\mu \tor\bet\rho\sig \rp;\label{sec_01_d-05}\\
0 & = \N\alp\bet\gam\mu +
\B\alp{(\bet}{\gam)}\mu.\label{sec_01_d-06}
\end{empheq}
\ew
The system \eqref{sec_01_d-05}-\eqref{sec_01_d-06} for $\bfA$ and
$\bfB$ exactly coincides with one in
\eqref{sec_01_c-08}-\eqref{sec_01_c-09}; therefore, its solution is given
 by \eqref{sec_01_d-07+}-\eqref{sec_01_d-08+}. As a
consequence of \eqref{sec_01_d-08+}, one has
\be\label{sec_01_d-09} \B\lam{[\rho}{\sig]}\mu = \frac{2}{3}
\N\lam\mu{[\rho}{\sig]}. \ee
Substituting \eqref{sec_01_d-07+}, \eqref{sec_01_d-09} into
\eqref{sec_01_d-04}, we obtain

\be\label{sec_01_d-13} \ba{l}
\Us\alp\mu = \lp \U\alp\mu - \bfrac{1}{3} \N\lam\mu\rho\sig \cur\lam\alp\rho\sig \rp\\
+ \sna\nu \lb \M\alp{[\mu}{\nu]} - \bfrac{2}{3} \lp \sna\lam \N\alp\lam{[\mu}{\nu]} + \bfrac{1}{2} \tor{[\mu}\rho\sig \N\alp{\nu]}\rho\sig \rp \rb\\
+ \bfrac{1}{2} \lb \M\alp{[\rho}{\sig]} - \bfrac{2}{3} \lp \sna\lam
\N\alp\lam{[\rho}{\sig]} + \bfrac{1}{2} \tor{[\rho}\veps\kap
\N\alp{\sig]}\veps\kap \rp \rb \tor\mu\rho\sig. \ea \ee
Note that the right hand side of \eqref{sec_01_d-13} exactly
coincides with the left hand side of the Klein identity
\eqref{app_01_d-26}. Therefore, one can write also
\be\label{sec_01_d-11} \Us\alp\mu = -\Ib\alp\mu. \ee
Comparing formulae \eqref{sec_01_d-10} with \eqref{sec_01_c-06}, one
finds that \emph{the generalized Belinfante tensor coincides with the
generalized canonical superpotential:}
\be\label{sec_01_d-14} \ba{l}
\potBdpara\mu\nu = \potdpara\mu\nu\\
= \lf -\M{\alp}{[\mu}{\nu]} + \bfrac{2}{3} \lp \sna\lam \N{\alp}{\lam}{[\mu}{\nu]} + \bfrac{1}{2} \tor{[\mu}{\rho}{\sig} \N{\alp}{\nu]}{\rho}{\sig} \rp \rf \dpara\alp\\
+ \lf -\bfrac{4}{3} \N{\alp}{\bet}{[\mu}{\nu]} \rf \na\bet
\dpara\alp. \ea \ee
Combining this equality with \eqref{sec_01_c-11},
\eqref{sec_01_c-02} and \eqref{sec_01_d-02}, one finds that the
\emph{generalized symmetrized superpotential} $\bfpotsdpara \Def \{
\potsdpara{[\mu}{\nu]} = \potsdpara\mu\nu \}$ corresponding to the
current $\bfJsdpara$ is equal to zero identically:
\be \potsdpara\mu\nu = \potdpara\mu\nu - \potBdpara\mu\nu = 0. \ee

Recall that
\be \Us\alp\mu = \sems\mu\alp, \ee
then relation \eqref{sec_01_d-11} is a proof of the claim that
\emph{symmetrized EMT $\bfsems$ does not depend on divergences in
the Lagrangian}. Indeed, the right hand side of \eqref{sec_01_d-11}
essentially is defined by the variational derivative of the action (see definition \eqref{sec_01_d-12}), compare this also with Refs.~\cite{Szabados_1991, Szabados_1992}.
Already in Ref.~\cite{Bak_Cangemi_Jackiw_1994}, it was stated that the
Belinfante procedure applied both to the Hilbert Lagrangian and to
the non-covariant Einstein Lagrangian (differed by a divergence)
give the same result. An analogous statement (that divergences in
Lagrangians do not influence the Belinfante operation) has been
proved for perturbations on a fixed curved background in GR in Refs.~\cite{Petrov_Katz_1999, Petrov_Katz_2002} and in metric theories in the review of Petrov \cite{Petrov_2008}.

Finally, we stress the following: The generalized Belinfante relation
\eqref{sec_01_d-02} with accounting for formulae \eqref{sec_01_d-01}, \eqref{sec_01_d-11} (the latter has been obtained with the use of the Klein identity \eqref{app_01_d-26}) and \eqref{sec_01_d-14} coincides with the boundary Klein-Noether theorem \eqref{sec_01_c-13}. Therefore the \emph{success} of the Belinfante
approach is based on the Klein-Noether system of identities
\eqref{sec_01_c-01}-\eqref{app_01_d-14} only.

\section*{Acknowledgments}

The authors are very grateful to the referee for useful recommendations and D.I. Bondar for correcting English.

\appendix

\section{Main relations of the Riemann-Cartan geometry}\label{app_01_a-00}

The goal of this Appendix is to introduce main formulae of the
Riemann-Cartan geometry, which are necessary
in the text, and to identify notations for a reader.

Let $\caM$ be a $(D+1)$-dimensional real manifold with a coordinate
system $x\Def\{x^\mu\}$ defined on it. The \emph{Riemann-Cartan
geometry} is given on $\caM$ if smooth fields
\bn
\item of a symmetric covariant tensor (\emph{metric})
\be\label{app_01_a-01} \bfmet\Def\{\met\mu\nu(x)\}, \qquad
\met{(\mu}{\nu)} = \met\mu\nu, \ee
and
\item of an \emph{affine connection} compatible with the metric
\eqref{app_01_a-01},
\be \bfcon\Def\{\con\lam\mu\nu(x)\}, \ee
are defined on $\caM$.
\en
Because one sets that $\caM$ presents a spacetime the metric tensor
$\bfmet$ is of the Lorentzian signature:
\be \sign\bfmet = (-1,\underbrace{1, 1, \dots, 1}_{D\mbox{ times}}).
\ee
The metric determinant is denoted as
\be g \Def \det \{\met\mu\nu\}. \ee
The connection $\bfcon$ is not symmetrical in lower indexes:
\be \con\lam{(\mu}{\nu)} \neq \con\lam\mu\nu \ee
and defines covariant derivatives of a vector $\bfV=\{V^\mu(x)\}$ and
an $1$-form $\bfW=\{W_\mu(x)\}$ by the rules
\be \na\lam V^\mu \Def \pa\lam V^\mu + \con\mu\alp\lam V^\alp, \ee
\be \na\lam W_\mu \Def \pa\lam W_\mu - \con\alp\mu\lam W_\alp. \ee
A compatible condition of a connection $\bfcon$ with a metric
$\bfmet$ is presented as
\be\label{app_01_a-02} \na\lam\met\mu\nu = \pa\lam\met\mu\nu -
\con\alp\mu\lam \met\alp\nu - \con\alp\nu\lam \met\mu\alp = 0 \ee
meaning that a \emph{tensor of non-metricity} $\bfQ \Def
\{Q_{\lam,\mu\nu}(x)\} \Def \{\na\lam\met\mu\nu(x)\}$ is equal to
zero. Usually a set $(\caM,\bfmet,\bfcon)$ is denoted as $\caC(1,D)$
and is called as the \emph{Riemann-Cartan manifold}.

The \emph{torsion tensor} $\bftor \Def \{\tor\lam\mu\nu(x)\}$ and
the \emph{curvature tensor} $\bfcur \Def \{\cur\kap\lam\mu\nu(x)\}$
are defined by the relation
\be\label{app_01_b-08} \lp \na\mu\na\nu - \na\nu\na\mu \rp V^\lam
\Def - \tor\alp\mu\nu \na\alp V^\lam + \cur\lam\alp\mu\nu V^\alp \ee
and are expressed through the connection as follows
\be \tor\lam\mu\nu = -2\con\lam{[\mu}{\nu]}; \ee
\be \cur\kap\lam\mu\nu = \pa\mu\con\kap\lam\nu -
\pa\nu\con\kap\lam\mu + \con\kap\alp\mu \con\alp\lam\nu -
\con\kap\alp\nu \con\alp\lam\mu. \ee
As is seen, the torsion tensor $\bftor$ is antisymmetric in lower
indexes:
\be \tor\lam{[\mu}{\nu]} = \tor\lam\mu\nu. \ee
Rising or lowering indexes for a torsion tensor $\bftor$, we remark
their places by coma. For example,
\be T^{\lam,\,\mu}{}_\nu \Def \metu\mu\alp \tor\lam\alp\nu; \qquad
T^{\lam,\,\mu\nu} \Def \metu\mu\alp \metu\nu\bet \tor\lam\alp\bet.
\ee

A compatible condition \eqref{app_01_a-02} permits to express a
connection $\bfcon$ trough both derivatives of the metric
$\{\pa\alp\met\mu\nu\}$ and the torsion tensor $\{\tor\lam\mu\nu\}$ in
an unique way. Thus,
\be \con\lam\mu\nu = \metu\lam\alp \cond\alp\mu\nu \ee
where
\be \ba{r}
\cond\alp\mu\nu = \bfrac{1}{2}\lp \pa\mu\met\alp\nu + \pa\nu\met\alp\mu - \pa\alp\met\mu\nu \rp\\
+ \bfrac{1}{2}\lp \tord\mu\alp\nu + \tord\nu\alp\mu -
\tord\alp\mu\nu \rp \ea \ee
and
\be \tord\lam\mu\nu \Def \met\lam\alp \tor\alp\mu\nu. \ee

In the Riemann-Cartan $(D+1)$-dimensional geometry, the fully
covariant curvature tensor
\be \curd\kap\lam\mu\nu \Def \met\kap\alp \cur\alp\lam\mu\nu \ee
has $(D+1)^2 D^2/4$ essential components, is antisymmetrical both
in the first pair of indexes:
\be\label{app_01_c-13} \curd{[\kap}{\lam]}\mu\nu =
\curd\kap\lam\mu\nu \ee
and in the second pair of indexes:
\be \curd\kap\lam{[\mu}{\nu]} = \curd\kap\lam\mu\nu, \ee
but, unlike Riemannian geometry, it is not symmetrical in
\emph{pairs} of indexes:
\be \curd\mu\nu\kap\lam \neq \curd\kap\lam\mu\nu. \ee
Tensor $\bfcur$ satisfies the generalized \emph{Ricci
identities}
\be\label{app_01_a-06} \cur\kap{[\lam}\mu{\nu]} \eq
\na{[\lam}\tor\kap\mu{\nu]} + \tor\kap\alp{[\lam} \tor\alp\mu{\nu]}
\ee
and the generalized \emph{Bianchi identities}
\be\label{app_01_a-06+}  \na{[\lam|}\cur\alp\bet{|\mu}{\nu]} \eq -
\cur\alp\bet\gam{[\lam} \tor\gam\mu{\nu]}. \ee

We need also in the \emph{torsion vector} $\{T_\mu\}$,
\emph{modified torsion tensor} $\bfstor\Def\{\stor\lam\mu\nu\}$ and
\emph{modified torsion vector} $\{\sT_\mu\}$ defined as
\be\label{app_01_a-04} T_\mu \Def \tor\alp\mu\alp; \ee
\be\label{app_01_a-03} \stor\lam\mu\nu \Def \tor\lam\mu\nu +
\kro\lam\mu T_\nu - \kro\lam\nu T_\mu; \ee
\be\label{app_01_a-05} \sT_\mu \Def \stor\alp\mu\alp, \ee
respectively, where $\kro\mu\nu$ is the Kronecker symbol. It is
easily to find a relation between the torsion vector and the
modified torsion vector:
\be \sT_\mu=-(D-1)T_\mu. \ee

The \emph{Ricci} and \emph{Einstein tensors}, and the
\emph{curvature scalar} are defined as usual:
\be \ric\mu\nu \Def \cur\alp\mu\alp\nu, \ee
\be\label{app_01_a-08} \ein\mu\nu \Def \ric\mu\nu -
\frac{1}{2}\met\mu\nu R, \ee
\be R \Def \metu\alp\bet \ric\alp\bet. \ee
The first two are not symmetrical now:
\be \ric{(\mu}{\nu)} \neq \ric\mu\nu; \qquad \ein{(\mu}{\nu)} \neq
\ein\mu\nu. \ee
Contracting the Ricci identities \eqref{app_01_a-06}, one easily
states that an antisymmetrical part of the Ricci tensor satisfies
the identity:
\be\label{app_01_c-06} \ric{[\mu}{\nu]} \eq -\frac{1}{2}
\sna\lam\stor\lam\mu\nu \ee
where
\be\label{app_01_a-07} \sna\lam \Def \na\lam + T_\lam \ee
is a \emph{modified covariant derivative}. From the definition
\eqref{app_01_a-07} it follows, first, that the commutator of
modified covariant derivatives is presented by
\be\label{app_01_c-05} \sna{[\mu}, \sna{\nu]} = \na{[\mu}, \na{\nu]}
+ \lp\na{[\mu}T_{\nu]}\rp, \ee
second, that for arbitrary two tensors
$\bfA=\{A^{\alp\dots}{}_{\bet\dots}\}$ and
$\bfB=\{B^{\gam\dots}{}_{\del\dots}\}$ the \emph{modified(!) formula
of differentiating their product (the modified Leibnitz rule)}:
\be\label{sec_01_b-03} \ba{l}
\sna\mu\lp A^{\alp\dots}{}_{\bet\dots} B^{\gam\dots}{}_{\del\dots} \rp\\
= \lp \sna\mu A^{\alp\dots}{}_{\bet\dots} \rp
B^{\gam\dots}{}_{\del\dots} + A^{\alp\dots}{}_{\bet\dots} \lp \na\mu
B^{\gam\dots}{}_{\del\dots} \rp \ea \ee
takes a place. It is also equivalent to
\be \ba{l}
\sna\mu\lp A^{\alp\dots}{}_{\bet\dots} B^{\gam\dots}{}_{\del\dots} \rp\\
= \lp \na\mu A^{\alp\dots}{}_{\bet\dots} \rp
B^{\gam\dots}{}_{\del\dots} + A^{\alp\dots}{}_{\bet\dots} \lp
\sna\mu B^{\gam\dots}{}_{\del\dots} \rp. \ea \ee
The last formula represented in the form
\be \ba{l}
A^{\alp\dots}{}_{\bet\dots} \lp \sna\mu B^{\gam\dots}{}_{\del\dots} \rp\\
= \sna\mu\lp A^{\alp\dots}{}_{\bet\dots} B^{\gam\dots}{}_{\del\dots}
\rp - \lp \na\mu A^{\alp\dots}{}_{\bet\dots} \rp
B^{\gam\dots}{}_{\del\dots} \ea \ee
we call as the \emph{formula of a differentiation by parts}, and we
use it actively.

The \emph{two times contracted} Bianchi identities
\eqref{app_01_a-06+} acquire the form:
\be
\sna\mu E^\mu{}_\nu \equiv - E^\mu{}_\lam \tor\lam\mu\nu + \frac12
\storu\pi\rho\sig \curd\rho\sig\pi\nu.
\ee
Thus, in the Riemann-Cartan geometry, unlike Riemannian geometry,
the Einstein tensor \eqref{app_01_a-08} is not conserved.

The \emph{contorsion tensor} $\bfctor=\{\ctor\lam\mu\nu\}$ defined
as
\bea
\ctor\lam\mu\nu \Def \metu\lam\alp \ctord\alp\mu\nu;\\
\ctord\lam\mu\nu \Def \frac{1}{2} \lp \tord\mu\lam\nu +
\tord\nu\lam\mu - \tord\lam\mu\nu \rp \eea
is useful also. The {\it contorsion vector} defined as
\be K_\mu \Def \ctor\alp\mu\alp \ee
is expressed trough the torsion vector \eqref{app_01_a-04} as
\be K_\mu = -T_\mu. \ee
The \emph{Gauss} and \emph{Stockes formulae} are modified
essentially with respect to ones in the Riemannian geometry and have
the form
\be \label{sec_01_a-01} \ba{c} \intOme\bfdome \sna\mu V^\mu = \intdOme
\bfdsig\mu V^\mu; \ea \ee
\be\label{sec_01_c-12} \ba{c} \intSig \bfdsig{[\mu} \sna{\nu]} W^{\mu\nu} +
{\bfrac{1}{2}} \intSig \bfdsig\lam \tor\lam\mu\nu W^{\mu\nu} =
{\bfrac{1}{2!}} \intdSig \bfds\mu\nu W^{\mu\nu}, \ea \ee
respectively. Here, $\{V^\mu\}$ and $\{W^{\mu\nu}\}$ are arbitrary
contravariant vector and antisymmetrical contravariant tensor,
$W^{[\mu\nu]}=W^{\mu\nu}$; $\Ome$ and $\partial\Ome$ are an
arbitrary $(D+1)$-dimensional domain in $\caC(1,D)$ and its
$D$-dimensional boundary; $\Sig$ and $\partial\Sig$ are an arbitrary
$D$-dimensional hypersurface in $\caC(1,D)$ and its
$(D-1)$-dimensional boundary; the notations
\be \bfdome \Def \frac{\rmet}{(D+1)!}
\veps_{{\alp_0}{\alp_1}{\dots}{\alp_D}}
\bfdx{\alp_0}\we\bfdx{\alp_1}\we\dots\bfdx{\alp_D}, \ee
\be \bfdsig\lam \Def \frac{\rmet}{D!}
\veps_{{\lam}{\alp_1}{\dots}{\alp_D}}
\bfdx{\alp_1}\we\bfdx{\alp_2}\we\dots\bfdx{\alp_D}, \ee
\be \bfds\mu\nu \Def \frac{\rmet}{(D-1)!}
\veps_{{\mu}{\nu}{\alp_1}{\alp_2}{\dots}{\alp_{D-1}}}
\bfdx{\alp_1}\we\bfdx{\alp_2}\we\dots\bfdx{\alp_{D-1}} \ee
mean $(D+1)$-, $D$- and $(D-1)$-forms of elementary volumes;
$\bfveps\Def\{\veps_{{\alp_0}{\alp_1}{\dots}{\alp_D}}\}$ is the
fully antisymmetrical $(D+1)$-dimensional Levi-Civita symbol,
\be \veps_{[{\alp_0}{\alp_1}{\dots}{\alp_D}]} =
\veps_{{\alp_0}{\alp_1}{\dots}{\alp_D}}, \qquad \veps_{012\dots D} =
+1; \ee
$\bfdx\alp$ are the basic $1$-forms, a symbol $\we$ means a wedge
product.

\section{Irreducible representations of a symmetric group for two- and
three-index quantities}\label{app_01_b-00}

To provide many of calculations we decompose tensors onto irreducible representations of a symmetric
group (group of index permutations). In this Appendix, we give main
properties of such representations for $2$- and $3$-index quantities and formulae necessary in the
text.
\subsection{The Young projectors for $2$-index quantities}\label{app_01_b-01}

For $2$-index quantities $\{A_{\dots}{}^{\bet\gam}\}$ one has $2$
\emph{Young diagrams} only:
\be \youtaa12 \qquad\mbox{and}\qquad \youtab12, \ee
to which the \emph{Young projectors}:
\be\label{app_01_b-05} \yous{\youtaa12} \Def \frac{1}{2} \lp (12) +
(21) \rp \ee
and
\be\label{app_01_b-13} \youa{\youtab12} \Def \frac{1}{2} \lp (12) -
(21) \rp \ee
correspond. Projectors \eqref{app_01_b-05} and \eqref{app_01_b-13}
act as follows
\be\label{app_01_b-07} \yous{\youtaa12} A_{\dots}{}^{\bet\gam} =
\frac{1}{2} \lp A_{\dots}{}^{\bet\gam} + A_{\dots}{}^{\gam\bet} \rp
= A_{\dots}{}^{(\bet\gam)} \ee
and
\be\label{app_01_b-14} \youa{\youtab12} A_{\dots}{}^{\bet\gam} =
\frac{1}{2} \lp A_{\dots}{}^{\bet\gam} - A_{\dots}{}^{\gam\bet} \rp
= A_{\dots}{}^{[\bet\gam]}. \ee
They are orthonormal:
\be \lf\ba{l}
\yous{\youtaa12} \yous{\youtaa12} = \yous{\youtaa12};\\
\youa{\youtab12} \youa{\youtab12} = \youa{\youtab12};\\
\yous{\youtaa12} \youa{\youtab12} = \youa{\youtab12}
\yous{\youtaa12} =0 \ea\rd \ee
and present themselves a full set:
\be\label{app_01_b-06} \yous{\youtaa12} + \youa{\youtab12} = 1. \ee
\subsection{A transformation of the expression $\M\alp\bet\gam \na\gam\na\bet \dpara\alp$}\label{app_01_b-02}

Using \eqref{app_01_b-06}, \eqref{app_01_b-07} and
\eqref{app_01_b-14}, one finds

\be\label{app_01_c-09} \ba{l}
\M\alp\bet\gam = 1\cdot\M\alp\bet\gam\\
= \lp \yous{\youtaa12} + \youa{\youtab12} \rp \M\alp\bet\gam\\
= \yous{\youtaa12}\M\alp\bet\gam + \youa{\youtab12}\M\alp\bet\gam\\
= \M{\alp}{(\bet}{\gam)} + \M{\alp}{[\bet}{\gam]}. \ea \ee
From here for an arbitrary vector $\dbfpara = \{\dpara\alp(x)\}$ one
has
\be\label{app_01_b-09} \ba{l}
\M\alp\bet\gam \na\gam\na\bet \dpara\alp = \M{\alp}{(\bet}{\gam)} \na\gam\na\bet \dpara\alp + \M{\alp}{[\bet}{\gam]} \na\gam\na\bet \dpara\alp\\
= \M\alp\bet\gam \na{(\gam}\na{\bet)}\dpara\alp + \M\alp\bet\gam
\na{[\gam}\na{\bet]}\dpara\alp. \ea \ee
Using in the last term the formula \eqref{app_01_b-08} for a
commutator of covariant derivatives, one obtains
\be \na{[\gam}\na{\bet]}\dpara\alp =  - \frac{1}{2}
\tor\veps\gam\bet \na\veps\dpara\alp + \frac{1}{2}
\cur\alp\veps\gam\bet \dpara\veps. \ee
Then the expression \eqref{app_01_b-09} transforms into
\be\label{sec_01_a-10} \ba{l}
\M\alp\bet\gam \na\gam\na\bet\dpara\alp = \lf \bfrac{1}{2} \cur\veps\alp\kap\lam \M\veps\lam\kap \rf \dpara\alp\\
+ \lf -\bfrac{1}{2} \tor\bet\kap\lam \M\alp\lam\kap \rf
\na\bet\dpara\alp + \lf \M\alp\bet\gam \rf
\na{(\gam}\na{\bet)}\dpara\alp. \ea \ee
\subsection{The Young projectors for $3$-index quantities}\label{app_01_b-03}

For $3$-index quantities $\{ A_{\dots}{}^{\bet\gam\del}\}$ one has
$4$ different Young diagram:
\be \youtba123, \qquad \youtbb213, \qquad \youtbb231
\qquad\mbox{and}\qquad \youtbc123. \ee
In this case, there are two \emph{different} full orthonormal sets
of the Young projectors. Their construction is carried out as
follows. For the \emph{sequence $I$} from the beginning one provides
an antisymmetrization in indexes in a column, only after that one
provides a symmetrization in indexes in a line; for the
\emph{sequence $II$}, inversely, from the beginning one provides a
symmetrization in indexes in a line, only after that one provides an
antisymmetrization in indexes in a column. Thus
\bw
\begin{empheq}[left=I\,\empheqlbrace]{flalign}
\yous{\youtba123} & \Def \bfrac{1}{6} \lp (123) + (132) + (312) +
(321) + (231) + (213) \rp
;\\
\yous{\youtbb213} & \Def \bfrac{1}{3} \lp (123) - (132) + (213) - (231) \rp;\label{app_01_d-15} \\
\yous{\youtbb231} & \Def \bfrac{1}{3} \lp (123) - (213) + (132) - (312) \rp; \\
\youa{\youtbc123} & \Def \bfrac{1}{6} \lp (123) - (132) + (312) -
(321) + (231) - (213) \rp
\end{empheq}
and
\begin{empheq}[left=II\,\empheqlbrace]{flalign}
\yous{\youtba123} & \Def \bfrac{1}{6} \lp (123) + (132) + (312) + (321) + (231) + (213) \rp;\label{app_01_d-05} \\
\youa{\youtbb213} & \Def \bfrac{1}{3} \lp (123) + (213) - (132) - (312) \rp;\label{app_01_d-06} \\
\youa{\youtbb231} & \Def \bfrac{1}{3} \lp (123) + (132) - (213) - (231) \rp;\label{app_01_d-07} \\
\youa{\youtbc123} & \Def \bfrac{1}{6} \lp (123) - (132) + (312) -
(321) + (231) - (213) \rp.\label{app_01_d-08}
\end{empheq}
\ew
As a result of action of the operator $\yous{\dots}$ onto a
$3$-index quantity, one obtains a \emph{symmetrical} in
correspondent indexes quantity, and, analogously, after action of
the operator $\youa{\dots}$ one obtains an \emph{antisymmetrical} in
correspondent indexes quantity.
\subsection{A transformation of the expression
$\N\alp\bet\gam\del \na\del\na\gam\na\bet \dpara\alp$}\label{app_01_b-04}

To transform the expression $\N\alp\bet\gam\del
\na\del\na\gam\na\bet \dpara\alp$ we use the set of projectors $II$.
Decompose the tensor $\N{\alp}{(\bet}{\gam)}{\del} =
\N\alp\bet\gam\del$ onto irreducible with respect to this set parts:
\bw
\be \ba{l}
\N\alp\bet\gam\del = 1\cdot\N\alp\bet\gam\del = \lp \yous{\youtba123} + \youa{\youtbb213} + \youa{\youtbc123} + \youa{\youtbb231} \rp \N\alp\bet\gam\del\\
= \yous{\youtba123}\N\alp\bet\gam\del + \youa{\youtbb213}\N\alp\bet\gam\del + \youa{\youtbc123}\N\alp\bet\gam\del + \youa{\youtbb231}\N\alp\bet\gam\del\\
\Def \Na\alp\bet\gam\del + \Nb\alp\bet\gam\del + \Nc\alp\bet\gam\del
+ \Nd\alp\bet\gam\del. \ea \ee
\ew
Here,
\bea \Na\alp\bet\gam\del \Def \yous{\youtba123}\N\alp\bet\gam\del =
\N{\alp}{(\bet}{\gam}{\del)}; \eea
\bea
\Nb\alp\bet\gam\del \Def \youa{\youtbb213}\N\alp\bet\gam\del\nonumber\\
= \bfrac{1}{3} \lp \N\alp\bet\gam\del + \N\alp\gam\bet\del - \N\alp\bet\del\gam - \N\alp\del\bet\gam \rp\nonumber\\
=\bfrac{2}{3} \lp \N\alp\bet\gam\del - \N\alp\bet\del\gam \rp =
\bfrac{4}{3} \N{\alp}{\bet}{[\gam}{\del]}; \eea
\bea \Nc\alp\bet\gam\del \Def \youa{\youtbc123}\N\alp\bet\gam\del =
\N{\alp}{[\bet}{\gam}{\del]} = 0; \eea
\bea
\Nd\alp\bet\gam\del \Def \youa{\youtbb231}\N\alp\bet\gam\del\nonumber\\
= \bfrac{1}{3} \lp \N\alp\bet\gam\del + \N\alp\bet\del\gam - \N\alp\gam\bet\del - \N\alp\gam\del\bet \rp\nonumber\\
=\bfrac{1}{3} \lp \N\alp\del\bet\gam - \N\alp\del\gam\bet \rp =
\bfrac{2}{3} \N{\alp}{\del}{[\bet}{\gam]}. \eea
Thus,
\be\label{app_01_c-12} \N\alp\bet\gam\del =
\N{\alp}{(\bet}{\gam}{\del)} + \bfrac{4}{3}
\N{\alp}{\bet}{[\gam}{\del]} + \bfrac{2}{3}
\N{\alp}{\del}{[\bet}{\gam]}. \ee
Then
\be\label{app_01_b-12} \ba{l}
\N\alp\bet\gam\del \na\del\na\gam\na\bet \dpara\alp = \N\alp\bet\gam\del \na{(\del}\na\gam\na{\bet)} \dpara\alp\\
+ \bfrac{4}{3} \N\alp\bet\gam\del \na{[\del}\na{\gam]}\na\bet
\dpara\alp + \bfrac{2}{3} \N\alp\del\bet\gam
\na\del\na{[\gam}\na{\bet]} \dpara\alp. \ea \ee
Here, to calculate $2$-nd and $3$-rd terms on the right hand side,
firstly, we apply the formulae for the commutator of the type
\eqref{app_01_b-08}, next, use a decomposition of a $2$-index
quantity of the type \eqref{sec_01_a-10}, at last, apply again the
formula \eqref{app_01_b-08}. After collecting similar terms we
obtain
\bw
\be\label{app_01_b-10} \ba{l}
\bfrac{4}{3} \N\alp\bet\gam\del \na{[\del}\na{\gam]}\na\bet \dpara\alp = \lf \bfrac{1}{3}
\N\kap\lam\mu\nu \tor\sig\mu\nu \cur\kap\alp\sig\lam \rf \dpara\alp\\
+ \lf \bfrac{2}{3} \N\alp\lam\mu\nu \lp \bfrac{1}{2} \tor\sig\mu\nu
\tor\bet\lam\sig + \cur\bet\lam\mu\nu \rp - \bfrac{2}{3}
\N\lam\bet\mu\nu \cur\lam\alp\mu\nu \rf \na\bet\dpara\alp + \lf
\bfrac{2}{3} \N\alp\bet\mu\nu \tor\gam\mu\nu \rf
\na{(\gam}\na{\bet)}\dpara\alp; \ea \ee
\be\label{app_01_b-11} \ba{l}
\bfrac{2}{3} \N\alp\del\bet\gam \na\del\na{[\gam}\na{\bet]} \dpara\alp = \lf \bfrac{1}{3} \N\kap\lam\mu\nu \lp \bfrac{1}{2} \tor\sig\mu\nu \cur\kap\alp\lam\sig - \na\lam \cur\kap\alp\mu\nu \rp \rf \dpara\alp\\
+ \lf \bfrac{1}{3} \N\alp\lam\mu\nu \lp \na\lam \tor\bet\mu\nu +
\bfrac{1}{2} \tor\sig\mu\nu \tor\bet\sig\lam \rp - \bfrac{1}{3}
\N\kap\bet\mu\nu \cur\kap\alp\mu\nu \rf \na\bet \dpara\alp + \lf
\bfrac{1}{3} \N\alp\gam\mu\nu \tor\bet\mu\nu \rf
\na{(\gam}\na{\bet)} \dpara\alp. \ea \ee

Substituting \eqref{app_01_b-10} and \eqref{app_01_b-11} into
\eqref{app_01_b-12}, we obtain finally
\be\label{sec_01_a-11} \ba{l}
\N\alp\bet\gam\del \na\del\na\gam\na\bet \dpara\alp = \lf -\bfrac{1}{3} \N\kap\lam\mu\nu \lp \na\lam \cur\kap\alp\mu\nu + \bfrac{1}{2} \tor\sig\mu\nu \cur\kap\alp\lam\sig \rp \rf \dpara\alp\\
+ \lf \bfrac{1}{3} \N\alp\lam\mu\nu \lp 2\cur\bet\lam\mu\nu + \na\lam \tor\bet\mu\nu + \bfrac{1}{2} \tor\sig\mu\nu \tor\bet\lam\sig \rp - \N\kap\bet\mu\nu \cur\kap\alp\mu\nu \rf \na\bet \dpara\alp\\
+ \lf \N\alp\bet\mu\nu \tor\gam\mu\nu \rf \na{(\gam}\na{\bet)}
\dpara\alp + \lf \N\alp\bet\gam\del \rf \na{(\del}\na\gam\na{\bet)}
\dpara\alp. \ea \ee
\ew

\section{Transformation of the Klein identities}\label{app_01_c-00}

\subsection{Three useful identities}\label{app_01_c-01}

For arbitrary tensors $\{\pota{[\mu}{\nu]}=\pota\mu\nu\}$,
$\{\potb{\alp}{[\mu}{\nu]}=\potb\alp\mu\nu\}$ and
$\{\potc{\alp}{\bet}{[\mu}{\nu]}=\potc\alp\bet\mu\nu\}$ the
identities

\be\label{app_01_c-07} \sna\mu \lb \sna\nu \pota\mu\nu +
\bfrac{1}{2} \pota\rho\sig \tor\mu\rho\sig \rb \eq 0; \ee

\be\label{sec_01_b-16} \sna\mu \lb \sna\nu \potb\alp\mu\nu +
\bfrac{1}{2}\potb\alp\rho\sig \tor\mu\rho\sig \rb \eq -\bfrac{1}{2}
\cur\lam\alp\rho\sig \potb\lam\rho\sig; \ee
\be\label{app_01_c-08} \ba{l}
\sna\mu \lb \sna\nu \potc\alp\bet\mu\nu + \bfrac{1}{2} \potc\alp\bet\rho\sig \tor\mu\rho\sig \rb\\
\qquad\qquad \eq \bfrac{1}{2} \lp -\cur\lam\alp\rho\sig
\potc\lam\bet\rho\sig + \cur\bet\lam\rho\sig \potc\alp\lam\rho\sig
\rp; \ea \ee
take a place. Let us prove, for example, the $1$-st one. Using the
formulae for commutators of the types \eqref{app_01_b-08} and
\eqref{app_01_c-05}, one has
\bse
\ba{l}
\sna\mu \lp \sna\nu \pota\mu\nu \rp = \sna{[\mu}\sna{\nu]}\pota\mu\nu =\\
\lp \na{[\mu}T_{\nu]} \rp \pota\mu\nu - \bfrac{1}{2} \tor\lam\mu\nu
\na\lam \pota\mu\nu + \bfrac{1}{2} \cur\mu\lam\mu\nu \pota\lam\nu
+ \bfrac{1}{2} \cur\nu\lam\mu\nu \pota\mu\lam\\
= \lp \sna{[\mu}T_{\nu]} \rp \pota\mu\nu - \sna\lam \lp \bfrac{1}{2} \tor\lam\mu\nu \pota\mu\nu \rp + \bfrac{1}{2} \lp \sna\lam \tor\lam\mu\nu \rp \pota\mu\nu\\
+ \ric{[\mu}{\nu]} \pota\mu\nu. \ea
\ese
The sum of the $1$-st and $3$-rd terms with taking into account
\eqref{app_01_a-03} is equal to
\bse
\frac{1}{2} \sna\lam \lp \kro\lam\mu T_\nu - \kro\lam\nu T_\mu +
\tor\lam\mu\nu \rp = \frac{1}{2} \sna\lam \stor\lam\mu\nu.
\ese
Then, keeping in mind the identity \eqref{app_01_c-06}, one obtains
finally
\bse
\sna\mu \lp \sna\nu \pota\mu\nu \rp \eq -\sna\lam \lp \frac{1}{2} \tor\lam\mu\nu \pota\mu\nu \rp
\ese
that coincides exactly with \eqref{app_01_c-07}. The identities
\eqref{sec_01_b-16} and \eqref{app_01_c-08} are proved
analogously.

\subsection{Representation of the Klein identity \eqref{sec_01_b-10} in the form \eqref{sec_01_b-15}}\label{app_01_c-02}

To transfer from the formula \eqref{sec_01_b-10} to the formula
\eqref{sec_01_b-15} it is enough in
$-\frac{1}{3} \N\kap\lam\mu\nu \sna\lam \cur\kap\alp\mu\nu$
to provide a differentiation by parts and collect similar
terms.

\subsection{Representation of the Klein identity \eqref{sec_01_b-14} in the form \eqref{sec_01_b-12}}\label{app_01_c-03}

We transform \eqref{sec_01_b-14} step by step as follows.
\bn
\item\label{app_01_c-10} In $\sna\mu \M\alp\bet\mu$ the quantity
$\M\alp\bet\mu$ is decomposed onto irreducible parts with
correspondence to the formula \eqref{app_01_c-09}:
    \bse
    \sna\mu \M\alp\bet\mu = \sna\mu \M{\alp}{[\bet}{\mu]} + \M{\alp}{(\bet}{\mu)}.
    \ese
\item In $\frac{1}{3} \N\alp\lam\mu\nu \na\lam \tor\bet\mu\nu$ a differentiation by parts is provided:
    \bse
    \ba{l}\frac{1}{3} \N\alp\lam\mu\nu \na\lam \tor\bet\mu\nu\\ = \sna\mu \lp \frac{1}{3} \N\alp\mu\rho\sig \tor\bet\rho\sig \rp + \frac{1}{2} \lp -\frac{2}{3} \sna\lam \N{\alp}{\lam}{[\rho}{\sig]} \rp \tor\bet\rho\sig.\ea
    \ese
\item The term $\frac{1}{3} \N\alp\lam\mu\nu \lp \frac{1}{2} \tor\sig\mu\nu \tor\bet\lam\sig \rp$ is rewritten in the way:
    \bse
    \frac{1}{3} \N\alp\lam\mu\nu \lp \frac{1}{2} \tor\sig\mu\nu \tor\bet\lam\sig \rp = \frac{1}{2} \lp \frac{1}{3} \N{\alp}{[\rho|}{\kap}{\lam} \tor{|\sig]}{\kap}{\lam} \rp \tor\bet\rho\sig.
    \ese
\item\label{app_01_c-11} The terms $\frac{2}{3} \N\alp\lam\mu\nu \cur\bet\lam\mu\nu - \N\lam\bet\mu\nu \cur\lam\alp\mu\nu$ are represented as
    \bse
    \ba{l}\bfrac{2}{3} \N\alp\lam\mu\nu \cur\bet\lam\mu\nu - \N\lam\bet\mu\nu \cur\lam\alp\mu\nu = -\bfrac{1}{3} \N\lam\bet\rho\sig \cur\lam\alp\rho\sig\\ + \bfrac{1}{2} \lp \bfrac{4}{3} \N\alp\lam\rho\sig \cur\bet\lam\rho\sig - \bfrac{4}{3} \N\lam\bet\rho\sig \cur\lam\alp\rho\sig \rp.\ea
    \ese
\item The above points \ref{app_01_c-10}--\ref{app_01_c-11} are taken into account in the identity
\eqref{sec_01_b-14}, and the expression
    \bse
    \ba{l}\sna\mu \lp -\frac{2}{3} \sna\lam \N{\alp}{\lam}{[\bet}{\mu]} + \frac{1}{3} \N{\alp}{[\bet|}{\rho}{\sig} \tor{|\mu]}{\rho}{\sig} \rp\\ + \sna\mu \lp \sna\lam \N\alp\bet\mu\lam + \N{\alp}{(\bet|}{\rho}{\sig} \tor{|\mu)}{\rho}{\sig} \rp\ea
    \ese
    is added and subtracted in the left hand side of \eqref{sec_01_b-14}. Then the left hand side of \eqref{sec_01_b-14} acquires the form:
    \bw
    \bse
    \ba{l} \lp \U\alp\bet - \bfrac{1}{3} \N\lam\bet\rho\sig \cur\lam\alp\rho\sig \rp\\ + \sna\mu \lp \M{\alp}{[\bet}{\mu]} - \bfrac{2}{3} \sna\lam \N{\alp}{\lam}{[\bet}{\mu]} + \bfrac{1}{3} \N{\alp}{[\bet|}{\rho}{\sig} \tor{|\mu]}{\rho}{\sig} \rp + \bfrac{1}{2} \lp \M{\alp}{[\rho}{\sig]} - \bfrac{2}{3} \sna\lam \N{\alp}{\lam}{[\rho}{\sig]} + \bfrac{1}{3} \N{\alp}{[\rho|}{\kap}{\lam} \tor{|\sig]}{\kap}{\lam} \rp \tor\bet\rho\sig\\
     + \sna\mu \lp \M{\alp}{(\bet}{\mu)} + \sna\lam \N\alp\bet\mu\lam + \N{\alp}{(\bet|}{\rho}{\sig} \tor{|\mu)}{\rho}{\sig} \rp + \bfrac{1}{2} \lp \bfrac{4}{3} \N\alp\lam\rho\sig \cur\bet\lam\rho\sig - \bfrac{4}{3} \N\alp\bet\rho\sig \cur\lam\alp\rho\sig \rp\\
     + \sna\mu \lp \bfrac{1}{3} \N\alp\mu\rho\sig \tor\bet\rho\sig \rp - \sna\mu \lp -\bfrac{2}{3} \sna\lam \N{\alp}{\lam}{[\bet}{\mu]} + \bfrac{1}{3} \N{\alp}{[\bet|}{\rho}{\sig} \tor{|\mu]}{\rho}{\sig} \rp - \sna\mu \lp \sna\lam \N\alp\bet\mu\lam + \N{\alp}{(\bet|}{\rho}{\sig} \tor{|\mu)}{\rho}{\sig} \rp. \ea
    \ese
    \ew
\item Here, the sum of the last three terms is transformed to the form

    \bse
    \ba{l} -\sna\mu \lb \sna\lam \lp \N\alp\bet\mu\lam - \bfrac{2}{3} \N{\alp}{\lam}{[\bet}{\mu]} \rp + \bfrac{1}{2} \lp \bfrac{4}{3} \N{\alp}{\bet}{[\rho}{\sig]} \rp \tor\mu\rho\sig \rb\\ = -\sna\mu \lp \sna\lam \N{\alp}{(\bet}{\mu}{\lam)} \rp\\ - \sna\mu \lb \sna\lam \lp \bfrac{4}{3} \N{\alp}{\bet}{[\mu}{\lam]} \rp + \bfrac{1}{2} \lp \bfrac{4}{3} \N{\alp}{\bet}{[\rho}{\sig]} \rp \tor\mu\rho\sig \rb \ea
    \ese
    where the decomposition \eqref{app_01_c-12} has been used.

\item At last, setting in the identity \eqref{app_01_c-08} $\potc\alp\bet\mu\nu = \frac{4}{3} \N{\alp}{\bet}{[\mu}{\nu]}$, one can see that
    \bse
    \ba{l}-\sna\mu \lb \sna\lam \lp \bfrac{4}{3} \N{\alp}{\bet}{[\mu}{\lam]} \rp + \bfrac{1}{2} \lp \bfrac{4}{3} \N{\alp}{\bet}{[\rho}{\sig]} \rp \tor\mu\rho\sig \rb\\
    \quad + \bfrac{1}{2} \lp \bfrac{4}{3} \N\alp\lam\rho\sig \cur\bet\lam\rho\sig - \bfrac{4}{3} \N\lam\bet\rho\sig \cur\lam\alp\rho\sig \rp \eq 0.\ea
    \ese
\en

In the result the Klein identity \eqref{sec_01_b-14} is represented
in the equivalent form:
\bw
\be
\ba{l}
\lp \U\alp\bet - \bfrac{1}{3} \N\lam\bet\rho\sig \cur\lam\alp\rho\sig \rp\\
+ \sna\mu \lp \M{\alp}{[\bet}{\mu]} - \bfrac{2}{3} \sna\lam
\N{\alp}{\lam}{[\bet}{\mu]} + \bfrac{1}{3}
\N{\alp}{[\bet|}{\rho}{\sig} \tor{|\mu]}{\rho}{\sig} \rp +
\bfrac{1}{2} \lp \M{\alp}{[\rho}{\sig]} - \bfrac{2}{3} \sna\lam
\N{\alp}{\lam}{[\rho}{\sig]} +
\bfrac{1}{3} \N{\alp}{[\rho|}{\kap}{\lam} \tor{|\sig]}{\kap}{\lam} \rp \tor\bet\rho\sig\\
+ \sna\mu \lp \M{\alp}{(\bet}{\mu)} + \sna\lam \N\alp\bet\mu\lam +
\N{\alp}{(\bet|}{\rho}{\sig} \tor{|\mu)}{\rho}{\sig} \rp -
\sna\mu\sna\lam \N{\alp}{(\bet}{\mu}{\lam)} \eq -\Ib\alp\bet;
\ea
\ee
\ew

\subsection{The proof of the formula \eqref{app_01_c-14}}\label{app_01_c-04}
The identity \eqref{sec_01_b-16} after using the antisymmetry
property of the curvature tensor \eqref{app_01_c-13} transforms to
\be \sna\mu \lb \sna\nu \potbu\lam\mu\nu +
\frac{1}{2}\potbu\lam\rho\sig \tor\mu\rho\sig \rb \eq \frac{1}{2}
\cur\lam\kap\mu\nu \potbu\kap\mu\nu. \ee
Using this identity for the expression $\Del \rsJ^\mu$
\eqref{app_01_c-15}, one obtains
\bse
\Del \rsJ^\mu = \frac{1}{2} \lp \cur\mu\kap\nu\lam - \na\kap \tor\mu\nu\lam - \tor\mu\rho\kap \tor\rho\nu\lam \rp \potbu\kap\nu\lam
\ese

\noindent Taking into account the property \eqref{app_01_c-16}, one
can write also

\bse
\Del \rsJ^\mu = \frac{1}{2} \lp \cur{\mu}{[\kap}{\nu}{\lam]} - \na{[\kap} \tor{\mu}{\nu}{\lam]} - \tor{\mu}{\rho}{[\kap} \tor{\rho}{\nu}{\lam]} \rp \potbu\kap\nu\lam.
\ese

\noindent But the expression in the parenthesis is equal to zero by
the Ricci identity \eqref{app_01_a-06}. Thus,

\bse
\Del \rsJ^\mu = 0.
\ese

\section{The solution to the system of equations \eqref{sec_01_c-07} -- \eqref{sec_01_c-09}}\label{app_01_d-00}

In this Appendix, essentially basing on the results of the Appendix
\ref{app_01_b-03}, we give the full solution to the system of
equations
\bw
\begin{empheq}[left=\empheqlbrace]{flalign}
\lp \sna\nu \A\alp\mu\nu + \bfrac{1}{2} \A\alp\rho\sig \tor\mu\rho\sig \rp + \bfrac{1}{2} \B\lam\rho\sig\mu \cur\lam\alp\rho\sig & = \U\alp\mu + \Ib\alp\mu;\label{app_01_d-23}\\
-\A\alp\bet\mu + \lp \sna\lam \B\alp\bet\mu\lam + \bfrac{1}{2} \B\alp\bet\rho\sig \tor\mu\rho\sig - \bfrac{1}{2} \B\alp\rho\sig\mu \tor\bet\rho\sig \rp & = \M\alp\bet\mu;\label{app_01_d-13}\\
-\B{\alp}{(\bet}{\gam)}{\mu} & =
\N\alp\bet\gam\mu.\label{app_01_d-24}
\end{empheq}
\ew

\subsection{Determination of the tensor $\bfB$}\label{app_01_d-01}

Using the full set of the Young projectors $II$ \eqref{app_01_d-05}
-- \eqref{app_01_d-08}, decompose the tensor $\{
\B{\alp}{\lam}{[\mu}{\nu]} = \B\alp\lam\mu\nu \}$ onto irreducible
with respect to its contravariant indexes parts:
\bse
\B\alp\lam\mu\nu = \Na\alp\lam\mu\nu + \Nb\alp\lam\mu\nu + \Nc\alp\lam\mu\nu + \Nd\alp\lam\mu\nu
\ese
where
\bse
\Na\alp\lam\mu\nu \Def \yous{\youtba123} \B\alp\lam\mu\nu = \B{\alp}{(\lam}{\mu}{\nu)} = 0;
\ese
\bse
\ba{l}\Nd\alp\lam\mu\nu \Def \youa{\youtbb231} \B\alp\lam\mu\nu \qquad\qquad\\
\quad\qquad\qquad\qquad = \bfrac{2}{3} \lp
\B{\alp}{\lam}{(\mu}{\nu)} - \B{\alp}{\mu}{(\lam}{\nu)} \rp =
0;\ea
\ese
\bse
\ba{l}\Nc\alp\lam\mu\nu \Def \yous{\youtbc123} \B\alp\lam\mu\nu\\
\quad = \B{\alp}{[\lam}{\mu}{\nu]} = \bfrac{1}{3} \lp
\B\alp\lam\mu\nu + \B\alp\mu\nu\lam + \B\alp\nu\lam\mu \rp;\ea
\ese
\bse
\Nb\alp\lam\mu\nu \Def \yous{\youtbb213} \B\alp\lam\mu\nu = \B\alp\lam\mu\nu - \Nc\alp\lam\mu\nu.
\ese
Consequently,
\be\label{app_01_d-09} \B\alp\lam\mu\nu = \Nb\alp\lam\mu\nu +
\Nc\alp\lam\mu\nu. \ee
By the construction, the tensors $\bfNb \Def \{ \Nb\alp\lam\mu\nu
\}$ and $\bfNc \Def \{ \Nc\alp\lam\mu\nu \}$ have the following
properties of symmetry:
\be\label{app_01_d-10} \Nb{\alp}{\lam}{[\mu}{\nu]} =
\Nb\alp\lam\mu\nu; \ee
\be\label{app_01_d-11} \Nb{\alp}{[\lam}{\mu}{\nu]} =
\Nb\alp\lam\mu\nu; \ee
\be\label{app_01_d-12} \Nc{\alp}{[\lam}{\mu}{\nu]} =
\Nc\alp\lam\mu\nu. \ee
Then from \eqref{app_01_d-09}, \eqref{app_01_d-11} and
\eqref{app_01_d-12} one has
\be \B{\alp}{(\bet}{\gam)}{\mu} = \Nb{\alp}{(\bet}{\gam)}{\mu}. \ee
Substituting this result into \eqref{app_01_d-24}, one obtains the
equation:
\be\label{app_01_d-16} \Nb{\alp}{(\bet}{\gam)}{\mu} = -
\N\alp\bet\gam\mu \ee
that has to determine tensor $\bfNb$.

Recall the symmetry properties \eqref{sec_01_b-05} and
\eqref{app_01_d-14} for the tensor $\bfN$:
\be\label{app_01_d-18} \N{\alp}{(\lam}{\mu)}{\nu} =
\N\alp\lam\mu\nu; \ee
\be \N{\alp}{(\lam}{\mu}{\nu)} = 0. \ee
Using them and the definition \eqref{app_01_d-15} of the Young
projector $\yous{\youtbb213}$, one finds that
\be\label{app_01_d-17} \yous{\youtbb213} \N\alp\lam\mu\nu =
\N\alp\lam\mu\nu. \ee
On the other hand, by the symmetry property \eqref{app_01_d-10},
\be\label{app_01_d-19} \yous{\youtbb213} \Nb\alp\lam\mu\nu =
\frac{4}{3} \Nb{\alp}{(\lam}{\mu)}{\nu}. \ee
Combining the last two formulae and \eqref{app_01_d-16} one
obtains
\be\label{app_01_d-28} \yous{\youtbb213} \Nb\alp\lam\mu\nu = -
\yous{\youtbb213} \frac{4}{3} \N\alp\lam\mu\nu. \ee
Now, act onto both sides of this equality  by the Young projector
$\youa{\youtbb213}$ \eqref{app_01_d-06}. Then, taking into account
\eqref{app_01_d-17}, \eqref{app_01_d-18}, \eqref{app_01_d-19} and
the property $\Nb{\alp}{(\lam}{\mu}{\nu)}=0$, one gets
\be\label{app_01_d-20} \Nb\alp\lam\mu\nu = -\frac{4}{3}
\N{\alp}{\lam}{[\mu}{\nu]}. \ee
Substituting this result into \eqref{app_01_d-09}, one finds finally
\be\label{app_01_d-29} \B\alp\lam\mu\nu = -\bfrac{4}{3}
\N{\alp}{\lam}{[\mu}{\nu]} + \Nc\alp\lam\mu\nu. \ee
One needs an antisymmetrical part of this quantity:
\be\label{app_01_d-21} \ba{l}
\B{\alp}{[\lam}{\mu]}{\nu} = -\bfrac{2}{3} \lp \N{\alp}{[\lam}{\mu]}{\nu} - \N{\alp}{\nu}{[\lam}{\mu]} \rp + \Nc{\alp}{[\lam}{\mu]}{\nu}\\
= \bfrac{2}{3} \N{\alp}{\nu}{[\lam}{\mu]} + \Nc\alp\lam\mu\nu, \ea
\ee
where properties \eqref{app_01_d-12} and
\eqref{app_01_d-18} have been used.

\subsection{Determination of the tensor $\bfA$}
Rewriting the equation \eqref{app_01_d-13} as
%
%
\bse
\ba{l}
\A\alp\mu\nu = -\M{\alp}{[\mu}{\nu]} -\M{\alp}{(\mu}{\nu)} + \sna\lam \B{\alp}{(\mu}{\nu)}{\lam}\\
+ \lp \sna\lam \B{\alp}{[\mu}{\nu]}{\lam} - \frac{1}{2}
\tor\mu\rho\sig \B{\alp}{[\rho}{\sig]}{\nu} + \frac{1}{2}
\B{\alp}{\mu}{[\rho}{\sig]} \tor\nu\rho\sig \rp, \ea
\ese
%
%
substituting here the expressions \eqref{app_01_d-24} and
\eqref{app_01_d-21}, and taking into account the identity
\eqref{app_01_d-22}, one obtains
\be\label{app_01_d-25} \ba{l}
\A\alp\mu\nu = -\M{\alp}{[\mu}{\nu]} + \bfrac{2}{3} \lp \sna\lam \N{\alp}{\lam}{[\mu}{\nu]} +
\bfrac{1}{2} \tor{[\mu}{\rho}{\sig} \N{\alp}{\nu]}{\rho}{\sig} \rp\\
+ \lp \sna\lam \Nc\alp\mu\nu\lam + \Nc{\alp}{[\mu|}{\rho}{\sig}
\tor{|\nu]}{\rho}{\sig} \rp. \ea \ee

\subsection{The use of the equation \eqref{app_01_d-23}}
Up to now only the equations \eqref{app_01_d-13} and
\eqref{app_01_d-24} from the system \eqref{app_01_d-23} --
\eqref{app_01_d-24} has been used. Now, turn to the equation
\eqref{app_01_d-23}. Substitution of the expressions
\eqref{app_01_d-29} and  \eqref{app_01_d-25} into
\eqref{app_01_d-23}, and taking into account the Klein identities
\eqref{app_01_d-26} -- \eqref{app_01_d-14} lead to

\bse
\ba{l}
\sna\mu \lb \sna\nu \Nc\alp\lam\mu\nu + \Nc{\alp}{[\lam|}{\rho}{\sig} \tor{|\mu]}{\rho}{\sig} \rb\\
+ \bfrac{1}{2} \tor\lam\rho\sig \lb \sna\mu \Nc\alp\rho\sig\mu +
\Nc{\alp}{[\rho|}{\veps}{\kap} \tor{|\sig]}{\veps}{\kap} \rb +
\frac{1}{2} \cur\veps\alp\rho\sig \Nc\veps\rho\sig\mu = 0. \ea
\ese
After taking into account the identity \eqref{app_01_c-08} and using
the symmetry property \eqref{app_01_d-12} this equality transfers to
\bse
\lp \cur{\mu}{[\pi}{\rho}{\sig]} - \na{[\pi} \tor{\mu}{\rho}{\sig]}
- \tor{\mu}{\veps}{[\pi} \tor{\veps}{\rho}{\sig]} \rp
\Nc\alp\pi\rho\sig = 0.
\ese
The last, by the Ricci identities \eqref{app_01_a-06}, is satisfied
identically with an \emph{arbitrary tensor} $\bfNc$.

Thus, the general solution to the system of equations
\eqref{app_01_d-23} -- \eqref{app_01_d-24} are presented by the
formulae \eqref{app_01_d-29} and \eqref{app_01_d-25} where $\{
\Nc{\alp}{[\lam}{\mu}{\nu]} = \Nc\alp\lam\mu\nu \}$ is undefined
tensor. In section \ref{sec_01_c-00}, we show why without loss of a
generality one can set $\Nc\alp\lam\mu\nu = 0$.

\nocite{*}
\bibliography{Lompay_Petrov_-_Part_1_-_Bibliography}

\providecommand{\noopsort}[1]{}\providecommand{\singleletter}[1]{#1}%
\begin{thebibliography}{145}%
\makeatletter
\providecommand \@ifxundefined [1]{%
 \@ifx{#1\undefined}
}%
\providecommand \@ifnum [1]{%
 \ifnum #1\expandafter \@firstoftwo
 \else \expandafter \@secondoftwo
 \fi
}%
\providecommand \@ifx [1]{%
 \ifx #1\expandafter \@firstoftwo
 \else \expandafter \@secondoftwo
 \fi
}%
\providecommand \natexlab [1]{#1}%
\providecommand \enquote  [1]{``#1''}%
\providecommand \bibnamefont  [1]{#1}%
\providecommand \bibfnamefont [1]{#1}%
\providecommand \citenamefont [1]{#1}%
\providecommand \href@noop [0]{\@secondoftwo}%
\providecommand \href [0]{\begingroup \@sanitize@url \@href}%
\providecommand \@href[1]{\@@startlink{#1}\@@href}%
\providecommand \@@href[1]{\endgroup#1\@@endlink}%
\providecommand \@sanitize@url [0]{\catcode `\\12\catcode `\$12\catcode
  `\&12\catcode `\#12\catcode `\^12\catcode `\_12\catcode `\%12\relax}%
\providecommand \@@startlink[1]{}%
\providecommand \@@endlink[0]{}%
\providecommand \url  [0]{\begingroup\@sanitize@url \@url }%
\providecommand \@url [1]{\endgroup\@href {#1}{\urlprefix }}%
\providecommand \urlprefix  [0]{URL }%
\providecommand \Eprint [0]{\href }%
\providecommand \doibase [0]{http://dx.doi.org/}%
\providecommand \selectlanguage [0]{\@gobble}%
\providecommand \bibinfo  [0]{\@secondoftwo}%
\providecommand \bibfield  [0]{\@secondoftwo}%
\providecommand \translation [1]{[#1]}%
\providecommand \BibitemOpen [0]{}%
\providecommand \bibitemStop [0]{}%
\providecommand \bibitemNoStop [0]{.\EOS\space}%
\providecommand \EOS [0]{\spacefactor3000\relax}%
\providecommand \BibitemShut  [1]{\csname bibitem#1\endcsname}%
\let\auto@bib@innerbib\@empty
\bibitem [{\citenamefont {Hinterbichler}(2012)}]{Hinterbichler_2012}%
  \BibitemOpen
  \bibfield  {author} {\bibinfo {author} {\bibfnamefont {K.}~\bibnamefont
  {Hinterbichler}},\ }\bibfield  {title} {\enquote {\bibinfo {title}
  {Theoretical aspects of massive gravity},}\ }\href@noop {} {\bibfield
  {journal} {\bibinfo  {journal} {Rev. Mod. Phys.}\ }\textbf {\bibinfo {volume}
  {84}},\ \bibinfo {pages} {671--710} (\bibinfo {year} {2012})}\BibitemShut
  {NoStop}%
\bibitem [{\citenamefont {Clifton}\ \emph {et~al.}(2012)\citenamefont
  {Clifton}, \citenamefont {Ferreira}, \citenamefont {Padilla},\ and\
  \citenamefont {Skordis}}]{Clifton_Ferreira_Padilla_Skordis_2012}%
  \BibitemOpen
  \bibfield  {author} {\bibinfo {author} {\bibfnamefont {T.}~\bibnamefont
  {Clifton}}, \bibinfo {author} {\bibfnamefont {P.~G.}\ \bibnamefont
  {Ferreira}}, \bibinfo {author} {\bibfnamefont {A.}~\bibnamefont {Padilla}}, \
  and\ \bibinfo {author} {\bibfnamefont {C.}~\bibnamefont {Skordis}},\
  }\bibfield  {title} {\enquote {\bibinfo {title} {Modified gravity and
  cosmology},}\ }\href@noop {} {\bibfield  {journal} {\bibinfo  {journal}
  {Phys. Rep.}\ }\textbf {\bibinfo {volume} {513}},\ \bibinfo {pages} {1--189}
  (\bibinfo {year} {2012})}\BibitemShut {NoStop}%
\bibitem [{\citenamefont {Capozziello}\ and\ \citenamefont
  {De~Laurentis}(2011)}]{Capozziello_DeLaurentis_2011}%
  \BibitemOpen
  \bibfield  {author} {\bibinfo {author} {\bibfnamefont {S.}~\bibnamefont
  {Capozziello}}\ and\ \bibinfo {author} {\bibfnamefont {M.}~\bibnamefont
  {De~Laurentis}},\ }\bibfield  {title} {\enquote {\bibinfo {title} {Extended
  theories of gravity},}\ }\href@noop {} {\bibfield  {journal} {\bibinfo
  {journal} {Phys. Rep.}\ }\textbf {\bibinfo {volume} {509}},\ \bibinfo {pages}
  {167--321} (\bibinfo {year} {2011})}\BibitemShut {NoStop}%
\bibitem [{\citenamefont {Hammond}(2002)}]{Hammond_2002}%
  \BibitemOpen
  \bibfield  {author} {\bibinfo {author} {\bibfnamefont {R.~T.}\ \bibnamefont
  {Hammond}},\ }\bibfield  {title} {\enquote {\bibinfo {title} {Torsion
  gravity},}\ }\href@noop {} {\bibfield  {journal} {\bibinfo  {journal} {Rep.
  Prog. Phys.}\ }\textbf {\bibinfo {volume} {65}},\ \bibinfo {pages} {599--649}
  (\bibinfo {year} {2002})}\BibitemShut {NoStop}%
\bibitem [{\citenamefont {Fujii}\ and\ \citenamefont
  {Maeda}(2004)}]{Fujii_Maeda_2004}%
  \BibitemOpen
  \bibfield  {author} {\bibinfo {author} {\bibfnamefont {Y.}~\bibnamefont
  {Fujii}}\ and\ \bibinfo {author} {\bibfnamefont {K.-I.}\ \bibnamefont
  {Maeda}},\ }\href@noop {} {\emph {\bibinfo {title} {The Scalar-Tensor Theory
  of Gravitation}}}\ (\bibinfo  {publisher} {Cambridge University Press},\
  \bibinfo {address} {Cambridge},\ \bibinfo {year} {2004})\BibitemShut
  {NoStop}%
\bibitem [{\citenamefont {Hehl}\ \emph {et~al.}(1976)\citenamefont {Hehl},
  \citenamefont {von~der Heyde}, \citenamefont {Kerlich},\ and\ \citenamefont
  {Nester}}]{Hehl_Heyde_Kerlick_Nester_1976}%
  \BibitemOpen
  \bibfield  {author} {\bibinfo {author} {\bibfnamefont {F.~W.}\ \bibnamefont
  {Hehl}}, \bibinfo {author} {\bibfnamefont {P.}~\bibnamefont {von~der Heyde}},
  \bibinfo {author} {\bibfnamefont {G.~D.}\ \bibnamefont {Kerlich}}, \ and\
  \bibinfo {author} {\bibfnamefont {J.~M.}\ \bibnamefont {Nester}},\ }\bibfield
   {title} {\enquote {\bibinfo {title} {General relativity with spin and
  torsion: Foundations and prospects},}\ }\href@noop {} {\bibfield  {journal}
  {\bibinfo  {journal} {Rev. Mod. Phys.}\ }\textbf {\bibinfo {volume} {48}},\
  \bibinfo {pages} {393--416} (\bibinfo {year} {1976})}\BibitemShut {NoStop}%
\bibitem [{\citenamefont {Lovelock}(1971)}]{Lovelock_1971}%
  \BibitemOpen
  \bibfield  {author} {\bibinfo {author} {\bibfnamefont {D.}~\bibnamefont
  {Lovelock}},\ }\bibfield  {title} {\enquote {\bibinfo {title} {The {E}instein
  tensor and its generalizations},}\ }\href@noop {} {\bibfield  {journal}
  {\bibinfo  {journal} {J. Math. Phys.}\ }\textbf {\bibinfo {volume} {12}},\
  \bibinfo {pages} {498--501} (\bibinfo {year} {1971})}\BibitemShut {NoStop}%
\bibitem [{\citenamefont {Zwiebach}(1985)}]{Zwiebach_1985}%
  \BibitemOpen
  \bibfield  {author} {\bibinfo {author} {\bibfnamefont {B.}~\bibnamefont
  {Zwiebach}},\ }\bibfield  {title} {\enquote {\bibinfo {title} {Curvature
  squared terms and string theories},}\ }\href@noop {} {\bibfield  {journal}
  {\bibinfo  {journal} {Phys. Lett. B}\ }\textbf {\bibinfo {volume} {156}},\
  \bibinfo {pages} {315--317} (\bibinfo {year} {1985})}\BibitemShut {NoStop}%
\bibitem [{\citenamefont {Boulware}\ and\ \citenamefont
  {Deser}(1985)}]{Boulware_Deser_1985}%
  \BibitemOpen
  \bibfield  {author} {\bibinfo {author} {\bibfnamefont {D.~G.}\ \bibnamefont
  {Boulware}}\ and\ \bibinfo {author} {\bibfnamefont {S.}~\bibnamefont
  {Deser}},\ }\bibfield  {title} {\enquote {\bibinfo {title} {String-generated
  gravity models},}\ }\href@noop {} {\bibfield  {journal} {\bibinfo  {journal}
  {Phys. Rev. Lett.}\ }\textbf {\bibinfo {volume} {55}},\ \bibinfo {pages}
  {2656--2660} (\bibinfo {year} {1985})}\BibitemShut {NoStop}%
\bibitem [{\citenamefont {Ponomarev}, \citenamefont {Barvinsky},\ and\
  \citenamefont {Obukhov}(1985)}]{Ponomarev_Barvinsky_Obukhov_1985_en}%
  \BibitemOpen
  \bibfield  {author} {\bibinfo {author} {\bibfnamefont {V.~N.}\ \bibnamefont
  {Ponomarev}}, \bibinfo {author} {\bibfnamefont {A.~O.}\ \bibnamefont
  {Barvinsky}}, \ and\ \bibinfo {author} {\bibfnamefont {Y.~N.}\ \bibnamefont
  {Obukhov}},\ }\href@noop {} {\emph {\bibinfo {title} {Geometrodynamical
  Methods and Gauge Approach to Theory of Gravitational Interactions}}}\
  (\bibinfo  {publisher} {Energoatomizdat},\ \bibinfo {address} {Moscow},\
  \bibinfo {year} {1985})\ \bibinfo {note} {(in {R}ussian)}\BibitemShut
  {NoStop}%
\bibitem [{\citenamefont {Hehl}\ \emph {et~al.}(1995)\citenamefont {Hehl},
  \citenamefont {McCrea}, \citenamefont {Mielke},\ and\ \citenamefont
  {Ne'eman}}]{Hehl_McCrea_Mielke_Neeman_1995}%
  \BibitemOpen
  \bibfield  {author} {\bibinfo {author} {\bibfnamefont {F.~W.}\ \bibnamefont
  {Hehl}}, \bibinfo {author} {\bibfnamefont {J.~D.}\ \bibnamefont {McCrea}},
  \bibinfo {author} {\bibfnamefont {E.~W.}\ \bibnamefont {Mielke}}, \ and\
  \bibinfo {author} {\bibfnamefont {Y.}~\bibnamefont {Ne'eman}},\ }\bibfield
  {title} {\enquote {\bibinfo {title} {Metric-affine gauge theory of gravity:
  field equations, world spinors, and breaking of dilaton invariance},}\
  }\href@noop {} {\bibfield  {journal} {\bibinfo  {journal} {Phys. Rep.}\
  }\textbf {\bibinfo {volume} {258}},\ \bibinfo {pages} {1--171} (\bibinfo
  {year} {1995})}\BibitemShut {NoStop}%
\bibitem [{\citenamefont {Bin\'{e}truy}, \citenamefont {Girardi},\ and\
  \citenamefont {Grimm}(2001)}]{Binetruy_Girardi_Grimm_2001}%
  \BibitemOpen
  \bibfield  {author} {\bibinfo {author} {\bibfnamefont {P.}~\bibnamefont
  {Bin\'{e}truy}}, \bibinfo {author} {\bibfnamefont {G.}~\bibnamefont
  {Girardi}}, \ and\ \bibinfo {author} {\bibfnamefont {R.}~\bibnamefont
  {Grimm}},\ }\bibfield  {title} {\enquote {\bibinfo {title} {Supergravity
  couplings: a geometric formulation},}\ }\href@noop {} {\bibfield  {journal}
  {\bibinfo  {journal} {Phys. Rep.}\ }\textbf {\bibinfo {volume} {343}},\
  \bibinfo {pages} {255--462} (\bibinfo {year} {2001})}\BibitemShut {NoStop}%
\bibitem [{\citenamefont {De~Felice}\ and\ \citenamefont
  {Tsujikawa}(2010)}]{DeFelice_Tsujikawa_2010}%
  \BibitemOpen
  \bibfield  {author} {\bibinfo {author} {\bibfnamefont {A.}~\bibnamefont
  {De~Felice}}\ and\ \bibinfo {author} {\bibfnamefont {S.}~\bibnamefont
  {Tsujikawa}},\ }\bibfield  {title} {\enquote {\bibinfo {title} {$f({R})$
  theories},}\ }\href@noop {} {\bibfield  {journal} {\bibinfo  {journal}
  {Living Rev. Relat.}\ }\textbf {\bibinfo {volume} {13}},\ \bibinfo {pages}
  {[161 pp]} (\bibinfo {year} {2010})}\BibitemShut {NoStop}%
\bibitem [{\citenamefont {Alexander}\ and\ \citenamefont
  {Yunes}(2009)}]{Alexander_Yunes_2009}%
  \BibitemOpen
  \bibfield  {author} {\bibinfo {author} {\bibfnamefont {S.}~\bibnamefont
  {Alexander}}\ and\ \bibinfo {author} {\bibfnamefont {N.}~\bibnamefont
  {Yunes}},\ }\bibfield  {title} {\enquote {\bibinfo {title} {{C}hern-{S}imons
  modified general relativity},}\ }\href@noop {} {\bibfield  {journal}
  {\bibinfo  {journal} {Phys. Rep.}\ }\textbf {\bibinfo {volume} {480}},\
  \bibinfo {pages} {1--55} (\bibinfo {year} {2009})}\BibitemShut {NoStop}%
\bibitem [{\citenamefont {Chandia}\ and\ \citenamefont
  {Zanelli}(1998)}]{Chandia_Zanelli_1998}%
  \BibitemOpen
  \bibfield  {author} {\bibinfo {author} {\bibfnamefont {O.}~\bibnamefont
  {Chandia}}\ and\ \bibinfo {author} {\bibfnamefont {J.}~\bibnamefont
  {Zanelli}},\ }\bibfield  {title} {\enquote {\bibinfo {title} {Torsional
  topological invariants (and their relevance for real life)},}\ }\href@noop {}
  {\bibfield  {journal} {\bibinfo  {journal} {AIP Conf. Proc.}\ }\textbf
  {\bibinfo {volume} {419}},\ \bibinfo {pages} {251--264} (\bibinfo {year}
  {1998})}\BibitemShut {NoStop}%
\bibitem [{\citenamefont {Deser}, \citenamefont {Jackiw},\ and\ \citenamefont
  {Templeton}(1982)}]{Deser_Jackiw_Templeton_1982_b_rep}%
  \BibitemOpen
  \bibfield  {author} {\bibinfo {author} {\bibfnamefont {S.}~\bibnamefont
  {Deser}}, \bibinfo {author} {\bibfnamefont {R.}~\bibnamefont {Jackiw}}, \
  and\ \bibinfo {author} {\bibfnamefont {S.}~\bibnamefont {Templeton}},\
  }\bibfield  {title} {\enquote {\bibinfo {title} {Topologically massive gauge
  theories},}\ }\href@noop {} {\bibfield  {journal} {\bibinfo  {journal} {Ann.
  Phys. (N.Y.)}\ }\textbf {\bibinfo {volume} {140}},\ \bibinfo {pages}
  {372--411} (\bibinfo {year} {1982})}\BibitemShut {NoStop}%
\bibitem [{\citenamefont {Deser}\ and\ \citenamefont
  {Kay}(1983)}]{Deser_Kay_1983}%
  \BibitemOpen
  \bibfield  {author} {\bibinfo {author} {\bibfnamefont {S.}~\bibnamefont
  {Deser}}\ and\ \bibinfo {author} {\bibfnamefont {J.~H.}\ \bibnamefont
  {Kay}},\ }\bibfield  {title} {\enquote {\bibinfo {title} {Topologically
  massive supergravity},}\ }\href@noop {} {\bibfield  {journal} {\bibinfo
  {journal} {Phys. Lett. B}\ }\textbf {\bibinfo {volume} {120}},\ \bibinfo
  {pages} {97--100} (\bibinfo {year} {1983})}\BibitemShut {NoStop}%
\bibitem [{\citenamefont {Bergshoeff}, \citenamefont {Hohm},\ and\
  \citenamefont {Townsend}(2009)}]{Bergshoeff_Hohm_Townsend_2009_b}%
  \BibitemOpen
  \bibfield  {author} {\bibinfo {author} {\bibfnamefont {E.~A.}\ \bibnamefont
  {Bergshoeff}}, \bibinfo {author} {\bibfnamefont {O.}~\bibnamefont {Hohm}}, \
  and\ \bibinfo {author} {\bibfnamefont {P.~K.}\ \bibnamefont {Townsend}},\
  }\bibfield  {title} {\enquote {\bibinfo {title} {More on massive $3{D}$
  gravity},}\ }\href@noop {} {\bibfield  {journal} {\bibinfo  {journal} {Phys.
  Rev. D}\ }\textbf {\bibinfo {volume} {79}},\ \bibinfo {pages} {124042 [13
  pp]} (\bibinfo {year} {2009})}\BibitemShut {NoStop}%
\bibitem [{\citenamefont {Deser}\ \emph {et~al.}(2011)\citenamefont {Deser},
  \citenamefont {Liu}, \citenamefont {L\"{u}}, \citenamefont {Pope},
  \citenamefont {\c{S}i\c{s}man},\ and\ \citenamefont
  {Tekin}}]{Deser_Liu_Lu_Pope_Sisman_Tekin_2011}%
  \BibitemOpen
  \bibfield  {author} {\bibinfo {author} {\bibfnamefont {S.}~\bibnamefont
  {Deser}}, \bibinfo {author} {\bibfnamefont {H.}~\bibnamefont {Liu}}, \bibinfo
  {author} {\bibfnamefont {H.}~\bibnamefont {L\"{u}}}, \bibinfo {author}
  {\bibfnamefont {C.~N.}\ \bibnamefont {Pope}}, \bibinfo {author}
  {\bibfnamefont {T.~C.}\ \bibnamefont {\c{S}i\c{s}man}}, \ and\ \bibinfo
  {author} {\bibfnamefont {B.}~\bibnamefont {Tekin}},\ }\bibfield  {title}
  {\enquote {\bibinfo {title} {Critical points of ${D}$-dimensional extended
  gravities},}\ }\href@noop {} {\bibfield  {journal} {\bibinfo  {journal}
  {Phys. Rev. D}\ }\textbf {\bibinfo {volume} {83}},\ \bibinfo {pages}
  {061502(R) [5 pp]} (\bibinfo {year} {2011})}\BibitemShut {NoStop}%
\bibitem [{\citenamefont {Li}, \citenamefont {Song},\ and\ \citenamefont
  {Strominger}(2008)}]{Li_Song_Strominger_2008}%
  \BibitemOpen
  \bibfield  {author} {\bibinfo {author} {\bibfnamefont {W.}~\bibnamefont
  {Li}}, \bibinfo {author} {\bibfnamefont {W.}~\bibnamefont {Song}}, \ and\
  \bibinfo {author} {\bibfnamefont {A.}~\bibnamefont {Strominger}},\ }\bibfield
   {title} {\enquote {\bibinfo {title} {Chiral gravity in three dimensions},}\
  }\href@noop {} {\bibfield  {journal} {\bibinfo  {journal} {J. High Energy
  Phys.}\ }\textbf {\bibinfo {volume} {04 (2008) 082}},\ \bibinfo {pages} {[15
  pp]} (\bibinfo {year} {2008})}\BibitemShut {NoStop}%
\bibitem [{\citenamefont {Chamseddine}(1990)}]{Chamseddine_1990}%
  \BibitemOpen
  \bibfield  {author} {\bibinfo {author} {\bibfnamefont {A.~H.}\ \bibnamefont
  {Chamseddine}},\ }\bibfield  {title} {\enquote {\bibinfo {title} {Topological
  gravity and supergravity in various dimensions},}\ }\href@noop {} {\bibfield
  {journal} {\bibinfo  {journal} {Nuclear Physics B}\ }\textbf {\bibinfo
  {volume} {346}},\ \bibinfo {pages} {213--234} (\bibinfo {year}
  {1990})}\BibitemShut {NoStop}%
\bibitem [{\citenamefont {Stelle}\ and\ \citenamefont
  {West}(1980)}]{Stelle_West_1980}%
  \BibitemOpen
  \bibfield  {author} {\bibinfo {author} {\bibfnamefont {K.~S.}\ \bibnamefont
  {Stelle}}\ and\ \bibinfo {author} {\bibfnamefont {P.~C.}\ \bibnamefont
  {West}},\ }\bibfield  {title} {\enquote {\bibinfo {title} {Spontaneously
  broken de {S}itter symmetry and the gravitational holonomy group},}\
  }\href@noop {} {\bibfield  {journal} {\bibinfo  {journal} {Phys. Rev. D}\
  }\textbf {\bibinfo {volume} {21}},\ \bibinfo {pages} {1466--1488} (\bibinfo
  {year} {1980})}\BibitemShut {NoStop}%
\bibitem [{\citenamefont {Salgado}, \citenamefont {Cataldo},\ and\
  \citenamefont {del Campo}(2002)}]{Salgado_Cataldo_delCampo_2002}%
  \BibitemOpen
  \bibfield  {author} {\bibinfo {author} {\bibfnamefont {P.}~\bibnamefont
  {Salgado}}, \bibinfo {author} {\bibfnamefont {M.}~\bibnamefont {Cataldo}}, \
  and\ \bibinfo {author} {\bibfnamefont {S.}~\bibnamefont {del Campo}},\
  }\bibfield  {title} {\enquote {\bibinfo {title} {Higher dimensional gravity
  invariant under the {P}oincar\'{e} group},}\ }\href@noop {} {\bibfield
  {journal} {\bibinfo  {journal} {Phys. Rev. D}\ }\textbf {\bibinfo {volume}
  {66}},\ \bibinfo {pages} {024013 [6 pp]} (\bibinfo {year}
  {2002})}\BibitemShut {NoStop}%
\bibitem [{\citenamefont {Troncoso}\ and\ \citenamefont
  {Zanelli}(2000)}]{Troncoso_Zanelli_2000}%
  \BibitemOpen
  \bibfield  {author} {\bibinfo {author} {\bibfnamefont {R.}~\bibnamefont
  {Troncoso}}\ and\ \bibinfo {author} {\bibfnamefont {J.}~\bibnamefont
  {Zanelli}},\ }\bibfield  {title} {\enquote {\bibinfo {title}
  {Higher-dimensional gravity, propagating torsion and {A}d{S} gauge
  invariance},}\ }\href@noop {} {\bibfield  {journal} {\bibinfo  {journal}
  {Class. Quantum Grav.}\ }\textbf {\bibinfo {volume} {17}},\ \bibinfo {pages}
  {4451--4446} (\bibinfo {year} {2000})}\BibitemShut {NoStop}%
\bibitem [{\citenamefont {Zanelli}(2008)}]{Zanelli_2008}%
  \BibitemOpen
  \bibfield  {author} {\bibinfo {author} {\bibfnamefont {J.}~\bibnamefont
  {Zanelli}},\ }\href@noop {} {\enquote {\bibinfo {title} {Lecture notes on
  {C}hern-{S}imons (super-)gravities},}\ } (\bibinfo {year} {2008}),\ \bibinfo
  {note} {e-print arXiv:hep-th/0502193}\BibitemShut {NoStop}%
\bibitem [{\citenamefont {Zanelli}(2012)}]{Zanelli_2012}%
  \BibitemOpen
  \bibfield  {author} {\bibinfo {author} {\bibfnamefont {J.}~\bibnamefont
  {Zanelli}},\ }\bibfield  {title} {\enquote {\bibinfo {title}
  {{C}hern-{S}imons forms in gravitation theories},}\ }\href@noop {} {\bibfield
   {journal} {\bibinfo  {journal} {Class. Quantum Grav.}\ }\textbf {\bibinfo
  {volume} {29}},\ \bibinfo {pages} {133001 [36 pp]} (\bibinfo {year}
  {2012})}\BibitemShut {NoStop}%
\bibitem [{\citenamefont {Lompay}\ and\ \citenamefont
  {Petrov}(2013{\natexlab{a}})}]{Lompay_Petrov_2013_a}%
  \BibitemOpen
  \bibfield  {author} {\bibinfo {author} {\bibfnamefont {R.~R.}\ \bibnamefont
  {Lompay}}\ and\ \bibinfo {author} {\bibfnamefont {A.~N.}\ \bibnamefont
  {Petrov}},\ }\bibfield  {title} {\enquote {\bibinfo {title} {Covariant
  differential identities and conservation laws in metric-torsion theories of
  gravitation. {I}. {G}eneral consideration},}\ }\href@noop {} {\bibfield
  {journal} {\bibinfo  {journal} {J. Math. Phys.}\ }\textbf {\bibinfo {volume}
  {54}},\ \bibinfo {pages} {062504 [30 pp]} (\bibinfo {year}
  {2013}{\natexlab{a}})}\BibitemShut {NoStop}%
\bibitem [{\citenamefont {Lompay}\ and\ \citenamefont
  {Petrov}(2013{\natexlab{b}})}]{Lompay_Petrov_2013_b}%
  \BibitemOpen
  \bibfield  {author} {\bibinfo {author} {\bibfnamefont {R.~R.}\ \bibnamefont
  {Lompay}}\ and\ \bibinfo {author} {\bibfnamefont {A.~N.}\ \bibnamefont
  {Petrov}},\ }\href@noop {} {\enquote {\bibinfo {title} {Covariant
  differential identities and conservation laws in metric-torsion theories of
  gravitation. {II}. {M}anifestly generally covariant theories},}\ } (\bibinfo
  {year} {2013}{\natexlab{b}}),\ \bibinfo {note} {(in preparation)}\BibitemShut
  {NoStop}%
\bibitem [{\citenamefont {Lompay}\ and\ \citenamefont
  {Petrov}(2013{\natexlab{c}})}]{Lompay_Petrov_2013_c}%
  \BibitemOpen
  \bibfield  {author} {\bibinfo {author} {\bibfnamefont {R.~R.}\ \bibnamefont
  {Lompay}}\ and\ \bibinfo {author} {\bibfnamefont {A.~N.}\ \bibnamefont
  {Petrov}},\ }\href@noop {} {\enquote {\bibinfo {title} {Covariant
  differential identities and conservation laws in metric-torsion theories of
  gravitation. {III}. {K}illing vectors, boundary terms, and applications},}\ }
  (\bibinfo {year} {2013}{\natexlab{c}}),\ \bibinfo {note} {(in
  preparation)}\BibitemShut {NoStop}%
\bibitem [{\citenamefont {Deser}(1972)}]{Deser_1972}%
  \BibitemOpen
  \bibfield  {author} {\bibinfo {author} {\bibfnamefont {S.}~\bibnamefont
  {Deser}},\ }\bibfield  {title} {\enquote {\bibinfo {title} {Note on current
  conservation, charge, and flux integrals},}\ }\href@noop {} {\bibfield
  {journal} {\bibinfo  {journal} {Am. J. Phys.}\ }\textbf {\bibinfo {volume}
  {40}},\ \bibinfo {pages} {1082--1084} (\bibinfo {year} {1972})}\BibitemShut
  {NoStop}%
\bibitem [{\citenamefont {Abbott}\ and\ \citenamefont
  {Deser}(1982)}]{Abbott_Deser_1982_b}%
  \BibitemOpen
  \bibfield  {author} {\bibinfo {author} {\bibfnamefont {L.~F.}\ \bibnamefont
  {Abbott}}\ and\ \bibinfo {author} {\bibfnamefont {S.}~\bibnamefont {Deser}},\
  }\bibfield  {title} {\enquote {\bibinfo {title} {Charge definition in
  non-{A}belian gauge theories},}\ }\href@noop {} {\bibfield  {journal}
  {\bibinfo  {journal} {Phys. Lett. B}\ }\textbf {\bibinfo {volume} {116}},\
  \bibinfo {pages} {259--263} (\bibinfo {year} {1982})}\BibitemShut {NoStop}%
\bibitem [{\citenamefont {Julia}\ and\ \citenamefont
  {Silva}(1998)}]{Julia_Silva_1998}%
  \BibitemOpen
  \bibfield  {author} {\bibinfo {author} {\bibfnamefont {B.}~\bibnamefont
  {Julia}}\ and\ \bibinfo {author} {\bibfnamefont {S.}~\bibnamefont {Silva}},\
  }\bibfield  {title} {\enquote {\bibinfo {title} {Currents and superpotentials
  in classical gauge-invariant theories: {I}. {L}ocal results with applications
  to perfect fluids and general relativity},}\ }\href@noop {} {\bibfield
  {journal} {\bibinfo  {journal} {Class. Quantum Grav.}\ }\textbf {\bibinfo
  {volume} {15}},\ \bibinfo {pages} {2173--2215} (\bibinfo {year}
  {1998})}\BibitemShut {NoStop}%
\bibitem [{\citenamefont {Silva}(1999)}]{Silva_1999}%
  \BibitemOpen
  \bibfield  {author} {\bibinfo {author} {\bibfnamefont {S.}~\bibnamefont
  {Silva}},\ }\bibfield  {title} {\enquote {\bibinfo {title} {On
  superpotentials and charge algebras of gauge theories},}\ }\href@noop {}
  {\bibfield  {journal} {\bibinfo  {journal} {Nuclear Physics B}\ }\textbf
  {\bibinfo {volume} {558}},\ \bibinfo {pages} {391--415} (\bibinfo {year}
  {1999})}\BibitemShut {NoStop}%
\bibitem [{\citenamefont {Barnich}\ and\ \citenamefont
  {Brandt}(2002)}]{Barnich_Brandt_2002}%
  \BibitemOpen
  \bibfield  {author} {\bibinfo {author} {\bibfnamefont {G.}~\bibnamefont
  {Barnich}}\ and\ \bibinfo {author} {\bibfnamefont {F.}~\bibnamefont
  {Brandt}},\ }\bibfield  {title} {\enquote {\bibinfo {title} {Covariant theory
  of asymptotic symmetries, conservation laws and central charges},}\
  }\href@noop {} {\bibfield  {journal} {\bibinfo  {journal} {Nuclear Physics
  B}\ }\textbf {\bibinfo {volume} {633}},\ \bibinfo {pages} {3--82} (\bibinfo
  {year} {2002})},\ \bibinfo {note} {e-print arXiv:hep-th/0111246}\BibitemShut
  {NoStop}%
\bibitem [{\citenamefont {Barnich}(2003)}]{Barnich_2003}%
  \BibitemOpen
  \bibfield  {author} {\bibinfo {author} {\bibfnamefont {G.}~\bibnamefont
  {Barnich}},\ }\bibfield  {title} {\enquote {\bibinfo {title} {Boundary
  charges in gauge theories: using {S}tokes theorem in the bulk},}\ }\href@noop
  {} {\bibfield  {journal} {\bibinfo  {journal} {Class. Quantum Grav.}\
  }\textbf {\bibinfo {volume} {20}},\ \bibinfo {pages} {3685--3697} (\bibinfo
  {year} {2003})}\BibitemShut {NoStop}%
\bibitem [{\citenamefont {Ray}(1968)}]{Ray_1968}%
  \BibitemOpen
  \bibfield  {author} {\bibinfo {author} {\bibfnamefont {J.~R.}\ \bibnamefont
  {Ray}},\ }\bibfield  {title} {\enquote {\bibinfo {title} {Covariant {N}oether
  identities in covariant field theories},}\ }\href@noop {} {\bibfield
  {journal} {\bibinfo  {journal} {Nuovo Cimento A}\ }\textbf {\bibinfo {volume}
  {56}},\ \bibinfo {pages} {189--196} (\bibinfo {year} {1968})}\BibitemShut
  {NoStop}%
\bibitem [{\citenamefont {Mitskevich}, \citenamefont {Efremov},\ and\
  \citenamefont {Nesterov}(1985)}]{Mitskievich_Efremov_Nesterov_1985}%
  \BibitemOpen
  \bibfield  {author} {\bibinfo {author} {\bibfnamefont {N.~V.}\ \bibnamefont
  {Mitskevich}}, \bibinfo {author} {\bibfnamefont {A.~P.}\ \bibnamefont
  {Efremov}}, \ and\ \bibinfo {author} {\bibfnamefont {A.~I.}\ \bibnamefont
  {Nesterov}},\ }\href@noop {} {\emph {\bibinfo {title} {The Field Dynamics in
  General Theory of Relativity}}}\ (\bibinfo  {publisher} {Energoatomizdat},\
  \bibinfo {address} {Moscow},\ \bibinfo {year} {1985})\ \bibinfo {note} {(in
  {R}ussian)}\BibitemShut {NoStop}%
\bibitem [{\citenamefont {Petrov}(2008)}]{Petrov_2008}%
  \BibitemOpen
  \bibfield  {author} {\bibinfo {author} {\bibfnamefont {A.~N.}\ \bibnamefont
  {Petrov}},\ }\bibfield  {title} {\enquote {\bibinfo {title} {Nonlinear
  perturbations and conservation laws on curved backgrounds in {G}{R} and other
  metric theories},}\ }in\ \href@noop {} {\emph {\bibinfo {booktitle}
  {Classical and Quantum Gravity Research}}},\ \bibinfo {editor} {edited by\
  \bibinfo {editor} {\bibfnamefont {M.~N.}\ \bibnamefont {Christiansen}}\ and\
  \bibinfo {editor} {\bibfnamefont {T.~K.}\ \bibnamefont {Rasmussen}}}\
  (\bibinfo  {publisher} {Nova Science Publishers},\ \bibinfo {address} {New
  York},\ \bibinfo {year} {2008})\ Chap.~\bibinfo {chapter} {2}, pp.\ \bibinfo
  {pages} {79--160},\ \bibinfo {note} {e-print arXiv:0705.0019
  [gr-qc]}\BibitemShut {NoStop}%
\bibitem [{\citenamefont {Petrov}(2011)}]{Petrov_2011}%
  \BibitemOpen
  \bibfield  {author} {\bibinfo {author} {\bibfnamefont {A.~N.}\ \bibnamefont
  {Petrov}},\ }\bibfield  {title} {\enquote {\bibinfo {title} {{N}oether and
  {B}elinfante corrected types of currents for perturbations in the
  {E}instein–{G}auss–{B}onnet gravity},}\ }\href@noop {} {\bibfield  {journal}
  {\bibinfo  {journal} {Class. Quantum Grav.}\ }\textbf {\bibinfo {volume}
  {28}},\ \bibinfo {pages} {215021 [17 pp]} (\bibinfo {year} {2011})},\
  \bibinfo {note} {e-print arXiv:1102.5636 [gr-qc]}\BibitemShut {NoStop}%
\bibitem [{\citenamefont {Petrov}\ and\ \citenamefont
  {Lompay}(2013)}]{Petrov_Lompay_2013}%
  \BibitemOpen
  \bibfield  {author} {\bibinfo {author} {\bibfnamefont {A.~N.}\ \bibnamefont
  {Petrov}}\ and\ \bibinfo {author} {\bibfnamefont {R.~R.}\ \bibnamefont
  {Lompay}},\ }\bibfield  {title} {\enquote {\bibinfo {title} {Covariantized
  {N}oether identities and conservation laws for perturbations in metric
  theories of gravity},}\ }\href@noop {} {\bibfield  {journal} {\bibinfo
  {journal} {Gen. Relativ. Gravit.}\ }\textbf {\bibinfo {volume} {45}},\
  \bibinfo {pages} {545--579} (\bibinfo {year} {2013})},\ \bibinfo {note}
  {e-print arXiv:1211.3268 [gr-qc] [33 pp]}\BibitemShut {NoStop}%
\bibitem [{Note1()}]{Note1}%
  \BibitemOpen
  \bibinfo {note} {In future we plan to close this gap in a separate work of
  the bibliographic character both for GR and for another theories of
  gravitation.}\BibitemShut {Stop}%
\bibitem [{\citenamefont {Szabados}(2009)}]{Szabados_2009}%
  \BibitemOpen
  \bibfield  {author} {\bibinfo {author} {\bibfnamefont {L.~B.}\ \bibnamefont
  {Szabados}},\ }\bibfield  {title} {\enquote {\bibinfo {title} {Quasi-local
  energy-momentum and angular momentum in general relativity},}\ }\href@noop {}
  {\bibfield  {journal} {\bibinfo  {journal} {Living Rev. Relat.}\ }\textbf
  {\bibinfo {volume} {12}},\ \bibinfo {pages} {[156 pp]} (\bibinfo {year}
  {2009})},\ \bibinfo {note}
  {http://www.livingreviews.org/lrr-2009-4}\BibitemShut {NoStop}%
\bibitem [{\citenamefont {Pitts}\ and\ \citenamefont
  {Schieve}(2001)}]{Pitts_Schive_2001_b}%
  \BibitemOpen
  \bibfield  {author} {\bibinfo {author} {\bibfnamefont {J.~B.}\ \bibnamefont
  {Pitts}}\ and\ \bibinfo {author} {\bibfnamefont {W.~C.}\ \bibnamefont
  {Schieve}},\ }\href@noop {} {\enquote {\bibinfo {title} {Null cones in
  {L}orentz-covariant general relativity},}\ } (\bibinfo {year} {2001}),\
  \bibinfo {note} {e-print arXiv:gr-qc/0111004}\BibitemShut {NoStop}%
\bibitem [{\citenamefont {Komar}(1959)}]{Komar_1959}%
  \BibitemOpen
  \bibfield  {author} {\bibinfo {author} {\bibfnamefont {A.}~\bibnamefont
  {Komar}},\ }\bibfield  {title} {\enquote {\bibinfo {title} {Covariant
  conservation laws in general relativity},}\ }\href@noop {} {\bibfield
  {journal} {\bibinfo  {journal} {Phys. Rev.}\ }\textbf {\bibinfo {volume}
  {113}},\ \bibinfo {pages} {934--936} (\bibinfo {year} {1959})}\BibitemShut
  {NoStop}%
\bibitem [{\citenamefont {Winicour}\ and\ \citenamefont
  {Tamburino}(1965{\natexlab{a}})}]{Winicour_Tamburino_1965_a}%
  \BibitemOpen
  \bibfield  {author} {\bibinfo {author} {\bibfnamefont {J.}~\bibnamefont
  {Winicour}}\ and\ \bibinfo {author} {\bibfnamefont {L.}~\bibnamefont
  {Tamburino}},\ }\bibfield  {title} {\enquote {\bibinfo {title}
  {Lorentz-covariant gravitational energy-momentum lincages},}\ }\href@noop {}
  {\bibfield  {journal} {\bibinfo  {journal} {Phys. Rev. Lett.}\ }\textbf
  {\bibinfo {volume} {15}},\ \bibinfo {pages} {601--605} (\bibinfo {year}
  {1965}{\natexlab{a}})},\ \bibinfo {note} {(Erratum, see
  Ref.~\cite{Winicour_Tamburino_1965_b})}\BibitemShut {NoStop}%
\bibitem [{\citenamefont {Winicour}\ and\ \citenamefont
  {Tamburino}(1965{\natexlab{b}})}]{Winicour_Tamburino_1965_b}%
  \BibitemOpen
  \bibfield  {author} {\bibinfo {author} {\bibfnamefont {J.}~\bibnamefont
  {Winicour}}\ and\ \bibinfo {author} {\bibfnamefont {L.}~\bibnamefont
  {Tamburino}},\ }\bibfield  {title} {\enquote {\bibinfo {title} {Erratum},}\
  }\href@noop {} {\bibfield  {journal} {\bibinfo  {journal} {Phys. Rev. Lett.}\
  }\textbf {\bibinfo {volume} {15}},\ \bibinfo {pages} {720--720} (\bibinfo
  {year} {1965}{\natexlab{b}})},\ \bibinfo {note} {(Erratum of
  Ref.~\cite{Winicour_Tamburino_1965_a})}\BibitemShut {NoStop}%
\bibitem [{\citenamefont {Winicour}(1980)}]{Winicour_1980}%
  \BibitemOpen
  \bibfield  {author} {\bibinfo {author} {\bibfnamefont {J.}~\bibnamefont
  {Winicour}},\ }\bibfield  {title} {\enquote {\bibinfo {title} {Angular
  momentum in general relativity},}\ }in\ \href@noop {} {\emph {\bibinfo
  {booktitle} {General Relativity and Gravitation. One Hundred Years After the
  Birth of {A}lbert {E}instein}}},\ Vol.~\bibinfo {volume} {2},\ \bibinfo
  {editor} {edited by\ \bibinfo {editor} {\bibfnamefont {A.}~\bibnamefont
  {Held}}}\ (\bibinfo  {publisher} {Plenum Press},\ \bibinfo {address} {New
  York - London},\ \bibinfo {year} {1980})\ pp.\ \bibinfo {pages}
  {71--96}\BibitemShut {NoStop}%
\bibitem [{\citenamefont {Lee}\ and\ \citenamefont
  {Wald}(1990)}]{Lee_Wald_1990}%
  \BibitemOpen
  \bibfield  {author} {\bibinfo {author} {\bibfnamefont {J.}~\bibnamefont
  {Lee}}\ and\ \bibinfo {author} {\bibfnamefont {R.~M.}\ \bibnamefont {Wald}},\
  }\bibfield  {title} {\enquote {\bibinfo {title} {Local symmetries and
  constraints},}\ }\href@noop {} {\bibfield  {journal} {\bibinfo  {journal} {J.
  Math. Phys.}\ }\textbf {\bibinfo {volume} {31}},\ \bibinfo {pages} {725 [19
  pp]} (\bibinfo {year} {1990})}\BibitemShut {NoStop}%
\bibitem [{\citenamefont {Wald}(1993)}]{Wald_1993}%
  \BibitemOpen
  \bibfield  {author} {\bibinfo {author} {\bibfnamefont {R.~M.}\ \bibnamefont
  {Wald}},\ }\bibfield  {title} {\enquote {\bibinfo {title} {Black hole entropy
  is the {N}oether charge},}\ }\href@noop {} {\bibfield  {journal} {\bibinfo
  {journal} {Phys. Rev. D}\ }\textbf {\bibinfo {volume} {48}},\ \bibinfo
  {pages} {R3427--R3431} (\bibinfo {year} {1993})}\BibitemShut {NoStop}%
\bibitem [{\citenamefont {Iyer}\ and\ \citenamefont
  {Wald}(1994)}]{Iyer_Wald_1994}%
  \BibitemOpen
  \bibfield  {author} {\bibinfo {author} {\bibfnamefont {V.}~\bibnamefont
  {Iyer}}\ and\ \bibinfo {author} {\bibfnamefont {R.~M.}\ \bibnamefont
  {Wald}},\ }\bibfield  {title} {\enquote {\bibinfo {title} {Some properties of
  the {N}oether charge and a proposal for dynamical black hole entropy},}\
  }\href@noop {} {\bibfield  {journal} {\bibinfo  {journal} {Phys. Rev. D}\
  }\textbf {\bibinfo {volume} {50}},\ \bibinfo {pages} {846--864} (\bibinfo
  {year} {1994})}\BibitemShut {NoStop}%
\bibitem [{\citenamefont {Iyer}\ and\ \citenamefont
  {Wald}(1995)}]{Iyer_Wald_1995}%
  \BibitemOpen
  \bibfield  {author} {\bibinfo {author} {\bibfnamefont {V.}~\bibnamefont
  {Iyer}}\ and\ \bibinfo {author} {\bibfnamefont {R.~M.}\ \bibnamefont
  {Wald}},\ }\bibfield  {title} {\enquote {\bibinfo {title} {Comparison of the
  {N}oether charge and {E}uclidean methods for computing the entropy of
  stationary black holes},}\ }\href@noop {} {\bibfield  {journal} {\bibinfo
  {journal} {Phys. Rev. D}\ }\textbf {\bibinfo {volume} {52}},\ \bibinfo
  {pages} {4430--4439} (\bibinfo {year} {1995})}\BibitemShut {NoStop}%
\bibitem [{\citenamefont {Wald}\ and\ \citenamefont
  {Zoupas}(2000)}]{Wald_Zoupas_2000}%
  \BibitemOpen
  \bibfield  {author} {\bibinfo {author} {\bibfnamefont {R.~M.}\ \bibnamefont
  {Wald}}\ and\ \bibinfo {author} {\bibfnamefont {A.}~\bibnamefont {Zoupas}},\
  }\bibfield  {title} {\enquote {\bibinfo {title} {General definition of
  "conserved quantities" in general relativity and other theories of
  gravity},}\ }\href@noop {} {\bibfield  {journal} {\bibinfo  {journal} {Phys.
  Rev. D}\ }\textbf {\bibinfo {volume} {61}},\ \bibinfo {pages} {084027 [16
  pp]} (\bibinfo {year} {2000})}\BibitemShut {NoStop}%
\bibitem [{\citenamefont {Giachetta}\ and\ \citenamefont
  {Sardanashvily}(1995)}]{Giachetta_Sardanashvily_1995_c}%
  \BibitemOpen
  \bibfield  {author} {\bibinfo {author} {\bibfnamefont {G.}~\bibnamefont
  {Giachetta}}\ and\ \bibinfo {author} {\bibfnamefont {G.}~\bibnamefont
  {Sardanashvily}},\ }\href@noop {} {\enquote {\bibinfo {title} {Stress-energy
  momentum of affine-metric graviy. {G}eneralized {K}omar superpotential},}\ }
  (\bibinfo {year} {1995}),\ \bibinfo {note} {e-print arXiv:gr-qc/9511008 [12
  pp]}\BibitemShut {NoStop}%
\bibitem [{\citenamefont {Giachetta}\ and\ \citenamefont
  {Sardanashvily}(1996)}]{Giachetta_Sardanashvily_1996}%
  \BibitemOpen
  \bibfield  {author} {\bibinfo {author} {\bibfnamefont {G.}~\bibnamefont
  {Giachetta}}\ and\ \bibinfo {author} {\bibfnamefont {G.}~\bibnamefont
  {Sardanashvily}},\ }\bibfield  {title} {\enquote {\bibinfo {title}
  {Stress-energy momentum of affine-metric graviy. {G}eneralized {K}omar
  superpotential},}\ }\href@noop {} {\bibfield  {journal} {\bibinfo  {journal}
  {Class. Quantum Grav.}\ }\textbf {\bibinfo {volume} {13}},\ \bibinfo {pages}
  {L67--L71} (\bibinfo {year} {1996})}\BibitemShut {NoStop}%
\bibitem [{\citenamefont {Aros}\ \emph
  {et~al.}(2000{\natexlab{a}})\citenamefont {Aros}, \citenamefont {Contreras},
  \citenamefont {Olea}, \citenamefont {Troncoso},\ and\ \citenamefont
  {Zanelli}}]{Aros_Contreras_Olea_Troncoso_Zanelli_2000_a}%
  \BibitemOpen
  \bibfield  {author} {\bibinfo {author} {\bibfnamefont {R.}~\bibnamefont
  {Aros}}, \bibinfo {author} {\bibfnamefont {M.}~\bibnamefont {Contreras}},
  \bibinfo {author} {\bibfnamefont {R.}~\bibnamefont {Olea}}, \bibinfo {author}
  {\bibfnamefont {R.}~\bibnamefont {Troncoso}}, \ and\ \bibinfo {author}
  {\bibfnamefont {J.}~\bibnamefont {Zanelli}},\ }\bibfield  {title} {\enquote
  {\bibinfo {title} {Conserved charges for gravity with locally anti–de
  {S}itter asymptotics},}\ }\href@noop {} {\bibfield  {journal} {\bibinfo
  {journal} {Phys. Rev. Lett.}\ }\textbf {\bibinfo {volume} {84}},\ \bibinfo
  {pages} {1647--1650} (\bibinfo {year} {2000}{\natexlab{a}})}\BibitemShut
  {NoStop}%
\bibitem [{\citenamefont {Aros}\ \emph
  {et~al.}(2000{\natexlab{b}})\citenamefont {Aros}, \citenamefont {Contreras},
  \citenamefont {Olea}, \citenamefont {Troncoso},\ and\ \citenamefont
  {Zanelli}}]{Aros_Contreras_Olea_Troncoso_Zanelli_2000_b}%
  \BibitemOpen
  \bibfield  {author} {\bibinfo {author} {\bibfnamefont {R.}~\bibnamefont
  {Aros}}, \bibinfo {author} {\bibfnamefont {M.}~\bibnamefont {Contreras}},
  \bibinfo {author} {\bibfnamefont {R.}~\bibnamefont {Olea}}, \bibinfo {author}
  {\bibfnamefont {R.}~\bibnamefont {Troncoso}}, \ and\ \bibinfo {author}
  {\bibfnamefont {J.}~\bibnamefont {Zanelli}},\ }\bibfield  {title} {\enquote
  {\bibinfo {title} {Conserved charges for even dimensional asymptotically
  {A}d{S} gravity theories},}\ }\href@noop {} {\bibfield  {journal} {\bibinfo
  {journal} {Phys. Rev. D}\ }\textbf {\bibinfo {volume} {62}},\ \bibinfo
  {pages} {044002 [7 pp]} (\bibinfo {year} {2000}{\natexlab{b}})}\BibitemShut
  {NoStop}%
\bibitem [{\citenamefont {Olea}(2007)}]{Olea_2007}%
  \BibitemOpen
  \bibfield  {author} {\bibinfo {author} {\bibfnamefont {R.}~\bibnamefont
  {Olea}},\ }\bibfield  {title} {\enquote {\bibinfo {title} {Regularization of
  odd-dimensional {A}d{S} gravity: Kounterterms},}\ }\href@noop {} {\bibfield
  {journal} {\bibinfo  {journal} {J. High Energy Phys.}\ }\textbf {\bibinfo
  {volume} {04 (2007) 073}},\ \bibinfo {pages} {[37 pp]} (\bibinfo {year}
  {2007})}\BibitemShut {NoStop}%
\bibitem [{\citenamefont {Kofinas}\ and\ \citenamefont
  {Olea}(2007)}]{Kofinas_Olea_2007}%
  \BibitemOpen
  \bibfield  {author} {\bibinfo {author} {\bibfnamefont {G.}~\bibnamefont
  {Kofinas}}\ and\ \bibinfo {author} {\bibfnamefont {R.}~\bibnamefont {Olea}},\
  }\bibfield  {title} {\enquote {\bibinfo {title} {Universal regularization
  prescription for {L}ovelock {A}d{S} gravity},}\ }\href@noop {} {\bibfield
  {journal} {\bibinfo  {journal} {J. High Energy Phys.}\ }\textbf {\bibinfo
  {volume} {11 (2007) 069}},\ \bibinfo {pages} {[20 pp]} (\bibinfo {year}
  {2007})}\BibitemShut {NoStop}%
\bibitem [{\citenamefont {Noether}(1918)}]{Noether_1918}%
  \BibitemOpen
  \bibfield  {author} {\bibinfo {author} {\bibfnamefont {E.}~\bibnamefont
  {Noether}},\ }\bibfield  {title} {\enquote {\bibinfo {title} {Invariante
  {V}ariationsprobleme},}\ }\href@noop {} {\bibfield  {journal} {\bibinfo
  {journal} {Nachr. K\"{o}nigl. Gesellsch. Wissensch. G\"{o}ttingen.
  Math.-phys. Klasse}\ ,\ \bibinfo {pages} {235--258}} (\bibinfo {year}
  {1918})},\ \bibinfo {note} {(in {G}erman) ({E}nglish translation, see
  Ref.~\cite{Noether_1918_en_c})}\BibitemShut {NoStop}%
\bibitem [{\citenamefont {Noether}(2011)}]{Noether_1918_en_c}%
  \BibitemOpen
  \bibfield  {author} {\bibinfo {author} {\bibfnamefont {E.}~\bibnamefont
  {Noether}},\ }\enquote {\bibinfo {title} {Invariant variational problems},}\
  in\ \href@noop {} {\emph {\bibinfo {booktitle} {The {N}oether Theorems.
  {I}nvariance and Conservation Laws in the Twentieth Century}}}\ (\bibinfo
  {publisher} {Springer},\ \bibinfo {address} {New York - Dordrecht -
  Heidelberg - London},\ \bibinfo {year} {2011})\ pp.\ \bibinfo {pages}
  {3--22},\ \bibinfo {note} {({E}nglish translation of the
  Ref.~\cite{Noether_1918})}\BibitemShut {NoStop}%
\bibitem [{\citenamefont {Konopleva}\ and\ \citenamefont
  {Popov}(1981)}]{Konopleva_Popov_1980_en}%
  \BibitemOpen
  \bibfield  {author} {\bibinfo {author} {\bibfnamefont {N.~P.}\ \bibnamefont
  {Konopleva}}\ and\ \bibinfo {author} {\bibfnamefont {V.~N.}\ \bibnamefont
  {Popov}},\ }\href@noop {} {\emph {\bibinfo {title} {Gauge Fields}}},\
  \bibinfo {edition} {2nd}\ ed.\ (\bibinfo  {publisher} {Harwood Academic
  Publisher},\ \bibinfo {address} {Chur - London - New York},\ \bibinfo {year}
  {1981})\BibitemShut {NoStop}%
\bibitem [{\citenamefont {Byers}(1998)}]{Byers_1998}%
  \BibitemOpen
  \bibfield  {author} {\bibinfo {author} {\bibfnamefont {N.}~\bibnamefont
  {Byers}},\ }\href@noop {} {\enquote {\bibinfo {title} {E. {N}oether's
  discovery of the deep connection between symmetries and conservation laws},}\
  } (\bibinfo {year} {1998}),\ \bibinfo {note} {e-print arXiv:physics/9807044
  [physics.hist-ph]}\BibitemShut {NoStop}%
\bibitem [{\citenamefont {Brading}\ and\ \citenamefont
  {Brown}(2000)}]{Brading_Brown_2000}%
  \BibitemOpen
  \bibfield  {author} {\bibinfo {author} {\bibfnamefont {K.}~\bibnamefont
  {Brading}}\ and\ \bibinfo {author} {\bibfnamefont {H.~R.}\ \bibnamefont
  {Brown}},\ }\href@noop {} {\enquote {\bibinfo {title} {{N}oether's theorems
  and gauge symmetries},}\ } (\bibinfo {year} {2000}),\ \bibinfo {note}
  {e-print arXiv:hep-th/0009058v1}\BibitemShut {NoStop}%
\bibitem [{\citenamefont {Brading}\ and\ \citenamefont
  {Brown}(2003)}]{Brading_Brown_2003}%
  \BibitemOpen
  \bibfield  {author} {\bibinfo {author} {\bibfnamefont {K.}~\bibnamefont
  {Brading}}\ and\ \bibinfo {author} {\bibfnamefont {H.~R.}\ \bibnamefont
  {Brown}},\ }\bibfield  {title} {\enquote {\bibinfo {title} {Symmetries and
  {N}oether's theorems},}\ }in\ \href@noop {} {\emph {\bibinfo {booktitle}
  {Symmetries in Physics. Philosophical Reflections}}},\ \bibinfo {editor}
  {edited by\ \bibinfo {editor} {\bibfnamefont {K.}~\bibnamefont {Brading}}\
  and\ \bibinfo {editor} {\bibfnamefont {E.}~\bibnamefont {Castellani}}}\
  (\bibinfo  {publisher} {Cambridge University Press},\ \bibinfo {address}
  {Cambridge},\ \bibinfo {year} {2003})\ pp.\ \bibinfo {pages}
  {89--109}\BibitemShut {NoStop}%
\bibitem [{\citenamefont {Brading}(2005)}]{Brading_2005}%
  \BibitemOpen
  \bibfield  {author} {\bibinfo {author} {\bibfnamefont {K.}~\bibnamefont
  {Brading}},\ }\bibfield  {title} {\enquote {\bibinfo {title} {A note on
  general relativity, energy conservation, and {N}oether's theorems},}\
  }\href@noop {} {\bibfield  {journal} {\bibinfo  {journal} {Einstein Studies}\
  }\textbf {\bibinfo {volume} {11}},\ \bibinfo {pages} {125--135} (\bibinfo
  {year} {2005})}\BibitemShut {NoStop}%
\bibitem [{\citenamefont {Hilbert}(1915)}]{Hilbert_1915}%
  \BibitemOpen
  \bibfield  {author} {\bibinfo {author} {\bibfnamefont {D.}~\bibnamefont
  {Hilbert}},\ }\bibfield  {title} {\enquote {\bibinfo {title} {Die
  {G}rundlagen der {P}hysik. ({E}rste {M}itteilung)},}\ }\href@noop {}
  {\bibfield  {journal} {\bibinfo  {journal} {Nachr. K\"{o}nigl. Gesellsch.
  Wissensch. G\"{o}ttingen. Math.-phys. Klasse}\ ,\ \bibinfo {pages}
  {395--407}} (\bibinfo {year} {1915})},\ \bibinfo {note} {(in {G}erman)
  ({E}nglish translation, see Ref.~\cite{Hilbert_1915_en})}\BibitemShut
  {NoStop}%
\bibitem [{\citenamefont {Hilbert}(2007)}]{Hilbert_1915_en}%
  \BibitemOpen
  \bibfield  {author} {\bibinfo {author} {\bibfnamefont {D.}~\bibnamefont
  {Hilbert}},\ }\bibfield  {title} {\enquote {\bibinfo {title} {The foundations
  of physics},}\ }in\ \href@noop {} {\emph {\bibinfo {booktitle} {The Genesis
  of General Relativity. Vol. 4. Gravitation in the Twilight of Classical
  Physics: The Promise of Mathematics}}},\ \bibinfo {editor} {edited by\
  \bibinfo {editor} {\bibfnamefont {J.}~\bibnamefont {Renn}}}\ (\bibinfo
  {publisher} {Springer},\ \bibinfo {address} {Dordrecht},\ \bibinfo {year}
  {2007})\ pp.\ \bibinfo {pages} {1003--1015},\ \bibinfo {note} {({E}nglish
  translation of the Ref.~\cite{Hilbert_1915})}\BibitemShut {NoStop}%
\bibitem [{\citenamefont {Klein}(1917)}]{Klein_1917}%
  \BibitemOpen
  \bibfield  {author} {\bibinfo {author} {\bibfnamefont {F.}~\bibnamefont
  {Klein}},\ }\bibfield  {title} {\enquote {\bibinfo {title} {Zu {H}ilberts
  erster {N}ote \"{u}ber die {G}rundlagen der {P}hysik},}\ }\href@noop {}
  {\bibfield  {journal} {\bibinfo  {journal} {Nachr. K\"{o}nigl. Gesellsch.
  Wissensch. G\"{o}ttingen. Math.-phys. Klasse}\ ,\ \bibinfo {pages}
  {469--482}} (\bibinfo {year} {1917})},\ \bibinfo {note} {(in
  {G}erman)}\BibitemShut {NoStop}%
\bibitem [{\citenamefont {Klein}(1918{\natexlab{a}})}]{Klein_1918_a}%
  \BibitemOpen
  \bibfield  {author} {\bibinfo {author} {\bibfnamefont {F.}~\bibnamefont
  {Klein}},\ }\bibfield  {title} {\enquote {\bibinfo {title} {\"{U}ber die
  {D}ifferentialgesetze f\"{u}r die {E}rhaltung von {I}mpuls und {E}nergie in
  der {E}nsteinschen {G}ravitationstheorie},}\ }\href@noop {} {\bibfield
  {journal} {\bibinfo  {journal} {Nachr. K\"{o}nigl. Gesellsch. Wissensch.
  G\"{o}ttingen. Math.-phys. Klasse}\ ,\ \bibinfo {pages} {171--189}} (\bibinfo
  {year} {1918}{\natexlab{a}})},\ \bibinfo {note} {(in {G}erman)}\BibitemShut
  {NoStop}%
\bibitem [{\citenamefont {Klein}(1918{\natexlab{b}})}]{Klein_1918_b}%
  \BibitemOpen
  \bibfield  {author} {\bibinfo {author} {\bibfnamefont {F.}~\bibnamefont
  {Klein}},\ }\bibfield  {title} {\enquote {\bibinfo {title} {\"{U}ber die
  {I}ntegralform der {E}rhaltungss\"{a}tze und die {T}heorie der
  r\"{a}umlich-geschlossenen {W}elt},}\ }\href@noop {} {\bibfield  {journal}
  {\bibinfo  {journal} {Nachr. K\"{o}nigl. Gesellsch. Wissensch. G\"{o}ttingen.
  Math.-phys. Klasse}\ ,\ \bibinfo {pages} {393--423}} (\bibinfo {year}
  {1918}{\natexlab{b}})},\ \bibinfo {note} {(in {G}erman)}\BibitemShut
  {NoStop}%
\bibitem [{\citenamefont {Bergmann}(1949)}]{Bergmann_1949}%
  \BibitemOpen
  \bibfield  {author} {\bibinfo {author} {\bibfnamefont {P.~G.}\ \bibnamefont
  {Bergmann}},\ }\bibfield  {title} {\enquote {\bibinfo {title} {Non-linear
  field theories},}\ }\href@noop {} {\bibfield  {journal} {\bibinfo  {journal}
  {Phys. Rev.}\ }\textbf {\bibinfo {volume} {75}},\ \bibinfo {pages} {680--685}
  (\bibinfo {year} {1949})}\BibitemShut {NoStop}%
\bibitem [{\citenamefont {Zatzkis}(1951)}]{Zatzkis_1951}%
  \BibitemOpen
  \bibfield  {author} {\bibinfo {author} {\bibfnamefont {H.}~\bibnamefont
  {Zatzkis}},\ }\bibfield  {title} {\enquote {\bibinfo {title} {Conservation
  laws in the general theory of relativity with electromagnetic field},}\
  }\href@noop {} {\bibfield  {journal} {\bibinfo  {journal} {Phys. Rev.}\
  }\textbf {\bibinfo {volume} {81}},\ \bibinfo {pages} {1023--1026} (\bibinfo
  {year} {1951})}\BibitemShut {NoStop}%
\bibitem [{\citenamefont {Bergmann}\ and\ \citenamefont
  {Schiller}(1953)}]{Bergmann_Schiller_1953}%
  \BibitemOpen
  \bibfield  {author} {\bibinfo {author} {\bibfnamefont {P.~G.}\ \bibnamefont
  {Bergmann}}\ and\ \bibinfo {author} {\bibfnamefont {R.}~\bibnamefont
  {Schiller}},\ }\bibfield  {title} {\enquote {\bibinfo {title} {Classical and
  quantum field theories in the {L}agrangian formalism},}\ }\href@noop {}
  {\bibfield  {journal} {\bibinfo  {journal} {Phys. Rev.}\ }\textbf {\bibinfo
  {volume} {89}},\ \bibinfo {pages} {4--16} (\bibinfo {year}
  {1953})}\BibitemShut {NoStop}%
\bibitem [{\citenamefont {Mizkjewitsch}(1957)}]{Mizkjewitsch_1957}%
  \BibitemOpen
  \bibfield  {author} {\bibinfo {author} {\bibfnamefont {N.}~\bibnamefont
  {Mizkjewitsch}},\ }\bibfield  {title} {\enquote {\bibinfo {title} {Zu den
  {I}nvariantzeigenschaften der {L}agrange-{F}unktionen der {F}elder},}\
  }\href@noop {} {\bibfield  {journal} {\bibinfo  {journal} {Ann. Phys.
  (Leipzig)}\ }\textbf {\bibinfo {volume} {456}},\ \bibinfo {pages} {319--333}
  (\bibinfo {year} {1957})},\ \bibinfo {note} {(in {G}erman)}\BibitemShut
  {NoStop}%
\bibitem [{\citenamefont {Mitzkievich}(1969)}]{Mitskievich_1969_en}%
  \BibitemOpen
  \bibfield  {author} {\bibinfo {author} {\bibfnamefont {N.~V.}\ \bibnamefont
  {Mitzkievich}},\ }\href@noop {} {\emph {\bibinfo {title} {Physical Fields in
  General Theory of Relativity}}}\ (\bibinfo  {publisher} {Nauka},\ \bibinfo
  {address} {Moscow},\ \bibinfo {year} {1969})\ \bibinfo {note} {(in
  {R}ussian)}\BibitemShut {NoStop}%
\bibitem [{\citenamefont {Trautman}(1962)}]{Trautman_1962_b}%
  \BibitemOpen
  \bibfield  {author} {\bibinfo {author} {\bibfnamefont {A.}~\bibnamefont
  {Trautman}},\ }\bibfield  {title} {\enquote {\bibinfo {title} {Conservation
  laws in general relativity},}\ }in\ \href@noop {} {\emph {\bibinfo
  {booktitle} {Gravitation: an introduction to current research}}},\ \bibinfo
  {editor} {edited by\ \bibinfo {editor} {\bibfnamefont {L.}~\bibnamefont
  {Witten}}}\ (\bibinfo  {publisher} {John Wiley and Sons},\ \bibinfo {address}
  {New York - London},\ \bibinfo {year} {1962})\ pp.\ \bibinfo {pages}
  {169--198}\BibitemShut {NoStop}%
\bibitem [{\citenamefont {Trautman}(1966)}]{Trautman_1966_en}%
  \BibitemOpen
  \bibfield  {author} {\bibinfo {author} {\bibfnamefont {A.}~\bibnamefont
  {Trautman}},\ }\bibfield  {title} {\enquote {\bibinfo {title} {The general
  theory of relativity},}\ }\href@noop {} {\bibfield  {journal} {\bibinfo
  {journal} {Usp. Fiz. Nauk [Sov. Phys. Usp.]}\ }\textbf {\bibinfo {volume}
  {89}},\ \bibinfo {pages} {319--339} (\bibinfo {year} {1966})}\BibitemShut
  {NoStop}%
\bibitem [{\citenamefont {Logunov}\ and\ \citenamefont
  {Folomeshkin}(1977{\natexlab{a}})}]{Logunov_Folomeshkin_1977_b_en}%
  \BibitemOpen
  \bibfield  {author} {\bibinfo {author} {\bibfnamefont {A.~A.}\ \bibnamefont
  {Logunov}}\ and\ \bibinfo {author} {\bibfnamefont {V.~N.}\ \bibnamefont
  {Folomeshkin}},\ }\bibfield  {title} {\enquote {\bibinfo {title}
  {Energy-momentum of gravitational waves in the general theory of
  relativity},}\ }\href@noop {} {\bibfield  {journal} {\bibinfo  {journal}
  {Theor. Math. Phys.}\ }\textbf {\bibinfo {volume} {32}},\ \bibinfo {pages}
  {667--672} (\bibinfo {year} {1977}{\natexlab{a}})}\BibitemShut {NoStop}%
\bibitem [{\citenamefont {Logunov}\ and\ \citenamefont
  {Folomeshkin}(1977{\natexlab{b}})}]{Logunov_Folomeshkin_1977_c_en}%
  \BibitemOpen
  \bibfield  {author} {\bibinfo {author} {\bibfnamefont {A.~A.}\ \bibnamefont
  {Logunov}}\ and\ \bibinfo {author} {\bibfnamefont {V.~N.}\ \bibnamefont
  {Folomeshkin}},\ }\bibfield  {title} {\enquote {\bibinfo {title} {The
  energy-momentum problem and the theory of gravitation},}\ }\href@noop {}
  {\bibfield  {journal} {\bibinfo  {journal} {Theor. Math. Phys.}\ }\textbf
  {\bibinfo {volume} {32}},\ \bibinfo {pages} {749--771} (\bibinfo {year}
  {1977}{\natexlab{b}})}\BibitemShut {NoStop}%
\bibitem [{\citenamefont {Trautman}(1996)}]{Trautman_1996}%
  \BibitemOpen
  \bibfield  {author} {\bibinfo {author} {\bibfnamefont {A.}~\bibnamefont
  {Trautman}},\ }\bibfield  {title} {\enquote {\bibinfo {title} {A metaphysical
  remark on variational principles},}\ }\href@noop {} {\bibfield  {journal}
  {\bibinfo  {journal} {Acta Phys. Polon. B}\ }\textbf {\bibinfo {volume}
  {27}},\ \bibinfo {pages} {839--848} (\bibinfo {year} {1996})}\BibitemShut
  {NoStop}%
\bibitem [{\citenamefont {Utiyama}(1956)}]{Utiyama_1956}%
  \BibitemOpen
  \bibfield  {author} {\bibinfo {author} {\bibfnamefont {R.}~\bibnamefont
  {Utiyama}},\ }\bibfield  {title} {\enquote {\bibinfo {title} {Invariant
  theoretical interpretation of interaction},}\ }\href@noop {} {\bibfield
  {journal} {\bibinfo  {journal} {Phys. Rev.}\ }\textbf {\bibinfo {volume}
  {101}},\ \bibinfo {pages} {1597--1607} (\bibinfo {year} {1956})}\BibitemShut
  {NoStop}%
\bibitem [{\citenamefont {Utiyama}(1959)}]{Utiyama_1959}%
  \BibitemOpen
  \bibfield  {author} {\bibinfo {author} {\bibfnamefont {R.}~\bibnamefont
  {Utiyama}},\ }\bibfield  {title} {\enquote {\bibinfo {title} {Theory of
  invariant variation and the generalized canonical dynamics},}\ }\href@noop {}
  {\bibfield  {journal} {\bibinfo  {journal} {Prog. Theor. Phys. Suppl.}\ ,\
  \bibinfo {pages} {19--44}} (\bibinfo {year} {1959})}\BibitemShut {NoStop}%
\bibitem [{\citenamefont {Kibble}(1961)}]{Kibble_1961}%
  \BibitemOpen
  \bibfield  {author} {\bibinfo {author} {\bibfnamefont {T.~W.~B.}\
  \bibnamefont {Kibble}},\ }\bibfield  {title} {\enquote {\bibinfo {title}
  {Lorentz invariance and the gravitational field},}\ }\href@noop {} {\bibfield
   {journal} {\bibinfo  {journal} {J. Math. Phys.}\ }\textbf {\bibinfo {volume}
  {2}},\ \bibinfo {pages} {212--221} (\bibinfo {year} {1961})}\BibitemShut
  {NoStop}%
\bibitem [{\citenamefont {Utiyama}\ and\ \citenamefont
  {Fukuyama}(1971)}]{Utiyama_Fukuyama_1971}%
  \BibitemOpen
  \bibfield  {author} {\bibinfo {author} {\bibfnamefont {R.}~\bibnamefont
  {Utiyama}}\ and\ \bibinfo {author} {\bibfnamefont {T.}~\bibnamefont
  {Fukuyama}},\ }\bibfield  {title} {\enquote {\bibinfo {title} {Gravitational
  field as a generalized gauge field},}\ }\href@noop {} {\bibfield  {journal}
  {\bibinfo  {journal} {Prog. Theor. Phys.}\ }\textbf {\bibinfo {volume}
  {45}},\ \bibinfo {pages} {612--627} (\bibinfo {year} {1971})}\BibitemShut
  {NoStop}%
\bibitem [{\citenamefont {Trautman}(1972{\natexlab{a}})}]{Trautman_1972_a}%
  \BibitemOpen
  \bibfield  {author} {\bibinfo {author} {\bibfnamefont {A.}~\bibnamefont
  {Trautman}},\ }\bibfield  {title} {\enquote {\bibinfo {title} {On the
  {E}instein-{C}artan equations. {I}},}\ }\href@noop {} {\bibfield  {journal}
  {\bibinfo  {journal} {Bull. l'Acad. Polon. Sci., S\'{e}r. math., astr. et
  phys.}\ }\textbf {\bibinfo {volume} {20}},\ \bibinfo {pages} {185--190}
  (\bibinfo {year} {1972}{\natexlab{a}})}\BibitemShut {NoStop}%
\bibitem [{\citenamefont {Trautman}(1972{\natexlab{b}})}]{Trautman_1972_b}%
  \BibitemOpen
  \bibfield  {author} {\bibinfo {author} {\bibfnamefont {A.}~\bibnamefont
  {Trautman}},\ }\bibfield  {title} {\enquote {\bibinfo {title} {On the
  {E}instein-{C}artan equations. {I}{I}},}\ }\href@noop {} {\bibfield
  {journal} {\bibinfo  {journal} {Bull. l'Acad. Polon. Sci., S\'{e}r. math.,
  astr. et phys.}\ }\textbf {\bibinfo {volume} {20}},\ \bibinfo {pages}
  {503--506} (\bibinfo {year} {1972}{\natexlab{b}})}\BibitemShut {NoStop}%
\bibitem [{\citenamefont {Trautman}(1972{\natexlab{c}})}]{Trautman_1972_c}%
  \BibitemOpen
  \bibfield  {author} {\bibinfo {author} {\bibfnamefont {A.}~\bibnamefont
  {Trautman}},\ }\bibfield  {title} {\enquote {\bibinfo {title} {On the
  {E}instein-{C}artan equations. {I}{I}{I}},}\ }\href@noop {} {\bibfield
  {journal} {\bibinfo  {journal} {Bull. l'Acad. Polon. Sci., S\'{e}r. math.,
  astr. et phys.}\ }\textbf {\bibinfo {volume} {20}},\ \bibinfo {pages}
  {895--896} (\bibinfo {year} {1972}{\natexlab{c}})}\BibitemShut {NoStop}%
\bibitem [{\citenamefont {Trautman}(1973)}]{Trautman_1973}%
  \BibitemOpen
  \bibfield  {author} {\bibinfo {author} {\bibfnamefont {A.}~\bibnamefont
  {Trautman}},\ }\bibfield  {title} {\enquote {\bibinfo {title} {On the
  {E}instein-{C}artan equations. {I}{V}},}\ }\href@noop {} {\bibfield
  {journal} {\bibinfo  {journal} {Bull. l'Acad. Polon. Sci., S\'{e}r. math.,
  astr. et phys.}\ }\textbf {\bibinfo {volume} {21}},\ \bibinfo {pages}
  {345--346} (\bibinfo {year} {1973})}\BibitemShut {NoStop}%
\bibitem [{\citenamefont {Henneaux}, \citenamefont {Julia},\ and\ \citenamefont
  {Silva}(1999)}]{Henneaux_Julia_Silva_1999}%
  \BibitemOpen
  \bibfield  {author} {\bibinfo {author} {\bibfnamefont {M.}~\bibnamefont
  {Henneaux}}, \bibinfo {author} {\bibfnamefont {B.}~\bibnamefont {Julia}}, \
  and\ \bibinfo {author} {\bibfnamefont {S.}~\bibnamefont {Silva}},\ }\bibfield
   {title} {\enquote {\bibinfo {title} {{N}oether superpotentials in
  supergravities},}\ }\href@noop {} {\bibfield  {journal} {\bibinfo  {journal}
  {Nuclear Physics B}\ }\textbf {\bibinfo {volume} {563}},\ \bibinfo {pages}
  {448--460} (\bibinfo {year} {1999})}\BibitemShut {NoStop}%
\bibitem [{\citenamefont {Brandt}(2002)}]{Brandt_2002}%
  \BibitemOpen
  \bibfield  {author} {\bibinfo {author} {\bibfnamefont {F.}~\bibnamefont
  {Brandt}},\ }\bibfield  {title} {\enquote {\bibinfo {title} {Lectures on
  supergravity},}\ }\href@noop {} {\bibfield  {journal} {\bibinfo  {journal}
  {Fortschr. Phys.}\ }\textbf {\bibinfo {volume} {50}},\ \bibinfo {pages}
  {1126--1172} (\bibinfo {year} {2002})}\BibitemShut {NoStop}%
\bibitem [{\citenamefont {Ferraris}\ and\ \citenamefont
  {Francaviglia}(1992)}]{Ferraris_Francaviglia_1992}%
  \BibitemOpen
  \bibfield  {author} {\bibinfo {author} {\bibfnamefont {M.}~\bibnamefont
  {Ferraris}}\ and\ \bibinfo {author} {\bibfnamefont {M.}~\bibnamefont
  {Francaviglia}},\ }\bibfield  {title} {\enquote {\bibinfo {title}
  {Conservation laws in general relativity},}\ }\href@noop {} {\bibfield
  {journal} {\bibinfo  {journal} {Class. Quantum Grav.}\ }\textbf {\bibinfo
  {volume} {9}},\ \bibinfo {pages} {S79--S95} (\bibinfo {year}
  {1992})}\BibitemShut {NoStop}%
\bibitem [{\citenamefont {Fatibene}, \citenamefont {Ferraris},\ and\
  \citenamefont {Francaviglia}(1994)}]{Fatibene_Ferraris_Francaviglia_1994}%
  \BibitemOpen
  \bibfield  {author} {\bibinfo {author} {\bibfnamefont {L.}~\bibnamefont
  {Fatibene}}, \bibinfo {author} {\bibfnamefont {M.}~\bibnamefont {Ferraris}},
  \ and\ \bibinfo {author} {\bibfnamefont {M.}~\bibnamefont {Francaviglia}},\
  }\bibfield  {title} {\enquote {\bibinfo {title} {N\"{o}ther formalism for
  conserved quantities in classical gauge field theories},}\ }\href@noop {}
  {\bibfield  {journal} {\bibinfo  {journal} {J. Math. Phys.}\ }\textbf
  {\bibinfo {volume} {35}},\ \bibinfo {pages} {1644 [14 pp]} (\bibinfo {year}
  {1994})},\ \bibinfo {note} {(Erratum, see
  Ref.~\cite{Fatibene_Ferraris_Francaviglia_1995})}\BibitemShut {NoStop}%
\bibitem [{\citenamefont {Fatibene}, \citenamefont {Ferraris},\ and\
  \citenamefont {Francaviglia}(1995)}]{Fatibene_Ferraris_Francaviglia_1995}%
  \BibitemOpen
  \bibfield  {author} {\bibinfo {author} {\bibfnamefont {L.}~\bibnamefont
  {Fatibene}}, \bibinfo {author} {\bibfnamefont {M.}~\bibnamefont {Ferraris}},
  \ and\ \bibinfo {author} {\bibfnamefont {M.}~\bibnamefont {Francaviglia}},\
  }\bibfield  {title} {\enquote {\bibinfo {title} {Erratum: N\"{o}ther
  formalism for conserved quantities in classical gauge field theories [{J}.
  {M}ath. {P}hys. {\bf 35}, 1644–1657 (1994)]},}\ }\href@noop {} {\bibfield
  {journal} {\bibinfo  {journal} {J. Math. Phys.}\ }\textbf {\bibinfo {volume}
  {36}},\ \bibinfo {pages} {3183 [3 pp]} (\bibinfo {year} {1995})},\ \bibinfo
  {note} {(Erratum of
  Ref.~\cite{Fatibene_Ferraris_Francaviglia_1994})}\BibitemShut {NoStop}%
\bibitem [{\citenamefont {Fatibene}, \citenamefont {Ferraris},\ and\
  \citenamefont {Francaviglia}(1997)}]{Fatibene_Ferraris_Francaviglia_1997}%
  \BibitemOpen
  \bibfield  {author} {\bibinfo {author} {\bibfnamefont {L.}~\bibnamefont
  {Fatibene}}, \bibinfo {author} {\bibfnamefont {M.}~\bibnamefont {Ferraris}},
  \ and\ \bibinfo {author} {\bibfnamefont {M.}~\bibnamefont {Francaviglia}},\
  }\bibfield  {title} {\enquote {\bibinfo {title} {N\"{o}ther formalism for
  conserved quantities in classical gauge field theories. {I}{I}. {T}he
  arbitrary bosonic matter case},}\ }\href@noop {} {\bibfield  {journal}
  {\bibinfo  {journal} {J. Math. Phys.}\ }\textbf {\bibinfo {volume} {38}},\
  \bibinfo {pages} {3953 [15 pp]} (\bibinfo {year} {1997})}\BibitemShut
  {NoStop}%
\bibitem [{\citenamefont {Fatibene}\ \emph {et~al.}(2002)\citenamefont
  {Fatibene}, \citenamefont {Ferraris}, \citenamefont {Francaviglia},\ and\
  \citenamefont {McLeneghan}}]{Fatibene_Ferraris_Francaviglia_McLeneghan_2002}%
  \BibitemOpen
  \bibfield  {author} {\bibinfo {author} {\bibfnamefont {L.}~\bibnamefont
  {Fatibene}}, \bibinfo {author} {\bibfnamefont {M.}~\bibnamefont {Ferraris}},
  \bibinfo {author} {\bibfnamefont {M.}~\bibnamefont {Francaviglia}}, \ and\
  \bibinfo {author} {\bibfnamefont {R.~G.}\ \bibnamefont {McLeneghan}},\
  }\bibfield  {title} {\enquote {\bibinfo {title} {Generalized symmetries in
  mechanics and field theories},}\ }\href@noop {} {\bibfield  {journal}
  {\bibinfo  {journal} {J. Math. Phys.}\ }\textbf {\bibinfo {volume} {43}},\
  \bibinfo {pages} {3147 [15 pp]} (\bibinfo {year} {2002})}\BibitemShut
  {NoStop}%
\bibitem [{\citenamefont {Fatibene}, \citenamefont {Ferraris},\ and\
  \citenamefont {Francaviglia}(2004)}]{Fatibene_Ferraris_Francaviglia_2004_b}%
  \BibitemOpen
  \bibfield  {author} {\bibinfo {author} {\bibfnamefont {L.}~\bibnamefont
  {Fatibene}}, \bibinfo {author} {\bibfnamefont {M.}~\bibnamefont {Ferraris}},
  \ and\ \bibinfo {author} {\bibfnamefont {M.}~\bibnamefont {Francaviglia}},\
  }\bibfield  {title} {\enquote {\bibinfo {title} {On the gauge natural
  structure of modern physics},}\ }\href@noop {} {\bibfield  {journal}
  {\bibinfo  {journal} {Int. J. Geom. Meth. Mod. Phys.}\ }\textbf {\bibinfo
  {volume} {1}},\ \bibinfo {pages} {443--466} (\bibinfo {year}
  {2004})}\BibitemShut {NoStop}%
\bibitem [{\citenamefont {Fatibene}\ and\ \citenamefont
  {Francaviglia}(2004)}]{Fatibene_Francaviglia_2004}%
  \BibitemOpen
  \bibfield  {author} {\bibinfo {author} {\bibfnamefont {L.}~\bibnamefont
  {Fatibene}}\ and\ \bibinfo {author} {\bibfnamefont {M.}~\bibnamefont
  {Francaviglia}},\ }\href@noop {} {\emph {\bibinfo {title} {Natural and Gauge
  Natural Formalism for Classical Field Theories}}}\ (\bibinfo  {publisher}
  {Kluwer Academic Publishers},\ \bibinfo {address} {Boston - Dordrecht -
  London},\ \bibinfo {year} {2004})\BibitemShut {NoStop}%
\bibitem [{\citenamefont {Bashkirov}\ \emph
  {et~al.}(2005{\natexlab{a}})\citenamefont {Bashkirov}, \citenamefont
  {Giachetta}, \citenamefont {Mangiarotti},\ and\ \citenamefont
  {Sardanashvily}}]{Bashkirov_Giachetta_Mangiarotti_Sardanashvily_2005_b}%
  \BibitemOpen
  \bibfield  {author} {\bibinfo {author} {\bibfnamefont {D.}~\bibnamefont
  {Bashkirov}}, \bibinfo {author} {\bibfnamefont {G.}~\bibnamefont
  {Giachetta}}, \bibinfo {author} {\bibfnamefont {L.}~\bibnamefont
  {Mangiarotti}}, \ and\ \bibinfo {author} {\bibfnamefont {G.}~\bibnamefont
  {Sardanashvily}},\ }\bibfield  {title} {\enquote {\bibinfo {title}
  {{N}oether's second theorem in a general setting: {R}educible gauge
  theories},}\ }\href@noop {} {\bibfield  {journal} {\bibinfo  {journal} {J.
  Phys. A}\ }\textbf {\bibinfo {volume} {38}},\ \bibinfo {pages} {5329--5344}
  (\bibinfo {year} {2005}{\natexlab{a}})}\BibitemShut {NoStop}%
\bibitem [{\citenamefont {Sardanashvily}(2009)}]{Sardanashvily_2009_a}%
  \BibitemOpen
  \bibfield  {author} {\bibinfo {author} {\bibfnamefont {G.}~\bibnamefont
  {Sardanashvily}},\ }\bibfield  {title} {\enquote {\bibinfo {title} {Gauge
  conservation laws in a general setting. {S}uperpotential},}\ }\href@noop {}
  {\bibfield  {journal} {\bibinfo  {journal} {Int. J. Geom. Meth. Mod. Phys.}\
  }\textbf {\bibinfo {volume} {6}},\ \bibinfo {pages} {1047--1056} (\bibinfo
  {year} {2009})},\ \bibinfo {note} {e-print arXiv:0906.1732v1
  [math-ph]}\BibitemShut {NoStop}%
\bibitem [{\citenamefont {Einstein}(1916)}]{Einstein_1916_b}%
  \BibitemOpen
  \bibfield  {author} {\bibinfo {author} {\bibfnamefont {A.}~\bibnamefont
  {Einstein}},\ }\bibfield  {title} {\enquote {\bibinfo {title} {Hamiltonsches
  {P}rinzip und allgemeine {R}elativit\"{a}tstheorie},}\ }\href@noop {}
  {\bibfield  {journal} {\bibinfo  {journal} {K\"{o}nigl. Preu{\ss}. Akad.
  Wissensch. (Berlin). Sitzungsber.}\ ,\ \bibinfo {pages} {1111--1116}}
  (\bibinfo {year} {1916})},\ \bibinfo {note} {(in {G}erman) ({E}nglish
  translation, see Ref.~\cite{Einstein_1916_b_en})}\BibitemShut {NoStop}%
\bibitem [{\citenamefont {Einstein}(1996)}]{Einstein_1916_b_en}%
  \BibitemOpen
  \bibfield  {author} {\bibinfo {author} {\bibfnamefont {A.}~\bibnamefont
  {Einstein}},\ }\bibfield  {title} {\enquote {\bibinfo {title} {Hamilton’s
  principle and the general theory of relativity},}\ }in\ \href@noop {} {\emph
  {\bibinfo {booktitle} {The Collected Papers of {A}lbert {E}instein. Volume 6.
  The Berlin Years: Writings, 1914–1917}}},\ \bibinfo {editor} {edited by\
  \bibinfo {editor} {\bibfnamefont {A.~J.}\ \bibnamefont {Kox}}, \bibinfo
  {editor} {\bibfnamefont {M.~J.}\ \bibnamefont {Klein}}, \ and\ \bibinfo
  {editor} {\bibfnamefont {R.}~\bibnamefont {Schulmann}}}\ (\bibinfo
  {publisher} {Princeton University Press},\ \bibinfo {address} {Princeton},\
  \bibinfo {year} {1996})\ pp.\ \bibinfo {pages} {409--416},\ \bibinfo {note}
  {({E}nglish translation of the Ref.~\cite{Einstein_1916_b})}\BibitemShut
  {NoStop}%
\bibitem [{\citenamefont {Obukhov}\ and\ \citenamefont
  {Rubilar}(2007)}]{Obukhov_Rubilar_2007}%
  \BibitemOpen
  \bibfield  {author} {\bibinfo {author} {\bibfnamefont {Y.~N.}\ \bibnamefont
  {Obukhov}}\ and\ \bibinfo {author} {\bibfnamefont {G.~F.}\ \bibnamefont
  {Rubilar}},\ }\bibfield  {title} {\enquote {\bibinfo {title} {Invariant
  conserved currents in gravity theories: Diffeomorphisms and local gauge
  symmetries},}\ }\href@noop {} {\bibfield  {journal} {\bibinfo  {journal}
  {Phys. Rev. D}\ }\textbf {\bibinfo {volume} {76}},\ \bibinfo {pages} {124030
  [13 pp]} (\bibinfo {year} {2007})}\BibitemShut {NoStop}%
\bibitem [{\citenamefont {Obukhov}\ and\ \citenamefont
  {Rubilar}(2008)}]{Obukhov_Rubilar_2008}%
  \BibitemOpen
  \bibfield  {author} {\bibinfo {author} {\bibfnamefont {Y.~N.}\ \bibnamefont
  {Obukhov}}\ and\ \bibinfo {author} {\bibfnamefont {G.~F.}\ \bibnamefont
  {Rubilar}},\ }\bibfield  {title} {\enquote {\bibinfo {title} {Invariant
  conserved currents for gravity},}\ }\href@noop {} {\bibfield  {journal}
  {\bibinfo  {journal} {Phys. Lett. B}\ }\textbf {\bibinfo {volume} {660}},\
  \bibinfo {pages} {240--246} (\bibinfo {year} {2008})}\BibitemShut {NoStop}%
\bibitem [{\citenamefont {Gratus}, \citenamefont {Obukhov},\ and\ \citenamefont
  {Tucker}(2012)}]{Gratus_Obukhov_Tucker_2012}%
  \BibitemOpen
  \bibfield  {author} {\bibinfo {author} {\bibfnamefont {J.}~\bibnamefont
  {Gratus}}, \bibinfo {author} {\bibfnamefont {Y.~N.}\ \bibnamefont {Obukhov}},
  \ and\ \bibinfo {author} {\bibfnamefont {R.~W.}\ \bibnamefont {Tucker}},\
  }\bibfield  {title} {\enquote {\bibinfo {title} {Conservation laws and
  stress-energy-momentum tensors for systems with background fields},}\
  }\href@noop {} {\bibfield  {journal} {\bibinfo  {journal} {Ann. Phys.
  (N.Y.)}\ }\textbf {\bibinfo {volume} {327}},\ \bibinfo {pages} {2560--2590}
  (\bibinfo {year} {2012})}\BibitemShut {NoStop}%
\bibitem [{\citenamefont {Baykal}\ and\ \citenamefont
  {Delice}(2011)}]{Baykal_Delice_2011}%
  \BibitemOpen
  \bibfield  {author} {\bibinfo {author} {\bibfnamefont {A.}~\bibnamefont
  {Baykal}}\ and\ \bibinfo {author} {\bibfnamefont {O.}~\bibnamefont
  {Delice}},\ }\bibfield  {title} {\enquote {\bibinfo {title} {A unified
  approach to variational derivatives of modified gravitational actions},}\
  }\href@noop {} {\bibfield  {journal} {\bibinfo  {journal} {Class. Quantum
  Grav.}\ }\textbf {\bibinfo {volume} {28}},\ \bibinfo {pages} {015014 [29 pp]}
  (\bibinfo {year} {2011})}\BibitemShut {NoStop}%
\bibitem [{\citenamefont {Sardanashvily}(2004)}]{Sardanashvily_2004}%
  \BibitemOpen
  \bibfield  {author} {\bibinfo {author} {\bibfnamefont {G.}~\bibnamefont
  {Sardanashvily}},\ }\href@noop {} {\enquote {\bibinfo {title} {On algebras of
  gauge transformations in a general setting},}\ } (\bibinfo {year} {2004}),\
  \bibinfo {note} {e-print arXiv:math/0411635 [math.QA]}\BibitemShut {NoStop}%
\bibitem [{\citenamefont {Bashkirov}\ \emph
  {et~al.}(2005{\natexlab{b}})\citenamefont {Bashkirov}, \citenamefont
  {Giachetta}, \citenamefont {Mangiarotti},\ and\ \citenamefont
  {Sardanashvily}}]{Bashkirov_Giachetta_Mangiarotti_Sardanashvily_2005_a}%
  \BibitemOpen
  \bibfield  {author} {\bibinfo {author} {\bibfnamefont {D.}~\bibnamefont
  {Bashkirov}}, \bibinfo {author} {\bibfnamefont {G.}~\bibnamefont
  {Giachetta}}, \bibinfo {author} {\bibfnamefont {L.}~\bibnamefont
  {Mangiarotti}}, \ and\ \bibinfo {author} {\bibfnamefont {G.}~\bibnamefont
  {Sardanashvily}},\ }\bibfield  {title} {\enquote {\bibinfo {title}
  {{N}oether's second theorem for {B}{R}{S}{T} symmetries},}\ }\href@noop {}
  {\bibfield  {journal} {\bibinfo  {journal} {J. Math. Phys.}\ }\textbf
  {\bibinfo {volume} {46}},\ \bibinfo {pages} {053517 [23 pp]} (\bibinfo {year}
  {2005}{\natexlab{b}})}\BibitemShut {NoStop}%
\bibitem [{\citenamefont {Giachetta}, \citenamefont {Mangiarotti},\ and\
  \citenamefont
  {Sardanashvily}(2005)}]{Giachetta_Mangiarotti_Sardanashvily_2005}%
  \BibitemOpen
  \bibfield  {author} {\bibinfo {author} {\bibfnamefont {G.}~\bibnamefont
  {Giachetta}}, \bibinfo {author} {\bibfnamefont {L.}~\bibnamefont
  {Mangiarotti}}, \ and\ \bibinfo {author} {\bibfnamefont {G.}~\bibnamefont
  {Sardanashvily}},\ }\bibfield  {title} {\enquote {\bibinfo {title}
  {{L}agrangian supersymmetries depending on derivatives. {G}lobal analysis and
  cohomology},}\ }\href@noop {} {\bibfield  {journal} {\bibinfo  {journal}
  {Commun. Math. Phys.}\ }\textbf {\bibinfo {volume} {259}},\ \bibinfo {pages}
  {103--128} (\bibinfo {year} {2005})}\BibitemShut {NoStop}%
\bibitem [{\citenamefont {Giachetta}, \citenamefont {Mangiarotti},\ and\
  \citenamefont
  {Sardanashvily}(2009)}]{Giachetta_Mangiarotti_Sardanashvily_2009_a}%
  \BibitemOpen
  \bibfield  {author} {\bibinfo {author} {\bibfnamefont {G.}~\bibnamefont
  {Giachetta}}, \bibinfo {author} {\bibfnamefont {L.}~\bibnamefont
  {Mangiarotti}}, \ and\ \bibinfo {author} {\bibfnamefont {G.}~\bibnamefont
  {Sardanashvily}},\ }\bibfield  {title} {\enquote {\bibinfo {title} {On the
  notion of gauge symmetries of generic {L}agrangian field theory},}\
  }\href@noop {} {\bibfield  {journal} {\bibinfo  {journal} {J. Math. Phys.}\
  }\textbf {\bibinfo {volume} {50}},\ \bibinfo {pages} {012903 [19 pp]}
  (\bibinfo {year} {2009})}\BibitemShut {NoStop}%
\bibitem [{\citenamefont {Fatibene}, \citenamefont {Francaviglia},\ and\
  \citenamefont {Palese}(2001)}]{Fatibene_Francaviglia_Palese_2001}%
  \BibitemOpen
  \bibfield  {author} {\bibinfo {author} {\bibfnamefont {L.}~\bibnamefont
  {Fatibene}}, \bibinfo {author} {\bibfnamefont {M.}~\bibnamefont
  {Francaviglia}}, \ and\ \bibinfo {author} {\bibfnamefont {M.}~\bibnamefont
  {Palese}},\ }\bibfield  {title} {\enquote {\bibinfo {title} {Conservation
  laws and variational sequences in gauge-natural theories},}\ }\href@noop {}
  {\bibfield  {journal} {\bibinfo  {journal} {Math. Proc. Camb. Phil. Soc.}\
  }\textbf {\bibinfo {volume} {130}},\ \bibinfo {pages} {555--569} (\bibinfo
  {year} {2001})}\BibitemShut {NoStop}%
\bibitem [{\citenamefont {Palese}\ and\ \citenamefont
  {Winterroth}(2004)}]{Palese_Winterroth_2004}%
  \BibitemOpen
  \bibfield  {author} {\bibinfo {author} {\bibfnamefont {M.}~\bibnamefont
  {Palese}}\ and\ \bibinfo {author} {\bibfnamefont {E.}~\bibnamefont
  {Winterroth}},\ }\bibfield  {title} {\enquote {\bibinfo {title} {Covariant
  gauge-natural conservation laws},}\ }\href@noop {} {\bibfield  {journal}
  {\bibinfo  {journal} {Rep. Math. Phys.}\ }\textbf {\bibinfo {volume} {54}},\
  \bibinfo {pages} {349--364} (\bibinfo {year} {2004})}\BibitemShut {NoStop}%
\bibitem [{\citenamefont {Fatibene}, \citenamefont {Francaviglia},\ and\
  \citenamefont {Mercadante}(2010)}]{Fatibene_Francaviglia_Mercadante_2010_a}%
  \BibitemOpen
  \bibfield  {author} {\bibinfo {author} {\bibfnamefont {L.}~\bibnamefont
  {Fatibene}}, \bibinfo {author} {\bibfnamefont {M.}~\bibnamefont
  {Francaviglia}}, \ and\ \bibinfo {author} {\bibfnamefont {S.}~\bibnamefont
  {Mercadante}},\ }\href@noop {} {\enquote {\bibinfo {title} {{N}oether
  symmetries and covariant conservation laws in classical, relativistic and
  quantum physics},}\ } (\bibinfo {year} {2010}),\ \bibinfo {note} {e-print
  arXiv:1001.2886 [gr-qc]}\BibitemShut {NoStop}%
\bibitem [{\citenamefont {Szabados}(1991)}]{Szabados_1991}%
  \BibitemOpen
  \bibfield  {author} {\bibinfo {author} {\bibfnamefont {L.~B.}\ \bibnamefont
  {Szabados}},\ }\href@noop {} {\enquote {\bibinfo {title} {Canonical
  pseudotensors, {S}parling's form and {N}oether currents},}\ } (\bibinfo
  {year} {1991}),\ \bibinfo {note} {preprint KFKI-1991-29/B. [45
  pp]}\BibitemShut {NoStop}%
\bibitem [{\citenamefont {Szabados}(1992)}]{Szabados_1992}%
  \BibitemOpen
  \bibfield  {author} {\bibinfo {author} {\bibfnamefont {L.~B.}\ \bibnamefont
  {Szabados}},\ }\bibfield  {title} {\enquote {\bibinfo {title} {On canonical
  pseudotensors, {S}parling's form and {N}oether currents},}\ }\href@noop {}
  {\bibfield  {journal} {\bibinfo  {journal} {Class. Quantum Grav.}\ }\textbf
  {\bibinfo {volume} {9}},\ \bibinfo {pages} {2521--2541} (\bibinfo {year}
  {1992})}\BibitemShut {NoStop}%
\bibitem [{\citenamefont {Pons}(2011)}]{Pons_2011}%
  \BibitemOpen
  \bibfield  {author} {\bibinfo {author} {\bibfnamefont {J.~M.}\ \bibnamefont
  {Pons}},\ }\bibfield  {title} {\enquote {\bibinfo {title} {{N}oether
  symmetries, energy-momentum tensors, and conformal invariance in classical
  field theory},}\ }\href@noop {} {\bibfield  {journal} {\bibinfo  {journal}
  {J. Math. Phys.}\ }\textbf {\bibinfo {volume} {52}},\ \bibinfo {pages}
  {012904 [21 pp]} (\bibinfo {year} {2011})}\BibitemShut {NoStop}%
\bibitem [{Note2()}]{Note2}%
  \BibitemOpen
  \bibinfo {note} {Remark that the position of indexes in definitions of
  $\protect \mathbf {M}$ \protect \textup {\hbox {\mathsurround \z@ \protect
  \normalfont (\ignorespaces \ref {sec_01_a-14}\unskip \@@italiccorr )}} and
  $\protect \mathbf {N}$ \protect \textup {\hbox {\mathsurround \z@ \protect
  \normalfont (\ignorespaces \ref {sec_01_a-12}\unskip \@@italiccorr )}}
  differs from that of Refs.~\cite {Petrov_2004_b_en, Petrov_2009_a,
  Petrov_2010_a, Petrov_2011, Petrov_Lompay_2013}. Thus $\protect \{M_{\alpha
  }{}^{\beta \mu }\protect \}$ and $\protect \{N_{\alpha }{}^{\beta \gamma \mu
  }\protect \}$ defined here correspond to $\protect \{\protect \mathaccentV
  {hat}05E{m}_\alpha {}^{\mu \beta }\protect \}$ and $\protect \{\protect
  \mathaccentV {hat}05E{n}_\alpha {}^{\mu \beta \gamma }\protect \}$ in the
  cited works. The definition of the tensor $\protect \mathbf {U}$ \protect
  \textup {\hbox {\mathsurround \z@ \protect \normalfont (\ignorespaces \ref
  {sec_01_a-13}\unskip \@@italiccorr )}} is also changed. Thus, the tensor
  $\protect \{U_{\alpha }{}^{\mu }+I_\alpha {}^\mu \protect \}$ corresponds to
  the tensor density $\protect \{\protect \mathaccentV {hat}05E{u}_\alpha
  {}^\mu \protect \}$. At last, the current $\protect \{\protect \mathrsfs
  {J}^\mu = J^\mu + I^\mu \protect \}$ \protect \textup {\hbox {\mathsurround
  \z@ \protect \normalfont (\ignorespaces \ref {sec_01_c-03}\unskip
  \@@italiccorr )}} corresponds to the current $\protect \{\protect
  \mathaccentV {hat}05E{i}^\mu \protect \}$.}\BibitemShut {Stop}%
\bibitem [{\citenamefont {Bergmann}\ \emph {et~al.}(1956)\citenamefont
  {Bergmann}, \citenamefont {Goldberg}, \citenamefont {Janis},\ and\
  \citenamefont {Newman}}]{Bergmann_Goldberg_Janis_Newman_1956}%
  \BibitemOpen
  \bibfield  {author} {\bibinfo {author} {\bibfnamefont {P.~G.}\ \bibnamefont
  {Bergmann}}, \bibinfo {author} {\bibfnamefont {I.}~\bibnamefont {Goldberg}},
  \bibinfo {author} {\bibfnamefont {A.}~\bibnamefont {Janis}}, \ and\ \bibinfo
  {author} {\bibfnamefont {E.}~\bibnamefont {Newman}},\ }\bibfield  {title}
  {\enquote {\bibinfo {title} {Canonical transformations and commutators in the
  {L}agrangian formalism},}\ }\href@noop {} {\bibfield  {journal} {\bibinfo
  {journal} {Phys. Rev.}\ }\textbf {\bibinfo {volume} {103}},\ \bibinfo {pages}
  {807--813} (\bibinfo {year} {1956})}\BibitemShut {NoStop}%
\bibitem [{\citenamefont {Bergmann}(1958)}]{Bergmann_1958}%
  \BibitemOpen
  \bibfield  {author} {\bibinfo {author} {\bibfnamefont {P.~G.}\ \bibnamefont
  {Bergmann}},\ }\bibfield  {title} {\enquote {\bibinfo {title} {Conservation
  laws in general relativity as the generators of coordinate
  transformations},}\ }\href@noop {} {\bibfield  {journal} {\bibinfo  {journal}
  {Phys. Rev.}\ }\textbf {\bibinfo {volume} {112}},\ \bibinfo {pages}
  {287--289} (\bibinfo {year} {1958})}\BibitemShut {NoStop}%
\bibitem [{\citenamefont {Schwinger}(1951)}]{Schwinger_1951}%
  \BibitemOpen
  \bibfield  {author} {\bibinfo {author} {\bibfnamefont {J.}~\bibnamefont
  {Schwinger}},\ }\bibfield  {title} {\enquote {\bibinfo {title} {The theory of
  quantized fields. {I}},}\ }\href@noop {} {\bibfield  {journal} {\bibinfo
  {journal} {Phys. Rev.}\ }\textbf {\bibinfo {volume} {82}},\ \bibinfo {pages}
  {914--927} (\bibinfo {year} {1951})}\BibitemShut {NoStop}%
\bibitem [{\citenamefont {Schwinger}(2000)}]{Schwinger_2000}%
  \BibitemOpen
  \bibfield  {author} {\bibinfo {author} {\bibfnamefont {J.}~\bibnamefont
  {Schwinger}},\ }\href@noop {} {\emph {\bibinfo {title} {Quantum Kinematics
  and Dynamics}}},\ Advanced Book Classics\ (\bibinfo  {publisher} {Westview
  Press},\ \bibinfo {address} {New York},\ \bibinfo {year} {2000})\BibitemShut
  {NoStop}%
\bibitem [{\citenamefont {Toms}(2007)}]{Toms_2007}%
  \BibitemOpen
  \bibfield  {author} {\bibinfo {author} {\bibfnamefont {D.~J.}\ \bibnamefont
  {Toms}},\ }\href@noop {} {\emph {\bibinfo {title} {The {S}chwinger Action
  Principle and Effective Action}}}\ (\bibinfo  {publisher} {Cambridge
  University Press},\ \bibinfo {address} {New York},\ \bibinfo {year}
  {2007})\BibitemShut {NoStop}%
\bibitem [{Note3()}]{Note3}%
  \BibitemOpen
  \bibinfo {note} {Formula \protect \textup {\hbox {\mathsurround \z@ \protect
  \normalfont (\ignorespaces \ref {sec_01_c-02}\unskip \@@italiccorr )}} is
  manifestly covariant generalization of the known Poincar\'e lemma to the case
  of the spacetime $\protect \mathcal {C}(1,D)$. The lemma has a local
  character. Attempts to extend the lemma to a global derivation meets
  difficulties connected with a topology of a spacetime $\protect \mathcal
  {C}(1,D)$, which is defined by homotopic and cohomologic properties of
  $\protect \mathcal {C}(1,D)$. In fact, this problem requires a further
  investigation. Here, for the sake of simplicity, we assume that formula
  \protect \textup {\hbox {\mathsurround \z@ \protect \normalfont
  (\ignorespaces \ref {sec_01_c-02}\unskip \@@italiccorr )}} is valid in the
  global sense also; thus, the current ${\protect \bm {\protect \mathrsfs
  {J}}}[\delta \protect \bm {\xi }]$ \protect \textup {\hbox {\mathsurround \z@
  \protect \normalfont (\ignorespaces \ref {sec_01_c-03}\unskip \@@italiccorr
  )}} is cohomological to zero.}\BibitemShut {Stop}%
\bibitem [{\citenamefont {Petrov}(2000)}]{Petrov_2000}%
  \BibitemOpen
  \bibfield  {author} {\bibinfo {author} {\bibfnamefont {A.~N.}\ \bibnamefont
  {Petrov}},\ }\href@noop {} {\enquote {\bibinfo {title} {Conservation laws in
  {G}{R} and their applications},}\ } (\bibinfo {year} {2000}),\ \bibinfo
  {note} {reported on the International Workshop on Geometrical Physics,
  National Center for Theoretical Sciences (NCTS) in Hsinchu, Taiwan, July 24
  -- 26, 2000; online version: http://www.astronet.ru/db/msg/1170672 (In
  {R}ussian)}\BibitemShut {NoStop}%
\bibitem [{\citenamefont {Bergmann}\ and\ \citenamefont
  {Brunings}(1949)}]{Bergmann_Brunings_1949}%
  \BibitemOpen
  \bibfield  {author} {\bibinfo {author} {\bibfnamefont {P.~G.}\ \bibnamefont
  {Bergmann}}\ and\ \bibinfo {author} {\bibfnamefont {J.~H.~M.}\ \bibnamefont
  {Brunings}},\ }\bibfield  {title} {\enquote {\bibinfo {title} {Non-linear
  field theories {I}{I}. {C}anonical equations and quantization},}\ }\href@noop
  {} {\bibfield  {journal} {\bibinfo  {journal} {Rev. Mod. Phys.}\ }\textbf
  {\bibinfo {volume} {21}},\ \bibinfo {pages} {480--487} (\bibinfo {year}
  {1949})}\BibitemShut {NoStop}%
\bibitem [{\citenamefont {Bergmann}\ \emph {et~al.}(1950)\citenamefont
  {Bergmann}, \citenamefont {Penfield}, \citenamefont {Schiller},\ and\
  \citenamefont {Zatzkis}}]{Bergmann_Penfield_Schiller_Zatzkis_1950}%
  \BibitemOpen
  \bibfield  {author} {\bibinfo {author} {\bibfnamefont {P.~G.}\ \bibnamefont
  {Bergmann}}, \bibinfo {author} {\bibfnamefont {R.}~\bibnamefont {Penfield}},
  \bibinfo {author} {\bibfnamefont {R.}~\bibnamefont {Schiller}}, \ and\
  \bibinfo {author} {\bibfnamefont {H.}~\bibnamefont {Zatzkis}},\ }\bibfield
  {title} {\enquote {\bibinfo {title} {The {H}amiltonian of the general theory
  of relativity with electromagnetic field},}\ }\href@noop {} {\bibfield
  {journal} {\bibinfo  {journal} {Phys. Rev.}\ }\textbf {\bibinfo {volume}
  {80}},\ \bibinfo {pages} {81--88} (\bibinfo {year} {1950})}\BibitemShut
  {NoStop}%
\bibitem [{\citenamefont {Regge}\ and\ \citenamefont
  {Teitelboim}(1974)}]{Regge_Teitelboim_1974_b}%
  \BibitemOpen
  \bibfield  {author} {\bibinfo {author} {\bibfnamefont {T.}~\bibnamefont
  {Regge}}\ and\ \bibinfo {author} {\bibfnamefont {C.}~\bibnamefont
  {Teitelboim}},\ }\bibfield  {title} {\enquote {\bibinfo {title} {Role of
  surface integrals in the {H}amiltonian formulation of general relativity},}\
  }\href@noop {} {\bibfield  {journal} {\bibinfo  {journal} {Ann. Phys.
  (N.Y.)}\ }\textbf {\bibinfo {volume} {88}},\ \bibinfo {pages} {286--318}
  (\bibinfo {year} {1974})}\BibitemShut {NoStop}%
\bibitem [{\citenamefont {Beig}\ and\ \citenamefont
  {Murchadha}(1987)}]{Beig_Murchadha_1987}%
  \BibitemOpen
  \bibfield  {author} {\bibinfo {author} {\bibfnamefont {R.}~\bibnamefont
  {Beig}}\ and\ \bibinfo {author} {\bibfnamefont {N.~O.}\ \bibnamefont
  {Murchadha}},\ }\bibfield  {title} {\enquote {\bibinfo {title} {The
  {P}oincar\'{e} group as the symmetry group of canonical general
  relativity},}\ }\href@noop {} {\bibfield  {journal} {\bibinfo  {journal}
  {Ann. Phys. (N.Y.)}\ }\textbf {\bibinfo {volume} {174}},\ \bibinfo {pages}
  {463--488} (\bibinfo {year} {1987})}\BibitemShut {NoStop}%
\bibitem [{\citenamefont {Jamsin}(2008)}]{Jamsin_2008}%
  \BibitemOpen
  \bibfield  {author} {\bibinfo {author} {\bibfnamefont {E.}~\bibnamefont
  {Jamsin}},\ }\bibfield  {title} {\enquote {\bibinfo {title} {A note on
  conserved charges of asymptotically flat and anti-de {S}itter spaces in
  arbitrary dimensions},}\ }\href@noop {} {\bibfield  {journal} {\bibinfo
  {journal} {Gen. Relativ. Gravit.}\ }\textbf {\bibinfo {volume} {40}},\
  \bibinfo {pages} {2569--2590} (\bibinfo {year} {2008})}\BibitemShut {NoStop}%
\bibitem [{\citenamefont {Solov'ev}(1985)}]{Soloviev_1985_en}%
  \BibitemOpen
  \bibfield  {author} {\bibinfo {author} {\bibfnamefont {V.~O.}\ \bibnamefont
  {Solov'ev}},\ }\bibfield  {title} {\enquote {\bibinfo {title} {Generator
  algebra of the asymptotic {P}oincare group in the general theory of
  relativity},}\ }\href@noop {} {\bibfield  {journal} {\bibinfo  {journal}
  {Theor. Math. Phys.}\ }\textbf {\bibinfo {volume} {65}},\ \bibinfo {pages}
  {1240--1249} (\bibinfo {year} {1985})}\BibitemShut {NoStop}%
\bibitem [{\citenamefont {Soloviev}(1993)}]{Soloviev_1993}%
  \BibitemOpen
  \bibfield  {author} {\bibinfo {author} {\bibfnamefont {V.~O.}\ \bibnamefont
  {Soloviev}},\ }\bibfield  {title} {\enquote {\bibinfo {title} {Boundary
  values as {H}amiltonian variables. {I}. {N}ew {P}oisson brackets},}\
  }\href@noop {} {\bibfield  {journal} {\bibinfo  {journal} {J. Math. Phys.}\
  }\textbf {\bibinfo {volume} {34}},\ \bibinfo {pages} {5747--5769} (\bibinfo
  {year} {1993})}\BibitemShut {NoStop}%
\bibitem [{\citenamefont {Soloviev}(1996)}]{Soloviev_1996_b}%
  \BibitemOpen
  \bibfield  {author} {\bibinfo {author} {\bibfnamefont {V.~O.}\ \bibnamefont
  {Soloviev}},\ }\bibfield  {title} {\enquote {\bibinfo {title} {Boundary terms
  and their {H}amiltonian dynamics},}\ }\href@noop {} {\bibfield  {journal}
  {\bibinfo  {journal} {Nucl. Phys. B (Proc. Suppl.)}\ }\textbf {\bibinfo
  {volume} {49}},\ \bibinfo {pages} {35--40} (\bibinfo {year}
  {1996})}\BibitemShut {NoStop}%
\bibitem [{\citenamefont {Soloviev}(1997{\natexlab{a}})}]{Soloviev_1997_a}%
  \BibitemOpen
  \bibfield  {author} {\bibinfo {author} {\bibfnamefont {V.~O.}\ \bibnamefont
  {Soloviev}},\ }\bibfield  {title} {\enquote {\bibinfo {title} {Difference
  between admissible and "differentiable" {H}amiltonians},}\ }\href@noop {}
  {\bibfield  {journal} {\bibinfo  {journal} {Phys. Rev. D}\ }\textbf {\bibinfo
  {volume} {55}},\ \bibinfo {pages} {7973--7976} (\bibinfo {year}
  {1997}{\natexlab{a}})}\BibitemShut {NoStop}%
\bibitem [{\citenamefont {Soloviev}(1997{\natexlab{b}})}]{Soloviev_1997_b_en}%
  \BibitemOpen
  \bibfield  {author} {\bibinfo {author} {\bibfnamefont {V.~O.}\ \bibnamefont
  {Soloviev}},\ }\bibfield  {title} {\enquote {\bibinfo {title} {The algebra
  independent of boundary conditions in the {A}shtekar formalism},}\
  }\href@noop {} {\bibfield  {journal} {\bibinfo  {journal} {Theor. Math.
  Phys.}\ }\textbf {\bibinfo {volume} {112}},\ \bibinfo {pages} {906--921}
  (\bibinfo {year} {1997}{\natexlab{b}})}\BibitemShut {NoStop}%
\bibitem [{\citenamefont {Soloviev}(1999)}]{Soloviev_1999}%
  \BibitemOpen
  \bibfield  {author} {\bibinfo {author} {\bibfnamefont {V.~O.}\ \bibnamefont
  {Soloviev}},\ }\bibfield  {title} {\enquote {\bibinfo {title} {Black hole
  entropy from {P}oisson brackets: Demystification of some calculations},}\
  }\href@noop {} {\bibfield  {journal} {\bibinfo  {journal} {Phys. Rev. D}\
  }\textbf {\bibinfo {volume} {61}},\ \bibinfo {pages} {027502 [4 pp]}
  (\bibinfo {year} {1999})}\BibitemShut {NoStop}%
\bibitem [{\citenamefont {Soloviev}(2002{\natexlab{a}})}]{Soloviev_2002_a}%
  \BibitemOpen
  \bibfield  {author} {\bibinfo {author} {\bibfnamefont {V.~O.}\ \bibnamefont
  {Soloviev}},\ }\bibfield  {title} {\enquote {\bibinfo {title} {Boundary
  values as {H}amiltonian variables. {I}{I}. {G}raded structures},}\
  }\href@noop {} {\bibfield  {journal} {\bibinfo  {journal} {J. Math. Phys.}\
  }\textbf {\bibinfo {volume} {43}},\ \bibinfo {pages} {3636--3654} (\bibinfo
  {year} {2002}{\natexlab{a}})}\BibitemShut {NoStop}%
\bibitem [{\citenamefont {Soloviev}(2002{\natexlab{b}})}]{Soloviev_2002_b}%
  \BibitemOpen
  \bibfield  {author} {\bibinfo {author} {\bibfnamefont {V.~O.}\ \bibnamefont
  {Soloviev}},\ }\bibfield  {title} {\enquote {\bibinfo {title} {Boundary
  values as {H}amiltonian variables. {I}{I}{I}. {I}deal fluid with a free
  surface},}\ }\href@noop {} {\bibfield  {journal} {\bibinfo  {journal} {J.
  Math. Phys.}\ }\textbf {\bibinfo {volume} {43}},\ \bibinfo {pages}
  {3655--3675} (\bibinfo {year} {2002}{\natexlab{b}})}\BibitemShut {NoStop}%
\bibitem [{\citenamefont {Belinfante}(1939)}]{Belinfante_1939}%
  \BibitemOpen
  \bibfield  {author} {\bibinfo {author} {\bibfnamefont {F.~J.}\ \bibnamefont
  {Belinfante}},\ }\bibfield  {title} {\enquote {\bibinfo {title} {On the spin
  angular momentum of mesons},}\ }\href@noop {} {\bibfield  {journal} {\bibinfo
   {journal} {Physica (Utrecht)}\ }\textbf {\bibinfo {volume} {6}},\ \bibinfo
  {pages} {887--898} (\bibinfo {year} {1939})}\BibitemShut {NoStop}%
\bibitem [{\citenamefont {Belinfante}(1940)}]{Belinfante_1940}%
  \BibitemOpen
  \bibfield  {author} {\bibinfo {author} {\bibfnamefont {F.~J.}\ \bibnamefont
  {Belinfante}},\ }\bibfield  {title} {\enquote {\bibinfo {title} {On the
  current and the density of the electric charge, the energy, the linear
  momentum and the angular momentum of arbitrary fields},}\ }\href@noop {}
  {\bibfield  {journal} {\bibinfo  {journal} {Physica (Utrecht)}\ }\textbf
  {\bibinfo {volume} {7}},\ \bibinfo {pages} {449--474} (\bibinfo {year}
  {1940})}\BibitemShut {NoStop}%
\bibitem [{Note4()}]{Note4}%
  \BibitemOpen
  \bibinfo {note} {Remark that Belinfante himself \cite {Belinfante_1939}
  credits Dr.~Podo1anski for construction of the symmetrization
  procedure}\BibitemShut {NoStop}%
\bibitem [{\citenamefont {Petrov}\ and\ \citenamefont
  {Katz}(1999)}]{Petrov_Katz_1999}%
  \BibitemOpen
  \bibfield  {author} {\bibinfo {author} {\bibfnamefont {A.~N.}\ \bibnamefont
  {Petrov}}\ and\ \bibinfo {author} {\bibfnamefont {J.}~\bibnamefont {Katz}},\
  }\bibfield  {title} {\enquote {\bibinfo {title} {Conservation laws for large
  perturbations on general backgrounds},}\ }in\ \href@noop {} {\emph {\bibinfo
  {booktitle} {Fundamental Interactions: From Symmetries to Black Holes}}},\
  \bibinfo {editor} {edited by\ \bibinfo {editor} {\bibfnamefont {J.~M.}\
  \bibnamefont {Frere}}, \bibinfo {editor} {\bibfnamefont {M.}~\bibnamefont
  {Henneaux}}, \bibinfo {editor} {\bibfnamefont {A.}~\bibnamefont {Sevrin}}, \
  and\ \bibinfo {editor} {\bibfnamefont {P.}~\bibnamefont {Spindel}}}\
  (\bibinfo  {publisher} {Universite de Bruxels},\ \bibinfo {address}
  {Belgium},\ \bibinfo {year} {1999})\ pp.\ \bibinfo {pages} {147--157},\
  \bibinfo {note} {e-print arXiv:gr-qc/9905088}\BibitemShut {NoStop}%
\bibitem [{\citenamefont {Petrov}\ and\ \citenamefont
  {Katz}(2002)}]{Petrov_Katz_2002}%
  \BibitemOpen
  \bibfield  {author} {\bibinfo {author} {\bibfnamefont {A.~N.}\ \bibnamefont
  {Petrov}}\ and\ \bibinfo {author} {\bibfnamefont {J.}~\bibnamefont {Katz}},\
  }\bibfield  {title} {\enquote {\bibinfo {title} {Conserved currents,
  superpotentials and cosmological perturbations},}\ }\href@noop {} {\bibfield
  {journal} {\bibinfo  {journal} {Proc. Roy. Soc. (London) A}\ }\textbf
  {\bibinfo {volume} {458}},\ \bibinfo {pages} {319--337} (\bibinfo {year}
  {2002})},\ \bibinfo {note} {e-print arXiv:gr-qc/9911025}\BibitemShut
  {NoStop}%
\bibitem [{\citenamefont {Bak}, \citenamefont {Cangemi},\ and\ \citenamefont
  {Jackiw}(1994)}]{Bak_Cangemi_Jackiw_1994}%
  \BibitemOpen
  \bibfield  {author} {\bibinfo {author} {\bibfnamefont {D.}~\bibnamefont
  {Bak}}, \bibinfo {author} {\bibfnamefont {D.}~\bibnamefont {Cangemi}}, \ and\
  \bibinfo {author} {\bibfnamefont {R.}~\bibnamefont {Jackiw}},\ }\bibfield
  {title} {\enquote {\bibinfo {title} {Energy-momentum conservation in gravity
  theories},}\ }\href@noop {} {\bibfield  {journal} {\bibinfo  {journal} {Phys.
  Rev. D}\ }\textbf {\bibinfo {volume} {49}},\ \bibinfo {pages} {5173--5181}
  (\bibinfo {year} {1994})}\BibitemShut {NoStop}%
\bibitem [{\citenamefont {Petrov}(2004)}]{Petrov_2004_b_en}%
  \BibitemOpen
  \bibfield  {author} {\bibinfo {author} {\bibfnamefont {A.~N.}\ \bibnamefont
  {Petrov}},\ }\bibfield  {title} {\enquote {\bibinfo {title} {Conserved
  currents in ${D}$-dimensional gravity and brane cosmology},}\ }\href@noop {}
  {\bibfield  {journal} {\bibinfo  {journal} {Moscow Univ. Phys. Bull.}\
  }\textbf {\bibinfo {volume} {59}},\ \bibinfo {pages} {11--15} (\bibinfo
  {year} {2004})},\ \bibinfo {note} {e-print
  arXiv:gr-qc/0401085v2.}\BibitemShut {Stop}%
\bibitem [{\citenamefont {Petrov}(2009)}]{Petrov_2009_a}%
  \BibitemOpen
  \bibfield  {author} {\bibinfo {author} {\bibfnamefont {A.~N.}\ \bibnamefont
  {Petrov}},\ }\bibfield  {title} {\enquote {\bibinfo {title} {Three types of
  superpotentials for perturbations in the {E}instein–{G}auss–{B}onnet
  gravity},}\ }\href@noop {} {\bibfield  {journal} {\bibinfo  {journal} {Class.
  Quantum Grav.}\ }\textbf {\bibinfo {volume} {26}},\ \bibinfo {pages} {135010
  [16 pp]} (\bibinfo {year} {2009})},\ \bibinfo {note} {(Corrigendum, see
  Ref.~\cite{Petrov_2010_a})}\BibitemShut {NoStop}%
\bibitem [{\citenamefont {Petrov}(2010)}]{Petrov_2010_a}%
  \BibitemOpen
  \bibfield  {author} {\bibinfo {author} {\bibfnamefont {A.~N.}\ \bibnamefont
  {Petrov}},\ }\bibfield  {title} {\enquote {\bibinfo {title} {Corrigendum},}\
  }\href@noop {} {\bibfield  {journal} {\bibinfo  {journal} {Class. Quantum
  Grav.}\ }\textbf {\bibinfo {volume} {27}},\ \bibinfo {pages} {069801 [2 pp]}
  (\bibinfo {year} {2010})},\ \bibinfo {note} {(Corrigendum of
  Ref.~\cite{Petrov_2009_a})}\BibitemShut {NoStop}%
\end{thebibliography}%

\end{document}